\documentclass[journal,twoside,final,letterpaper,twocolumn]{IEEEtran}

\ifCLASSINFOpdf
\usepackage[pdftex]{graphicx}
\else

\fi

\usepackage{epstopdf}
\usepackage{epsfig}
\usepackage{float}
\usepackage{amssymb,amsbsy,subfigure,amsmath,amsfonts,multirow,color,url,bm,cite,cases,eqparbox}
\usepackage{algorithmic}
\usepackage{algorithm}
\usepackage[mathscr]{eucal}
\usepackage{enumitem}
\usepackage{amssymb}
\usepackage{soul}
\usepackage{xcolor}
\usepackage{bbding}

\newcommand{\tablesize}{\fontsize{6.5pt}{12pt}\selectfont}

\def\Xcal{\mathcal{X}}
\def\Ycal{\mathcal{Y}}
\def\Scal{\mathcal{S}}
\def\Ecal{\mathcal{E}}
\def\Fcal{\mathcal{F}}
\def\Ccal{\mathcal{C}}
\def\Acal{\mathcal{A}}
\def\Hcal{\mathcal{H}}
\def\Rcal{\mathcal{R}}

\def\A{\mathbf{A}}
\def\B{\mathbf{B}}

\def\F{\mathbf{F}}

\def\H{\mathbf{H}}
\def\K{\mathbf{K}}

\def\Q{\mathbf{Q}}

\def\S{\mathbf{S}}

\def\V{\mathbf{V}}
\def\W{\mathbf{W}}

\def\b{\mathbf{b}}
\def\c{\mathbf{c}}
\def\d{\mathbf{d}}

\def\r{\mathbf{r}}

\def\u{\mathbf{u}}
\def\v{\mathbf{v}}

\def\x{\mathbf{x}}

\def\cited#1{{\iffalse #1 \fi}}
\def\1#1{\textcolor{red}{\textbf{#1}}}
\def\2#1{\textcolor{blue}{\textbf{#1}}}
\def\3#1{\textcolor{green}{\textbf{#1}}}

\begin{document}
	
	\title{Hyperspectral Image Denoising via Spatial-Spectral Recurrent Transformer}


	\author{Guanyiman Fu, Fengchao Xiong, \IEEEmembership{Member,~IEEE}, Jianfeng Lu, \IEEEmembership{Member,~IEEE}, Jun Zhou, \IEEEmembership{Senior Member,~IEEE}, Jiantao Zhou, \IEEEmembership{Senior Member,~IEEE}, and Yuntao Qian, \IEEEmembership{Senior Member,~IEEE}

\thanks {This work was supported in part by the National Natural Science Foundation of China under Grant 62002169, 62071421  and 62371237 and  the Fundamental Research Funds for the Central Universities Under Grant No.30923010213. (Corresponding author: Fengchao Xiong.)}
\thanks{Guanyiman Fu and Jianfeng Lu are with the School of Computer Science and Engineering, Nanjing University of Science and Technology, Nanjing 210094, China (e-mail: lujf@njust.edu.cn). }
\thanks{Fengchao Xiong is  with the School of Computer Science and Engineering, Nanjing University of Science and Technology, Nanjing 210094,  China and also  with the State Key Laboratory of Internet of Things for Smart City, Department of Computer and Information Science, University of Macau,  Macau 999078, China (e-mail: fcxiong@njust.edu.cn). }
\thanks{Jun Zhou is with the School of Information and Communication Technology, Griffith University, Nathan, Australia (e-mail: jun.zhou@griffith.edu.au).}
\thanks{Jiantao Zhou is with the State Key Laboratory of Internet of Things for
Smart City, Department of Computer and Information Science, University of
Macau, Macau 999078, China (e-mail: jtzhou@umac.mo).}

\thanks{Yuntao Qian  is with the College of Computer Science, Zhejiang University, Hangzhou 310027, China (e-mail: ytqian@zju.edu.cn).}
}
\maketitle
\begin{abstract}
Hyperspectral images (HSIs) often suffer from noise arising from both intra-imaging mechanisms and environmental factors. Leveraging domain knowledge specific to HSIs, such as global spectral correlation (GSC) and non-local spatial self-similarity (NSS), is crucial for effective denoising. Existing methods tend to independently utilize each of these knowledge components with multiple blocks, overlooking the inherent 3D nature of HSIs where domain knowledge is strongly interlinked, resulting in suboptimal performance. To address this challenge, this paper introduces a spatial-spectral recurrent transformer U-Net (SSRT-UNet) for HSI denoising. The proposed SSRT-UNet integrates NSS and GSC properties within a single SSRT block. This block consists of a spatial branch and a spectral branch. The spectral branch employs a combination of transformer and recurrent neural network to perform recurrent computations across bands, allowing for GSC exploitation beyond a fixed number of bands. Concurrently, the spatial branch encodes NSS for each band by sharing \emph{keys} and \emph{values} with the spectral branch under the guidance of GSC. This interaction between the two branches enables the joint utilization of NSS and GSC, avoiding their independent treatment. Experimental results demonstrate that our method outperforms several alternative approaches.  The source code will be available at https://github.com/lronkitty/SSRT.
  \end{abstract}

	%

\section{Introduction}
Hyperspectral images (HSIs), being 3D data, offer not only spatial but also detailed material information due to the captured continuous spectral band. This wealth of material information significantly enhances the discriminative capabilities of HSIs, leading to their widespread application in various tasks such as object detection~\cite{Yang2023,Gao2023,Dong2023}, tree species  classification~\cite{Li2024}, and change detection~\cite{Luo2023}. However, during the acquisition of hyperspectral data, various factors, such as insufficient exposure time, mechanical vibrations of imaging platforms, atmospheric perturbations, stochastic errors in photon counting, and other intrinsic and extrinsic elements, inevitably introduce noise to the images. Therefore, HSI denoising becomes an essential preprocessing step to enhance the quality and practical value of HSIs.


Domain knowledge of HSIs is inherently advantageous for achieving a superior  HSI recovery. Global spectral correlation (GSC) and non-local spatial self-similarity (NSS) are the most popularly used knowledge~\cite{Pang2022,Wang2022}. The GSC refers to the correlation between the spectral bands of an HSI, while the NSS measures the similarity between different spatial patches within the HSI.  Given the tridimensional nature of HSIs,  the spatial and spectral information is tightly coupled within HSIs. Therefore, it is imperative to exploit both the GSC and NSS jointly for the accurate recovery of the HSIs.

\begin{figure}
   \centering
\includegraphics[width=0.9\linewidth]{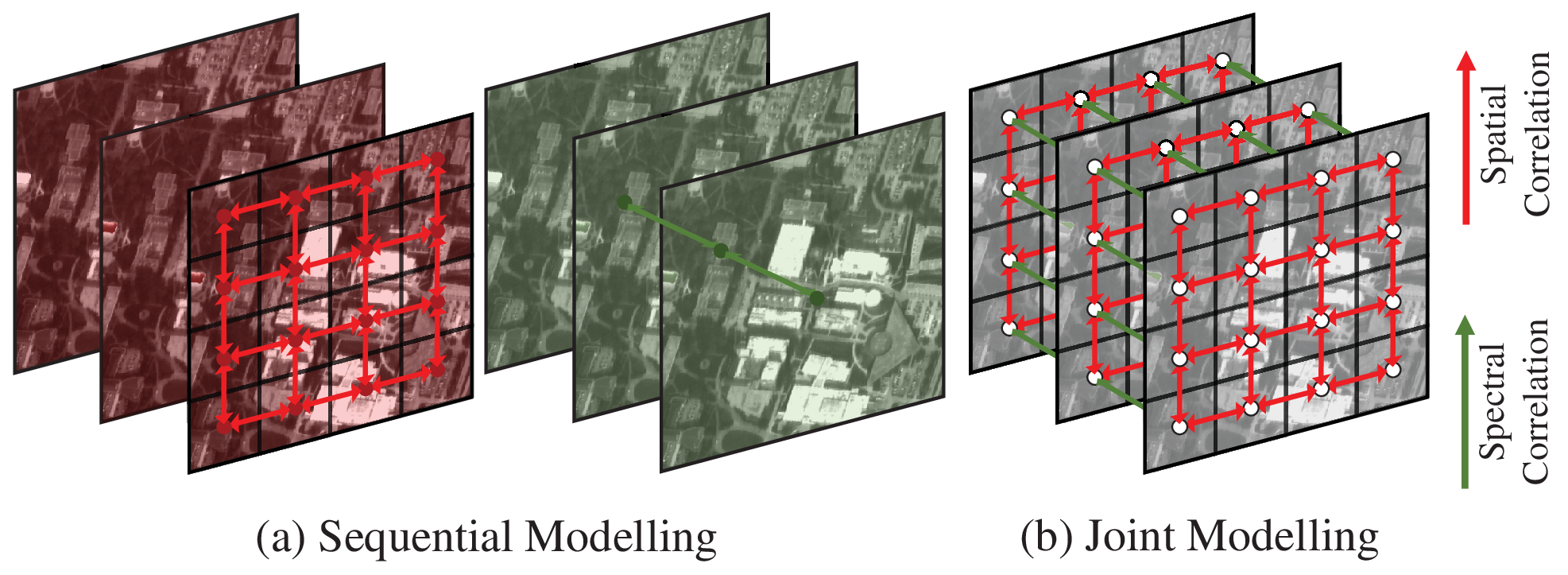}
   \caption{Comparison of the modelling schemes of the existing methods (a) and ours (b). The existing methods use separate modules to explore NSS (non-local spatial self-similarity) and GSC (global spectral correlation) in an alternating and sequential manner. Our proposed method employs a single spatial-spectral recurrent transformer block for joint spatial-spectral modelling of the GSC and NSS to improve the recovery of HSIs.}\label{fig:interact}
\end{figure}

Unfortunately, existing methodologies exhibit a simplistic approach by modelling spatial and spectral properties in a sequential and alternating fashion, as depicted in Fig.~\ref{fig:interact} (a).  Constrained by the local nature of convolution, Convolutional Neural Networks (CNNs) resort to utilizing separate modules, such as Recurrent Neural Networks (RNNs), attention mechanisms, and multi-scale units, concatenated to encapsulate the aforementioned domain knowledge~\cite{wang2019_2,xiong2022smds,yuan2019,Wei2021}.  In the pursuit of global context awareness, Transformers have demonstrated remarkable efficacy across various computer vision tasks~\cite{Liu2021,tang2022,Liu2022}.  Nonetheless, prevailing methodologies employing Transformers still rely on  separate modules for  distinct domain knowledge, sequentially assembling multiple transformer modules for spatial and spectral characterization~\cite{Chen2022_hider,li2022spatial}.  This sequential exploitation necessitates a perpetual trade-off between diverse knowledge domains, introducing complexity to the learning process and rendering it challenging to comprehensively capture the intrinsic characteristics of  HSIs. In addition, the shortcomings of the transformers in remembering past states~\cite{Dai2019}  limit them to handle only a predetermined fixed number of bands after the network trained. Given the high cost of acquiring HSIs, it is always desirable that a single network that can handle an arbitrary number of bands.


%


%

Against these issues, this paper introduces a novel SSRT-UNet  that strategically  exploits the NSS and GSC with a single spatial-spectral recurrent transformer (SSRT) block. The SSRT block comprises strongly interconnected non-local spatial correlation and global spectral correlation exploitation, as illustrated in Fig.~\ref{fig:interact}(b).  By treating HSIs as sequential data along bands, the spectral branch conceptualizes each band as a state within a Recurrent Neural Network (RNN) to model GSC. Under the framework of RNN, the spectral branch employs self-attention and cross-attention mechanisms to amalgamate the state from the preceding bands with spatial information from the spatial branch. This RNN-based transformer scheme adeptly facilitates the effective exploration of long-range spectral correlations beyond a fixed number of bands.  Concurrently sharing \emph{keys} and \emph{values} with the spectral branch, the spatial branch integrates the state from the previous band in the spectral branch with the information in the current band to exploit NSS. This enables the network to express the distinctive characteristics of each band while being guided by the spectral branch. Based on the SSRT block, we further construct an SSRT-UNet to exploit multi-scale information in HSIs.  Extensive experiments show that our SSRT-UNet achieves better  performance against the compared methods. The contribution of this paper can be summarised as follows:

\begin{itemize}
\item The SSRT-UNet  models GSC and NSS through a single SSRT block, where the spectral and spatial branches interact by sharing \emph{keys} and \emph{values}. This cohesive strategy enhances the spatial-spectral characterization of 3D HSIs.
\item By embedding the transformer within the RNN structure, our spectral branch  adeptly captures global spectral correlations in HSIs beyond predefined band limitations, a capability not realized by other transformer-based methods.
\item Utilizing the SSRT as the foundational module, we establish the SSRT-UNet for HSI denoising. Experiments show that the proposed SSRT-UNet achieves the state-of-the-art denoising.
\end{itemize}

The remainder of this paper is organized as follows: Section \ref{sec:related_work} reviews related works. Section \ref{sec:method} introduces the details of the proposed SSRT-UNet. Section~\ref{sec:experiments} presents quantitative and visual experiments, along with ablation studies. Finally, Section~\ref{sec:conclusion} concludes the paper.

\section{Related Work}\label{sec:related_work}

In this section, we provide a comprehensive review of HSI denoising, encompassing both model-based and deep learning-based methods. Additionally, we discuss the application of transformers for HSI modelling. 

\subsection{Hyperspectral Image Denoising}

Traditional model-based methods take advantage of sparse and low-rank representations to describe the physical properties of the underlying clean HSI for denoising~\cite{Liu2023,Fu2023,Zhang2014a,ye2014multitask,xiong2019, Zha2023,Su2023}.   As an extension of BM3D~\cite{Dabov2007}, BM4D~\cite{maggioni2012nonlocal} employs a collective sparse representation to delineate the similarity among non-local space-spectral cubes. Similarly, Peng~\emph{et al.}~\cite{Peng2014} introduced a tensor dictionary learning (TDL) method to capture the non-local spatial-spectral similarity, where the dictionary is learned adaptively from the data. The incorporation of an analysis-based hyper-Laplacian prior into the low-rank model, as demonstrated by LLRT~\cite{Chang2017}, facilitates the effective restoration of HSI while preserving its spectral structure. NGMeet~\cite{He2020} leverages a low-rank Tucker decomposition and a matrix rank minimization approach to articulate global spectral correlation and non-local spatial self-similarity, respectively. Zhuang \emph{et al.}\cite{zhuang2018fast} introduce the FastHyDe method, which fully exploits spatial-spectral representations in HSI linked with their sparse, low-rank, and self-similarity characteristics in subspaces. Similarly, Chen~\emph{et al.}~\cite{Chen2023} performed total variation in the spectral subspace and achieved fast HSI denoising. Enhanced 3DTV (E-3DTV)\cite{peng2020} discerns global correlations and local differences across all bands of HSIs by evaluating the sparsity of gradient maps within these bands in subspaces. By learning a spatial-spectral transform from the data,~\cite{Zhang2023a} introduced the  low-TR-rank regularizer and  sparse regularizer to characterize the global low-rankness and the sparsity of the transformed HSI, respectively for denoising. However, the model-based approaches usually require iterative optimization and manual fine-tuning of hyperparameters, rendering them computationally inefficient and possibly unsuitable for practical HSI denoising.

Deep learning has been a hotspot in HSI denoising, owing to its robust learning capabilities~\cite{Zhang2023, Maffei2020,zhang2021,Zhang2022,Miao2023}. Techniques involving 3D convolution and its variants~\cite{liu20193,Wang2022,Wei2021} have been instrumental in capturing spatial-spectral correlations for HSI denoising. QRNN3D~\cite{Wei2021} incorporates a recurrent unit following 3D convolutions to exploit both spatial-spectral and global spectral correlations. T3SC~\cite{Bodrito2021} and MAC-Net~\cite{xiong2021mac} transform the optimization of physical models of HSIs into neural architectures, enhancing denoising efficacy. Attention mechanism can guide the network focus on the predominant information and have also been used for HSI denoising~\cite{Xiao2023}. NSSNN~\cite{guanyiman2022} combines a recurrent unit with attention mechanisms to extract both global spectral correlation and non-local spatial features, achieving promising denoising performance. In contrast, approaches such as \cite{Wang2022} employ stacked band-wise CSWin transformer blocks to exploit non-local spatial self-similarity, neglecting global spectral correlation. Spectral and spatial transformer blocks introduced in~\cite{Chen2022_hider} and ~\cite{li2022spatial}, respectively, aim to unearth global spectral correlation and local-global spatial information. Nevertheless, the exploitation of non-local spatial self-similarity, global spectral correlation, and spatial-spectral correlation occurs independently through multiple blocks rather than being jointly addressed by a single block, resulting in suboptimal exploitation of 3D HSIs.

\begin{figure*}[ht]
   \centering
   \includegraphics[width=1\linewidth]{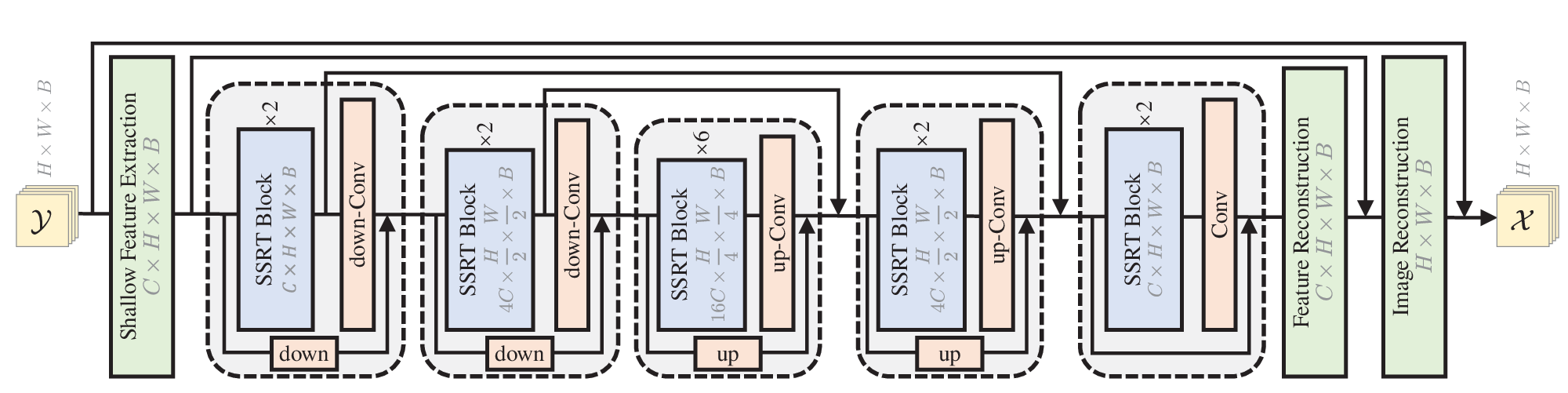}
   \caption{The overall architecture of the SSRT-UNet, where down and up denote $\text{downsampling}(\cdot)$ and $\text{upsampling}(\cdot)$, respectively.}\label{fig:ssrtnet}
 \end{figure*}

\subsection{Transformer for Hyperspectral Image Modelling}

Transformers have demonstrated notable success  in hyperspectral information processing, raning from classification~\cite{Hong2022,Zhang2023b,Qiu2023} to low-level recovery~\cite{ Wang2022a,Li2023,Chen2023a}. HSI inherently entails three-dimensional data, where all dimensions exhibit strong correlations. While 3D CNNs have traditionally been employed for HSI modelling, the 3D transformer~\cite{Liu_2022_CVPR} presents itself as an intuitive alternative. However, the computational intensity of the 3D transformer in HSIs, where certain pixels may be unrelated~\cite{cai2022}, poses a challenge.   TRQ3D~\cite{Pang2022} addresses this challenge by incorporating a shifted windowing scheme to model non-local spatial relations. Transformers also find applications in modelling spectral correlation~\cite{he2021}. HSI-BERT~\cite{he2020_hsi-bert} directly utilizes the bidirectional encoder representation from transformers (BERT) to encode semantic context-aware representations within each pixel, leveraging multiple bands to capture global spectral correlation. Approaches such as Hider~\cite{Chen2022_hider} and SST~\cite{li2022spatial} reshape the height and width dimensions into one, obtaining multiple spectral vectors. Spectral correlation is then derived by calculating attention across all spectral vectors. MST~\cite{Cai_2022_CVPR} employs a spatial transformer to guide the modelling of the spectral transformer, facilitating the comprehensive exploitation of spatial and spectral relationships. However, the absence of the memory mechanism in current transformers limits their ability to handle fixed-length contexts~\cite{Dai2019}. Consequently, current transformer-based methods can only model fixed-length spectral correlations once the network is trained. Effectively adapting transformers to capture global spectral correlations in HSIs beyond the predefined number of bands remains an unexplored frontier, aligning with one of the goals of our work.




\section{Method}\label{sec:method}
This section provides a detailed exposition of the proposed spatial-spectral recurrent transformer UNet (SSRT-UNet), including the overall architecture, the SSRT block,  the bidirectional SSRT and also the loss function.

\subsection{Overall Architecture}

 Let $\Xcal$ be a clean HSI with $H \times W$ pixels and $B$ bands. When degraded with additive noises $\Ecal$, such as the Gaussian noise, impulse noise, strips, and deadlines, the observed HSI $\Ycal$ can be mathematically modeled by
 \begin{equation}
 	\Ycal=\Xcal+\Ecal.    \label{eq:noise}
 \end{equation}
In this paper, we resort to SSRT-UNet to learn the mapping from $\Ycal$ to $\Xcal$ for noise reduction.

As illustrated in Fig.~\ref{fig:ssrtnet}, the SSRT-UNet exhibits a U-shaped architecture, encompassing a shallow feature extraction module, a multi-scale encoder-decoder module, a feature reconstruction module, and an image reconstruction module. To enhance denoising performance and facilitate network training, residual connections are incorporated. The shallow feature extraction module employs 3D convolution to derive initial low-level features denoted as $\Fcal \in \mathbb{R}^{C\times B \times H\times W}$ from the HSI $\Ycal$, where $C$ represents the number of channels. Subsequently, the extracted features are input into the encoder-decoder modules comprising five layers. Each layer contains multiple SSRT blocks to progressively extract deeper spatial-spectral features via
\begin{equation}
   \begin{alignedat}{2}
 &\Fcal^{i+1}&=&~\text{SSRT-Block}(\Fcal^{i}), ~~~i=0,1, \cdots, L-1,\\
 &\Fcal_{out}&=&~G(\text{Conv}(\Fcal^{L}))+G(\Fcal^{0}), \label{eq:one_layer}
 \end{alignedat}
 \end{equation} 
where $L$ is the total number of the SSRT blocks in a layer and set as $[2,2,6,2,2]$ from the first to the fifth layer. 
The operation $G(\cdot)$ represents either downsampling or upsampling and is excluded in the last decoder. Following~\cite{Liang_2021_ICCV}, the convolution operation $\text{Conv}(\cdot)$ is implemented before the residual connection in each layer to enjoy the shift-invariance of the convolution, which enhances the translational equivalence of the network. Upon completion of the encoder-decoder stages, the spatial-spectral features obtained undergo filtering by a 3D convolution within the feature reconstruction module. The outcomes of this module are then combined with the shallow features. Subsequently, the image reconstruction module employs a 3D convolution to  recover the clean HSI $\Xcal$ from the aggregated features. In the subsequent sections, we elaborate on the details of the introduced SSRT block.

\begin{figure*}[ht]
   \centering
   \includegraphics[width=1\linewidth]{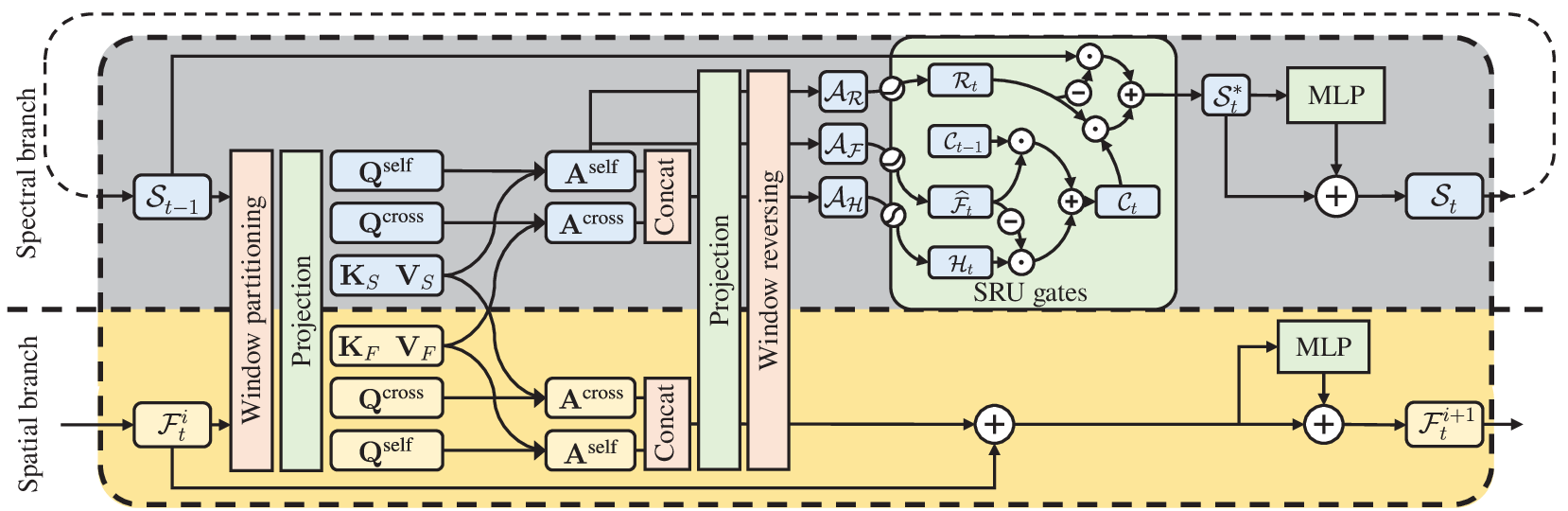}
   \caption{The structure of a typical SSRT. It consists of a spectral branch and a spatial branch, which interact with each other to achieve joint exploitation of the spatial and spectral properties of the HSI.}\label{fig:ssrt}
 \end{figure*}

\subsection{Spatial-Spectral Recurrent Transformer Block}\label{sec:ssrt}

Fig.~\ref{fig:ssrt} illustrates our proposed SSRT block when processing the $t$-th band, including a spectral branch and a spatial branch.  In the following, we introduce two branches in more detail. 

\subsubsection{Spectral  Branch} The spectral branch conceptualizes HSIs as sequential data along bands, integrating the transformer architecture into the recurrent neural network (RNN) framework to harness their combined advantages for capturing long-range spectral correlations.  Transformers excel in learning dependencies across multiple tokens yet lack the ability to retain the memory of past context, resulting in a focus on a fixed number of tokens and limited handling of sequences with predetermined lengths post-training. Conversely, RNNs condense the entire preceding sequence into a single state vector but encounter modelling constraints when processing extensive sequences.  

Within the proposed spectral branch, the transformer collaborates with the RNN to enhance the encoding of each state related to the preceding bands. This  enables the spectral branch to encode each state more effectively while facilitating the flow of correlation information across all bands, thus enabling a more robust modelling of global spectral correlation. Simultaneously, the RNN contributes to modelling spectral correlation within HSIs with varying numbers of bands, which is always desirable for HSI denoising, especially when it is costly to acquire numerous training HSIs with the same number of bands as the HSI to be recovered.

%



As shown in Fig.~\ref{fig:ssrt},  the spectral branch models the  correlation from band 1 to $t-1$ as the state $\Scal_{t-1}\in\mathbb{R}^{C \times H \times W}$ in a RNN.  Given $\Scal_{t-1}$ and the spatial feature map $\Fcal_{t}$ at the current band,  the spectral branch processes them in the window partitioning under the shifted windowing scheme \cite{Liu2021}. Specifically, $\frac{HW}{M^2}$ pairs of non-overlapping patches with size of $M^2\times C$ are extracted from  $\Scal_{t-1}$ and $\Fcal_t$, \emph{i.e.}, $\{\S_n, \F_n\}_{n=1, \cdots ,N}$, where $N=\frac{HW}{M^2}$. The patches are portioned in the same way so that the  patches in $\Scal_{t-1}$ and $\Fcal_t$  are spatially aligned. In the following, $n$ is dropped out for simplicity. Each patch pair  is linearly projected to yield the \emph{queries}, \emph{keys}, and \emph{values}, \emph{i.e.},
\begin{alignat}{6}
   &\Q^{\text{cross}}&=&~\S\W_{\text{cross}}^{Q}~&,&~&\Q^{\text{self}}&=&~\S&\W_{\text{self}}^{Q}&, \label{eq:qsqs}\\
	&\K_S&=&~\S\W_S^{K}&,&~&\V_S&=&~\S&\W_S^{V}&, \label{eq:ksvs}\\
   &\K_F&=&~\F\W_F^{K}&,&~&\V_F&=&~\F&\W_F^{V}&, \label{eq:kfvf}
\end{alignat}
where  $\{\W_{\text{cross}}^{Q}, \W_{\text{self}}^{Q}, \W_S^{K}, \W_S^{V}, \W_F^{K}, \W_F^{V} \} \in \mathbb{R}^{C\times \frac{C}{2}}$ are the learnable projection matrices. 

We then use the cross-attention to combine  $\F$ and $\S$, \emph{i.e.},
\begin{equation}
 	\A^\text{cross}=\text{softmax}( \Q^{\text{cross}}\K^\top_F/\sqrt{d}+\B^\text{cross})\V_F, \label{eq:cross}
\end{equation}
where  $\B^\text{cross}$ is the learnable relative positional encoding. It is well-known that each band in a HSI has the NSS property. When linked by GSC, the spectral correlation state $\Scal_{t-1}$ should still retain the NSS property. To this end, we adopt the self-attention to enhance its  non-local spatial structure  via
\begin{equation}	\A^\text{self}=\text{softmax}(\Q^{\text{self}}\K^\top_S/\sqrt{d}+\B^\text{self})\V_S,\\ \label{eq:self}
\end{equation}
where $\B^\text{self}$ is also the learnable relative positional encoding. $\{\Q^{\text{self}}, \K_S, \V_S\}$  are \emph{query}, \emph{key}, and \emph{value} matrices generated from $\S$.

Based on $\A^\text{cross}$ and $\A^\text{self}$, we can obtain the intermediate spectral correlation information  $\Acal_\Hcal \in \mathbb{R}^{C \times H \times W}$ via
\begin{equation}
	\Acal_\Hcal =\text{WR}(\{\text{cat}(\A_n^\text{self}, \A_n^\text{cross})\W_\Hcal\}), n=1,\cdots ,N, \label{eq:fuse} 
\end{equation}
where $\W_\Hcal \in \mathbb{R}^{C\times C}$ is  a  learnable projection matrix to fuse self-attention and cross-attention further.  $\text{WR}(\cdot)$ is the window reversing operator to  put the extracted patches back to the original position.

As described in \cite{greff2017,lei2018simple,zhang2020}, gates ensure effective long-range modelling in recurrent neural networks. Therefore, we implement modified simple recurrent unit (SRU) gates \cite{lei2018simple}  to  encode the GSC, \emph{i.e.},
 \begin{alignat}{2}
    &\Hcal_t &&= \tanh(\Acal_\Hcal),\\
    &\widehat{\Fcal}_t &&= \tanh(\text{Relu}(\Acal_\Fcal)),\label{subequ:sru_H}\\
    \Ccal_t =& \widehat{\Fcal}_t &&\odot \Ccal_{t-1} + (1-\widehat{\Fcal}_t) \odot \Hcal_t,\label{subequ:light_recurrence}\\
    &\Rcal_t &&= \tanh(\text{Relu}(\Acal_\Rcal)),\\
    \Scal^*_t =& \Rcal_t &&\odot \Ccal_t + (1-\Rcal_t) \odot \Scal_{t-1},\label{subequ:highway_network}
 \end{alignat}
 where $\widehat{\Fcal}_t$ and $\Rcal_t$ are the forget gate and the reset gate, respectively. Here $ \Acal_\Fcal = \text{WR}(\{\A_n^\text{self}\W_\Fcal\})$ and $ \Acal_\Rcal = \text{WR}(\{\A_n^\text{self}\W_\Rcal\})$ with   $\{ \W_\Fcal, \W_\Rcal\} \in \mathbb{R}^{\frac{C}{2} \times C}$  being the projection matrices  to make the dimension of $\{\Acal_\Hcal, \Acal_\Fcal, \Acal_\Rcal\} \in \mathbb{R}^{C\times H\times W}$ match.   Eq.~(\ref{subequ:light_recurrence}) formulates a light recurrence to selectively combine the correlation information between states in the previous $t$$\sim$$1$ bands ($\Ccal_{t-1}$) with $\Hcal_t$. Eq.~(\ref{subequ:highway_network}) formulates a highway network to refine the light recurrence and handles scenarios where there are significant variations in the data of nearing bands. In processing the $t$-th band, we can obtain an updated state $\Scal_t^*$ containing the spectral correlation within bands 1$\sim$$t$. By processing all the bands with the spectral branch,  the information can be  propagated over long distances to depict the global spectral correlation accurately.

 Following the Swin Transformer~\cite{Liu2021}, a  multilayer perceptron (MLP) with residual connection is applied on $\Scal^*_{t}$  for further feature transformations, \emph{i.e.},
\begin{equation}
\begin{alignedat}{2}
   &\Scal_t &&= \text{MLP}(\Scal^*_{t})+ \Scal^*_{t},\\
\end{alignedat}
\end{equation}
where $\Scal_t$ is the updated state for the next band.




\subsubsection{Spatial Branch}  
HSIs are linked by the global spectral correlation but differentiated by the band-specific properties. Therefore, we introduce a spatial branch, which focuses on encoding the feature map of each band under the guidance of the spectral correlation information achieved in the spectral branch. Like the spectral  branch, the spatial branch also takes  $\Fcal_t$ and $\Scal_{t-1}$ as the input. For patches $\F$ in the spatial branch,  we compute the \emph{queries} with 
\begin{equation}
   \quad \Q^{\text{cross}}=\F\W_{\text{cross}}^{Q}~,\Q^{\text{self}}=\F\W_{\text{self}}^{Q},
\end{equation}
where $\{\W_{\text{cross}}^{Q}, \W_{\text{self}}^{Q}\}$ are projection matrices  size of $C\times \frac{C}{2}$. The \emph{keys}, \emph{i.e.}, $\{\K_S, \K_F\}$, and \emph{values}, \emph{i.e.}, $\{\V_S, \V_F\}$,  are shared with the spectral  branch. In this way, the spectral branch and the spatial branch are tightly coupled rather than independent, more effectively capturing the spatial and spectral properties of the 3D HSI.


 Based on $\{\Q^{\text{self}}, \K_F, \V_F\}$ from  $\F$, the self-attention $\A^\text{self}$ is built to exploit NSS within the current patch using Eq. ~(\ref{eq:self}). With $\Q^{\text{cross}}$  and $\{\K_S, \V_S\}$, we can compute the cross-attention map $\A^\text{cross}$ with  Eq. ~(\ref{eq:cross}) to further guide the learning of the NSS property.  After $\A^\text{cross}$ and $\A^\text{self}$  obtained, we can obtain the intermediate spatial feature map $\Fcal^{'} \in \mathbb{R}^{C \times H \times W}$ via
\begin{equation}
	\Fcal^{'} =\text{WR}(\{\text{cat}(\A_n^\text{self}, \A_n^\text{cross})\W_\Fcal\}), n=1,\cdots ,N, 
\end{equation}
where $\W_\Fcal \in \mathbb{R}^{C\times C}$ is  a  learnable projection matrix.  Finally, the spatial feature map of the current band $\Fcal_t$ is updated by
\begin{equation}
\begin{split}
	   &\Fcal^*_t = \Fcal^{'}_t + \Fcal_t,\\
	   &\Fcal_t^{i+1}=\text{MLP}(\Fcal^*_t)+\Fcal^*_t.
\end{split}
\end{equation}


\begin{table*}[htbp]
   \caption{Comparison of Different Methods on 50 Testing HSIs from ICVL Dataset. The Top Three Values Are Marked as \1{Red}, \2{Blue}, and \3{Green}.}\label{tab:icvl}
   \centering
   \resizebox{\linewidth}{!}{\tablesize{
   \begin{tabular}{c|c|c|c|c|c|c|c|c|c|c|c|c|c|c|c|c}
      \hline
   &&&\multicolumn{8}{c|}{\textbf{Model-based methods}}&\multicolumn{5}{c}{\textbf{Deep learning-based methods}}\\
   \hline
   \multirow{2}*{\makebox[0.02\textwidth][c]{$\sigma$}}&\multirow{2}*{\makebox[0.02\textwidth][c]{Index}}&\multirow{2}*{\makebox[0.032\textwidth][c]{Noisy}}&\multirow{1}*{\makebox[0.032\textwidth][c]{BM4D}}&\multirow{1}*{\makebox[0.032\textwidth][c]{MTSNMF}}&\multirow{1}*{\makebox[0.032\textwidth][c]{LLRT}}&\multirow{1}*{\makebox[0.032\textwidth][c]{NGMeet}}&\multirow{1}*{\makebox[0.032\textwidth][c]{LRMR}}&\multirow{1}*{\makebox[0.032\textwidth][c]{FastHyDe}}&\multirow{1}*{\makebox[0.032\textwidth][c]{LRTF$L_0$}}&\multirow{1}*{\makebox[0.032\textwidth][c]{E-3DTV}}&\multirow{1}*{\makebox[0.032\textwidth][c]{T3SC}}&\multirow{1}*{\makebox[0.032\textwidth][c]{MAC-Net}}&\multirow{1}*{\makebox[0.032\textwidth][c]{NSSNN}}&\multirow{1}*{\makebox[0.032\textwidth][c]{TRQ3D}}&\multirow{1}*{\makebox[0.032\textwidth][c]{SST}}&\makebox[0.04\textwidth][c]{\textbf{SSRT-UNet}}\\
   &&&\cite{maggioni2012nonlocal}&\cite{ye2014multitask}&\cite{Chang2017}&\cite{He2020}&\cite{Zhang2014a}& \cite{zhuang2018fast}&\cite{xiong2019}& \cite{peng2020}& \cite{Bodrito2021}& \cite{xiong2021mac}& \cite{guanyiman2022}& \cite{Pang2022}& \cite{li2022spatial}& (ours)\\
   \hline
   \multirow{3}*{\makebox[0.02\textwidth][c]{[0,15]}}
   & \makebox[0.02\textwidth][c]{PSNR$\uparrow$}& 33.18 & 44.39 & 45.39 & 45.74 & 39.63 & 41.50 & 48.08 &43.41& 46.05   & 49.68  & 48.21   & \3{49.83}  & 46.43  & \2{50.87}  & \1{52.12}  \\
   & \makebox[0.02\textwidth][c]{SSIM$\uparrow$}& .6168  & .9683  & .9592  & .9657  & .8612  & .9356  & .9917 & .9315& .9811   & .9912 & .9915  & \3{.9934} & .9878 & \2{.9938} & \1{.9950} \\
   & \makebox[0.02\textwidth][c]{SAM$\downarrow$} & .3368  & .0692  & .0845  & .0832  & .2144  & .1289  & .0404  & .0570	& .0560   & .0486 & .0387  & \3{.0302} & .0437 & \2{.0298} & \1{.0225}\\
   \hline 		
   \multirow{3}*{\makebox[0.02\textwidth][c]{[0,55]}}
   & \makebox[0.02\textwidth][c]{PSNR$\uparrow$}& 21.72 & 37.63 & 38.02 & 36.80 & 31.53 & 31.50 & 42.86 &35.63 &	40.20   & 45.15  & 43.74   & \3{46.27}  & 44.64  & \2{46.39}  & \1{47.85}  \\
   & \makebox[0.02\textwidth][c]{SSIM$\uparrow$}& .2339  & .9008  & .8586  & .8285  & .6785  & .6233  & .9800 & .8125& .9505   & .9810 & .9768  & \3{.9868} & .9840 & \2{.9872} & \1{.9894} \\
   & \makebox[0.02\textwidth][c]{SAM$\downarrow$} & .7012  & .1397  & .234   & .2316  & .4787  & .3583  & .063  & .1914	& .0993   & .0652 & .0582  & \2{.0393} & .0487 & \3{.0457} & \1{.0319}\\
   \hline
   \multirow{3}*{\makebox[0.02\textwidth][c]{[0,95]}}
   & \makebox[0.02\textwidth][c]{PSNR$\uparrow$}& 17.43 & 34.71 & 34.81 & 31.89 & 27.62 & 27.00 & 40.84 &32.83	&37.80   & 43.10  & 41.24   & \3{44.42}  & 43.54  & \2{44.83}  & \1{46.20}  \\
   & \makebox[0.02\textwidth][c]{SSIM$\uparrow$}& .1540   & .8402  & .7997  & .6885  & .5363  & .4208  & .9734  & .7482	& .9279   & .9734 & .9577  & \3{.9809} & .9806 & \2{.9838} & \1{.9862} \\
   & \makebox[0.02\textwidth][c]{SAM$\downarrow$} & .8893  & .1906  & .3266  & .3444  & .642   & .5142  & .0771  & .3014	& .1317   & .0747 & .0841  & .0524 & \3{.0523} & \2{.0513} & \1{.0375} \\
   \hline 	
   \multirow{3}*{\makebox[0.02\textwidth][c]{Mixture}}
   & \makebox[0.02\textwidth][c]{PSNR$\uparrow$}& 13.21 & 23.36 & 27.55 & 18.23 & 23.61 & 23.10 & 27.58 &30.93	&34.90   & 34.09  & 28.44   & \2{40.54}  & \3{39.73}  & 39.22  & \1{42.52}  \\
   & \makebox[0.02\textwidth][c]{SSIM$\uparrow$}& .0841  & .4275  & .6743  & .1731  & .4448  & .3463  & .7250 & .8378	& .9041   & .9052 & .7393  & \3{.9560} & .9491 & \1{.9626} & \2{.9570} \\
   & \makebox[0.02\textwidth][c]{SAM$\downarrow$} & .9124  & .5476  & .5326  & .6873  & .6252  & .5144  & .4534 & .3613	& .1468   & .2340 & .4154  & .1097 & \2{.0869} & \1{.0743} & \3{.0986}\\
   \hline
   \end{tabular}}}
   \end{table*}

%
%

%
%
\subsection{Bidirectional SSRT and Shifted Windowing SSRT}\label{sec:bi-ssrt}


\begin{figure}[ht]
   \centering
   \includegraphics[width=1\linewidth]{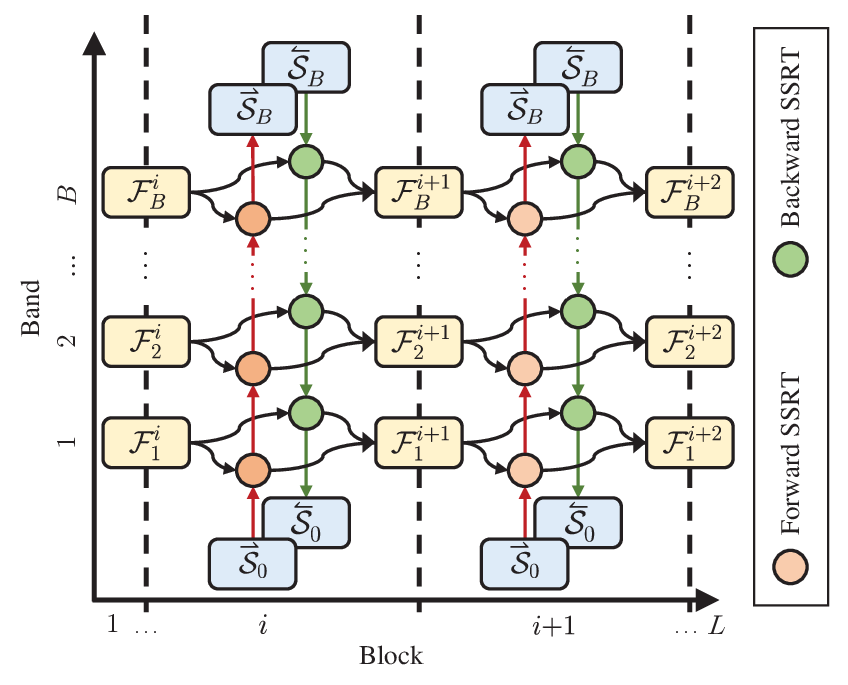}
   \caption{The bidirectional  SSRT block where $L$ is the total number of the SSRT blocks in a layer.}\label{fig:2ssrt}
 \end{figure}

 A single-direction SSRT block can model the spectral correlation in one direction and obtain feature maps that cache the spectral correlation of bands in the previous of the current band. For example, feature map $\Fcal^{i+1}_{t}$ is updated under the guidance of the spectral correlation from the first band to the ($t$$-$1)-th band but neglects spectral correlation from $t$$+$1 to the last band. As shown in Fig.~\ref{fig:2ssrt}, we extend the SSRT block to a bidirectional one to fully exploit the global spectral correlation across all the bands. The bidirectional SSRT  can be formulated as
\begin{equation}
   \begin{alignedat}{2}
      {\mathop{\Scal}\limits^{\rightharpoonup}}\,^i_t ~,&~ {\mathop{\Fcal}\limits^{\rightharpoonup}}\,^{i+1}_t &=~& \textit{Forward}({\mathop{\Scal}\limits^{\rightharpoonup}}\,^i_{t-1}, \Fcal\,^i_t),\\
      {\mathop{\Scal}\limits^{\leftharpoonup}}\,^i_{t-1} ,&~ {\mathop{\Fcal}\limits^{\leftharpoonup}}\,^{i+1}_t &=~& \textit{Backward}({\mathop{\Scal}\limits^{\leftharpoonup}}\,^i_{t}~,~ \Fcal\,^i_t),\\
      &\Fcal\,^{i+1}_t &=~& {\mathop{\Fcal}\limits^{\rightharpoonup}}\,^{i+1}_t + {\mathop{\Fcal}\limits^{\leftharpoonup}}\,^{i+1}_t,
   \end{alignedat}
\end{equation}
where the $\textit{Forward}(\cdot)$ and the $\textit{Backward}(\cdot)$ are the forward SSRT and the backward SSRT, respectively. They share the same architecture, except that the order of the input bands  is reversed. The spatial feature map at the current band  $\Fcal^{i+1}_t$ is computed by  adding  $\textit{Forward}(\cdot)$ and the $\textit{Backward}(\cdot)$.

According to Eq.~(\ref{eq:one_layer}), each layer consists of $L$ blocks.  Following~\cite{Liu2021}, we utilize $[1,1,3,1,1]$ pairs of windowing and shift windowing SSRT blocks from the first to the fifth layers to  progressively exploit the long-range spatial self-similarity.   By employing bidirectional SSRT and the shifted windowing scheme, the proposed SSRT-UNet can simultaneously capture the GSC and NSS within HSIs to achieve preferable denoising.

\subsection{Loss Function}
The training loss of our SSRT-UNet is set as the Euclidean distance between the predicted HSI and the ground truth:
\begin{equation}
   \mathcal{L}=\frac{1}{N}\sum_{i=1}^{N}\|\text{SSRT-UNet}(\Ycal_i)-\Xcal_i\|_F^2,
\end{equation}
where $\{\mathcal{Y}_i, \mathcal{X}_i\}$ is a noisy-clean HSI pair and $N$ is the total number of training samples.

\section{Experiments}\label{sec:experiments}

In this section, we showcase the denoising capabilities of the proposed SSRT-UNet through experiments conducted on both synthetic and real-world datasets. Subsequently, we conduct additional ablation experiments to analyze the performance of the individual modules within the network.
\subsection{Experiment Settings}

\subsubsection{Training and Testing Setting}  Following ~\cite{Wei2021,xiong2021mac}, we selected 100 HSIs from the ICVL HSI dataset for training. All the HSIs were acquired with a Specim PS Kappa DX4 hyperspectral camera with  a size of $1392 \times 1300$ and 31 spectral bands from 400 to 700 nm. The HSIs were split into patches with the size of $64 \times 64 \times 31$ and augmented by random flipping, cropping, and resizing.  The testing set comprises both synthetic datasets and real-world datasets. The synthetic datasets consist of the ICVL testing set, Houston 2018 HSI, and Pavia city center HSI. The real-world datasets encompass Gaofen-5 (GF-5) Shanghai HSI, and Earth Observing-1 (EO-1) HSI.  The number of bands also increases from ICVL to EO-1 HSI, providing a comprehensive evaluation set to test all the methods with respect to a varying number of bands.

\subsubsection{Synthetic Data Generation}   According to~\cite{chen2018}, the real-world noises in HSIs always follow non-independent and identical distribution (non-i.i.d.).  In addition, remote sensing  HSIs are usually contaminated with mixture noises. Therefore, non-i.i.d. Gaussian noise and mixture noise were considered in our experiments. For the non-i.i.d. Gaussian noise,  the standard deviation of noises was set as $\sigma \in$ [0, 15], [0, 55], and [0, 95]. The mixture noise consisted of: 1) non-i.i.d. Gaussian noise with $\sigma \in$ [0,95]; 2) impulse noise on 1/3 of the bands, with intensities ranging from 10\% to 70\%; 3) strips on 5\%-15\% of columns on 1/3 of the bands; 4) deadlines on 5\%-15\% of columns on 1/3 of the bands.
\subsubsection{Evaluation Index} Peak signal-to-noise ratio (PSNR), structure similarity (SSIM), and spectral angle mapper (SAM) are utilized to quantitatively compare the  denoising performance of all methods. Larger PSNR and SSIM and smaller SAM imply higher denoising effectiveness.

\subsubsection{Compared Methods} Thirteen methods were used for comparison, including 8 model-based methods, \emph{i.e.}, BM4D~\cite{maggioni2012nonlocal}, MTSNMF~\cite{ye2014multitask}, LLRT~\cite{Chang2017}, NGMeet~\cite{He2020}, LRMR~\cite{Zhang2014a}, FastHyDe~\cite{zhuang2018fast}, LRTF$L_0$~\cite{xiong2019}, and E-3DTV~\cite{peng2020}, and 5 deep learning-based methods, \emph{i.e.}, T3SC~\cite{Bodrito2021}, MAC-Net~\cite{xiong2021mac}, NSSNN~\cite{guanyiman2022}, TRQ3D~\cite{Pang2022}, and SST~\cite{li2022spatial}. For a fair comparison, we retrained one model for all the deep learning-based methods in each noise case.

\subsubsection{Implementation Details} Our SSRT-UNet was implemented with PyTorch on NVIDIA TESLA V100 GPU. We trained one model on each noise case for 15 epochs with the ADAM optimizer. The learning rate was initially set to $1\times 10^{-4}$ and decreased at the 8th and 12th epochs with a factor of $0.3$.  The training set was augmented by random flipping, cropping, and resizing. We fine-tuned the model trained on the Gaussian noise case for mixture noise removal.

\subsection{Comparision on Synthetic Datasets}


We first present the denoising performance on the  ICVL testing HSIs,  Houston 2018 and Pavia city center HSIs. 
\subsubsection{ICVL Dataset} 

The ICVL testing dataset includes 50  HSIs from the rest of the ICVL dataset. All of these HSIs were cropped into the size of $512\times 512\times 31$, given  the high computational burden of some model-based methods such as NGMeet. The ICVL dataset contains a larger number of  HSIs, providing a comprehensive evaluation platform to asses the  performance of all the methods, under various noise conditions. Therefore, both Gaussian and mixture noises were added for assessment.  Table~\ref{tab:icvl} shows the quantitative comparison, where the top 3 results are highlighted with bold \1{red}, bold \2{blue}, and bold \3{green}, respectively. FastHyDe achieves the best performance among the model-based methods in Gaussian noisy cases thanks to the simultaneous modelling of the GSC and NSS of HSIs. Considering various noise types, such as stripes, LRTF$L_0$ and E-3DTV perform better in the case of mixed noise.
Deep learning-based methods are equipped with powerful representation ability by learning from data, so they perform better than model-based methods.  T3SC and MAC-Net benefit from the physical model and data-driven learning, gaining a strong ability to remove Gaussian noise. Except for our SSRT-UNet, all other networks model the domain knowledge separately with a combination of independent modules, so they have inherent limitations in fully depicting the 3D spatial-spectral characteristics of HSIs. By modelling global spectral correlations with deeply coupled spatial and spectral branches and benefiting from the hybrid advantages of transformers and RNNs, our SSRT-UNet is more capable of exploiting multiple domain knowledge.  As a result, our SSRT-UNet stands out from all other approaches in all cases.

\begin{table*}[htbp]
   \caption{Comparison of Different Methods on Houston 2018 HSI. The Top Three Values Are Marked as \1{Red}, \2{Blue}, and \3{Green}.}\label{tab:houston}
   \centering
   \resizebox{\linewidth}{!}{\tablesize{
   \begin{tabular}{c|c|c|c|c|c|c|c|c|c|c|c|c|c|c|c}
      \hline
   &&\multicolumn{8}{c|}{\textbf{Model-based methods}}&\multicolumn{5}{c}{\textbf{Deep learning-based methods}}\\
   \hline
   \multirow{2}*{\makebox[0.02\textwidth][c]{Index}}&\multirow{2}*{\makebox[0.032\textwidth][c]{Noisy}}&\multirow{1}*{\makebox[0.032\textwidth][c]{BM4D}}&\multirow{1}*{\makebox[0.032\textwidth][c]{MTSNMF}}&\multirow{1}*{\makebox[0.032\textwidth][c]{LLRT}}&\multirow{1}*{\makebox[0.032\textwidth][c]{NGMeet}}&\multirow{1}*{\makebox[0.032\textwidth][c]{LRMR}}&\multirow{1}*{\makebox[0.032\textwidth][c]{FastHyDe}}&\multirow{1}*{\makebox[0.032\textwidth][c]{LRTF$L_0$}}&\multirow{1}*{\makebox[0.032\textwidth][c]{E-3DTV}}&\multirow{1}*{\makebox[0.032\textwidth][c]{T3SC}}&\multirow{1}*{\makebox[0.032\textwidth][c]{MAC-Net}}&\multirow{1}*{\makebox[0.032\textwidth][c]{NSSNN}}&\multirow{1}*{\makebox[0.032\textwidth][c]{TRQ3D}}&\multirow{1}*{\makebox[0.032\textwidth][c]{SST}}&\makebox[0.04\textwidth][c]{\textbf{SSRT-UNet}}\\
   &&\cite{maggioni2012nonlocal}&\cite{ye2014multitask}&\cite{Chang2017}&\cite{He2020}&\cite{Zhang2014a}& \cite{zhuang2018fast}&\cite{xiong2019}& \cite{peng2020}& \cite{Bodrito2021}& \cite{xiong2021mac}& \cite{guanyiman2022}& \cite{Pang2022}& \cite{li2022spatial}& (ours)\\
    \hline

  \makebox[0.02\textwidth][c]{PSNR$\uparrow$} & 11.72 & 22.76 & 25.86 & 15.58 & 22.36 & 21.84 & 27.07&28.75&	30.64 &29.84&28.93&\2{33.77}&\3{32.55}&31.08&\1{34.71} \\
  \makebox[0.02\textwidth][c]{SSIM$\uparrow$} & .0843 & .4762 & .6933 & .1386 & .5169 & .3914 & .7757&.8038& .8570 &.8751&.8080&\2{.9296}&\3{.9194}&.9166&\1{.9387} \\
  \makebox[0.02\textwidth][c]{SAM$\downarrow$} & .9778 & .5168 & .4977 & .7652 & .5728 & .4857 & .4518&.2221& .1323 &.1943&.2537&\2{.1094}&\3{.1241}&.1390&\1{.0915} \\
 \hline
 \end{tabular}}}
 \end{table*}

\begin{figure*}[htbp]
   \centering
   \subfigure[Clean]{\label{fig:houston_clean}\includegraphics[width=0.124\linewidth]{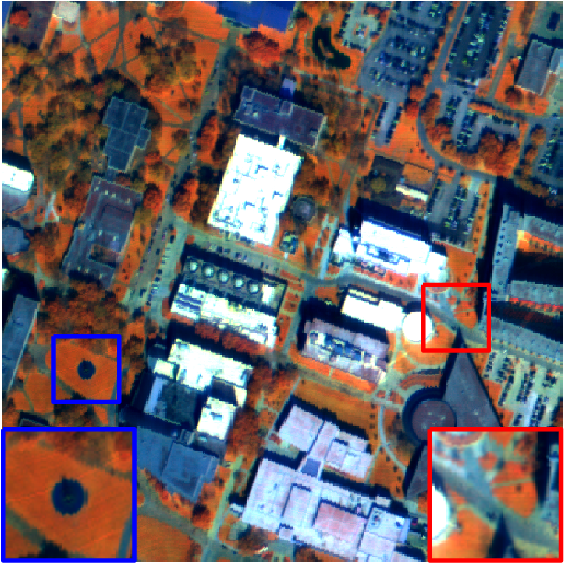}}
   \hspace{-2.1mm}
   \subfigure[Noisy]{\label{fig:houston_noisy}\includegraphics[width=0.124\linewidth]{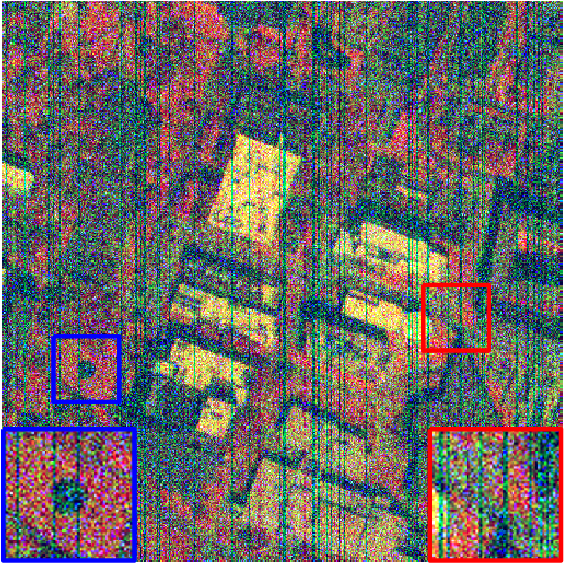}}
   \hspace{-2.1mm}
   \subfigure[BM4D \cite{maggioni2012nonlocal}]{\label{fig:houston_BM4D}\includegraphics[width=0.124\linewidth]{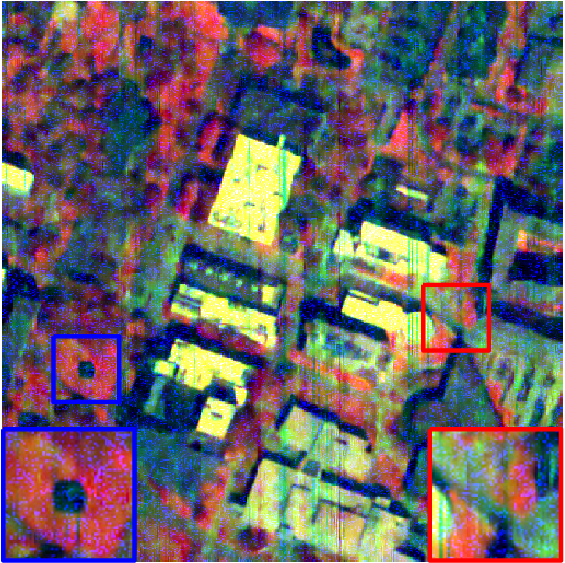}}
   \hspace{-2.1mm}
   \subfigure[MTSNMF \cite{ye2014multitask}]{\label{fig:houston_MTSNMF}\includegraphics[width=0.124\linewidth]{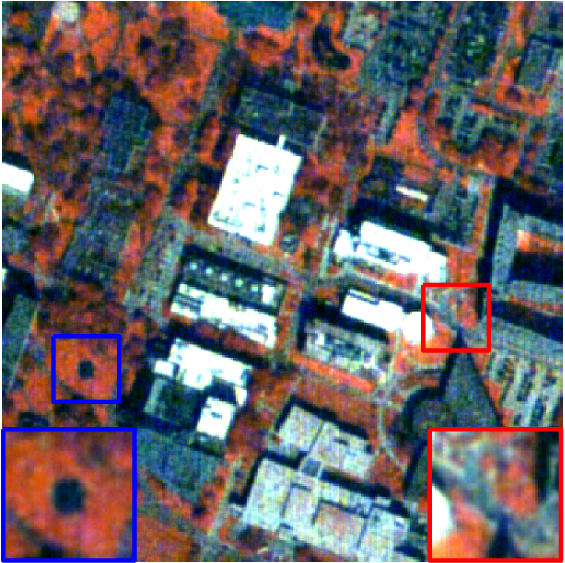}}
   \hspace{-2.1mm}
   \subfigure[LLRT \cite{Chang2017}]{\label{fig:houston_LLRT}\includegraphics[width=0.124\linewidth]{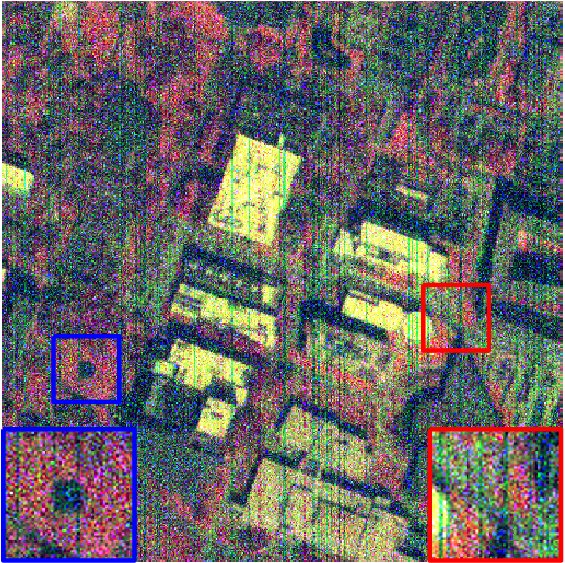}}
   \hspace{-2.1mm}
   \subfigure[NGMeet \cite{He2020}]{\label{fig:houston_NGMeet}\includegraphics[width=0.124\linewidth]{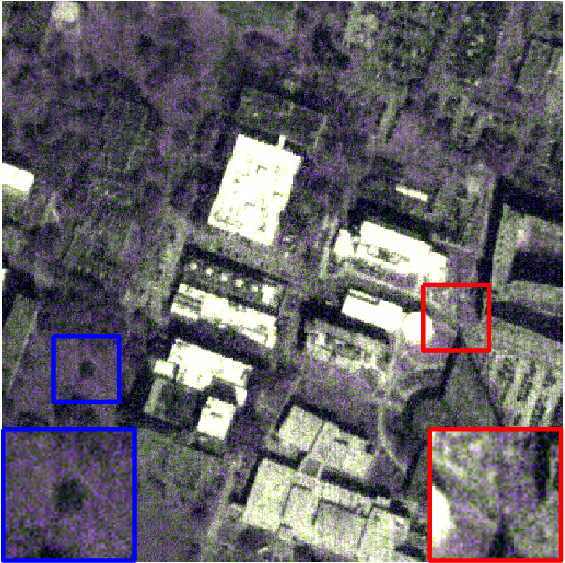}}
   \hspace{-2.1mm}
   \subfigure[LRMR \cite{Zhang2014a}]{\label{fig:houston_LRMR}\includegraphics[width=0.124\linewidth]{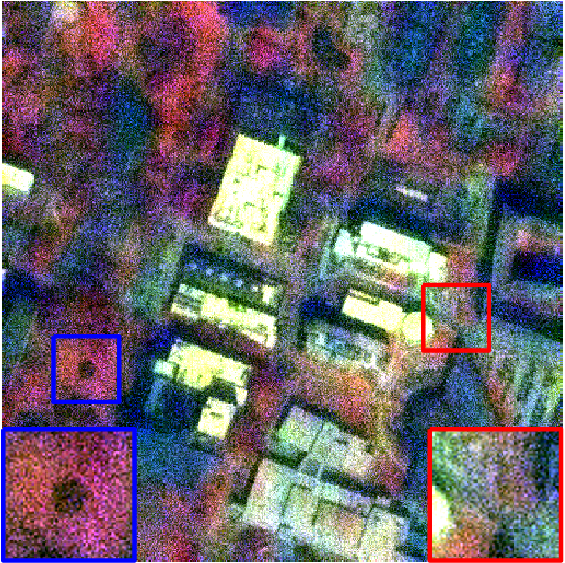}}
   \hspace{-2.1mm}
   \subfigure[FastHyDe \cite{zhuang2018fast}]{\label{fig:houston_FastHyDe}\includegraphics[width=0.124\linewidth]{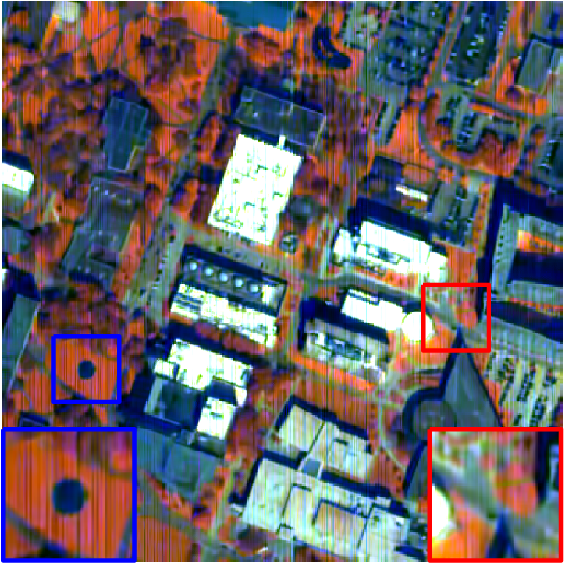}}\\
   \subfigure[LRTF$L_0$ \cite{xiong2019}]{\label{fig:houston_lrtfl0}\includegraphics[width=0.124\linewidth]{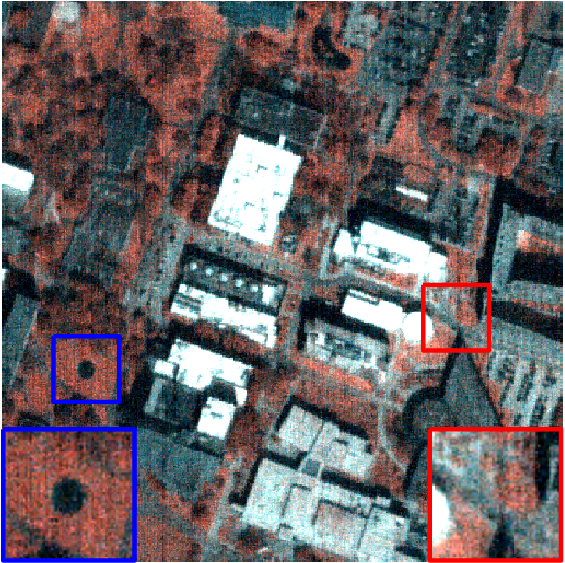}}
   \hspace{-2.1mm}
   \subfigure[E-3DTV \cite{peng2020}]{\label{fig:houston_e3dtv}\includegraphics[width=0.124\linewidth]{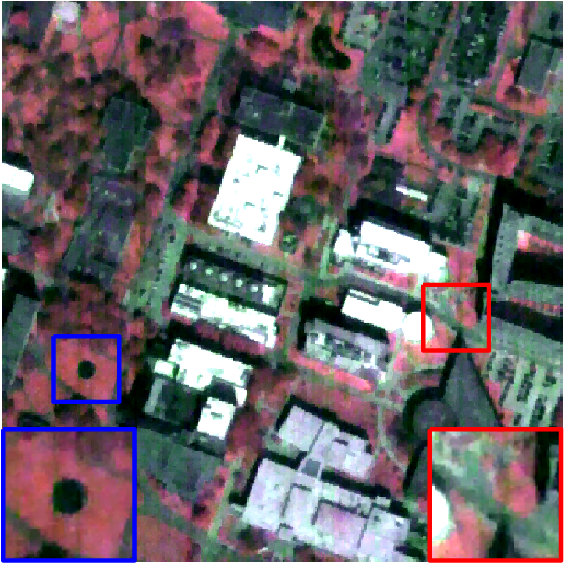}}
   \hspace{-2.1mm}
   \subfigure[T3SC \cite{Bodrito2021}]{\label{fig:houston_T3SC}\includegraphics[width=0.124\linewidth]{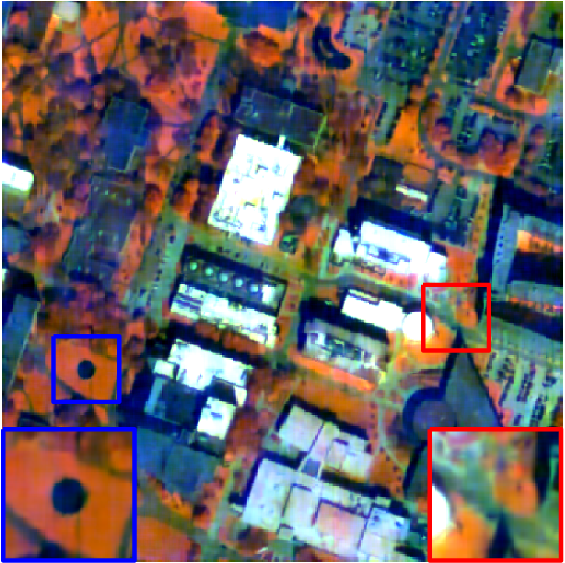}}
   \hspace{-2.1mm}
   \subfigure[MAC-Net \cite{xiong2021mac}]{\label{fig:houston_MAC-Net}\includegraphics[width=0.124\linewidth]{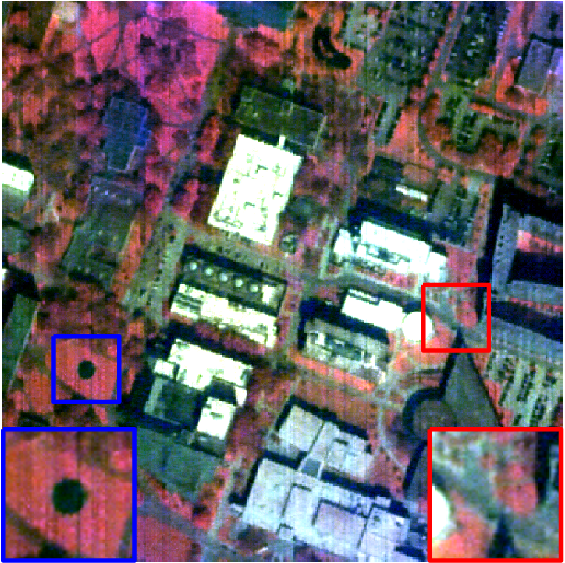}}
   \hspace{-2.1mm}
   \subfigure[NSSNN \cite{guanyiman2022}]{\label{fig:houston_NSSNN}\includegraphics[width=0.124\linewidth]{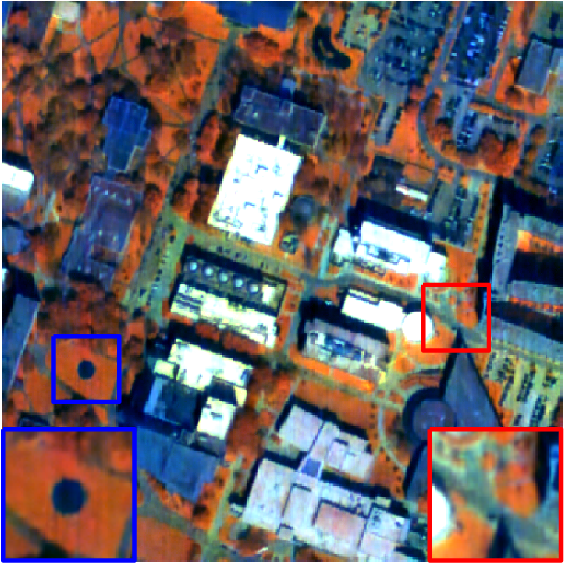}}
   \hspace{-2.1mm}
   \subfigure[TRQ3D \cite{Pang2022}]{\label{fig:houston_TRQ3D}\includegraphics[width=0.124\linewidth]{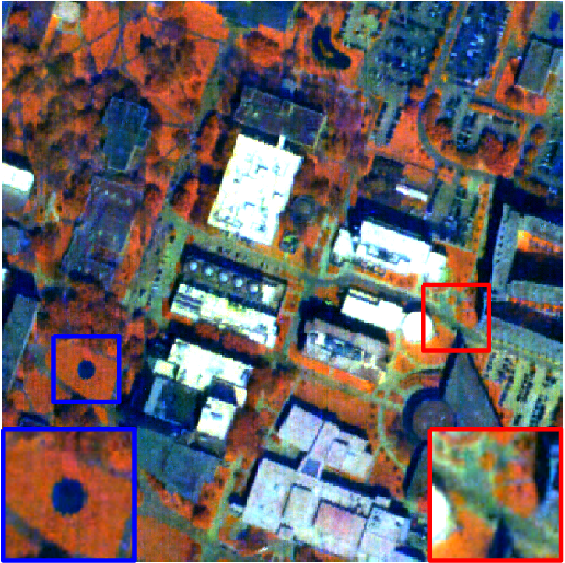}}
   \hspace{-2.1mm}
   \subfigure[SST \cite{li2022spatial}]{\label{fig:houston_SST}\includegraphics[width=0.124\linewidth]{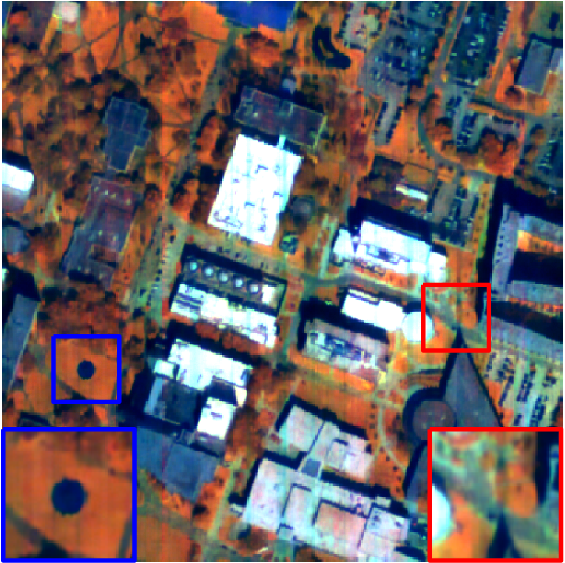}}
   \hspace{-2.1mm}
   \subfigure[\textbf{SSRT-UNet}]{\label{fig:houston_SSRT}\includegraphics[width=0.124\linewidth]{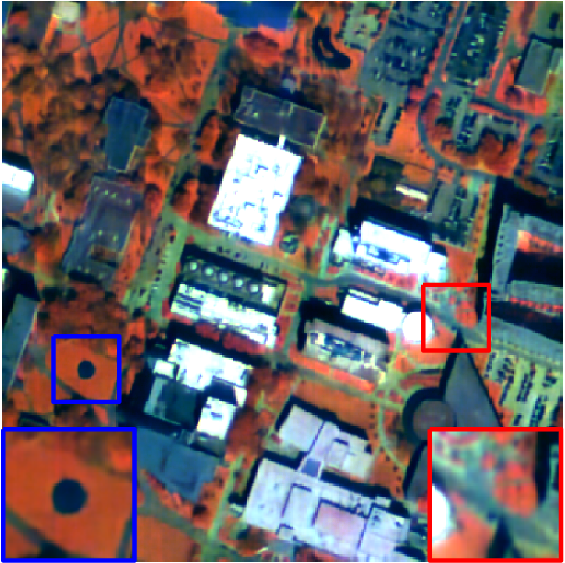}}
      \caption{Denoising results on the Houston 2018 HSI with the mixture noise. The false-color images are generated by combining bands 46, 23, and 1.} \label{fig:houston_visual}
\end{figure*}

\begin{figure*}[!ht]
   \centering
   \subfigure[Clean]{\label{fig:houston_pixel_clean}\includegraphics[width=0.124\linewidth]{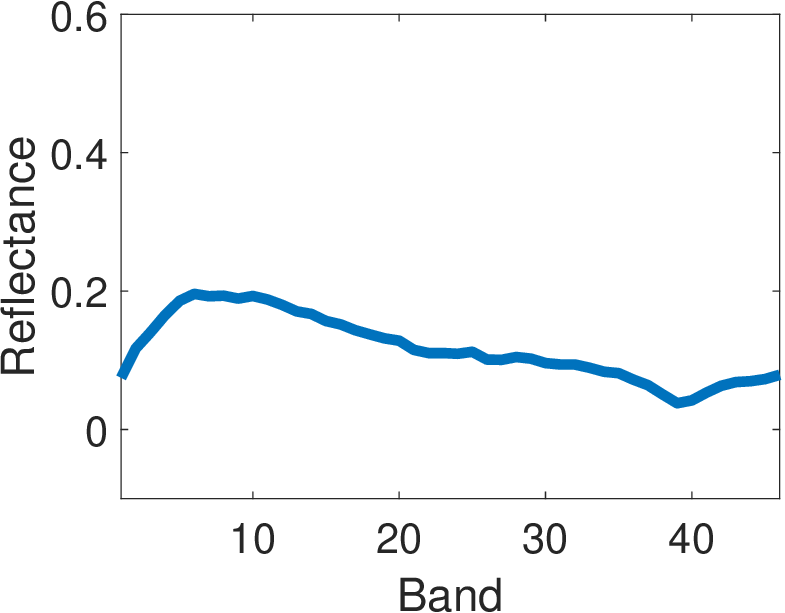}}
   \hspace{-2.1mm}
   \subfigure[Noisy]{\label{fig:houston_pixel_noisy}\includegraphics[width=0.124\linewidth]{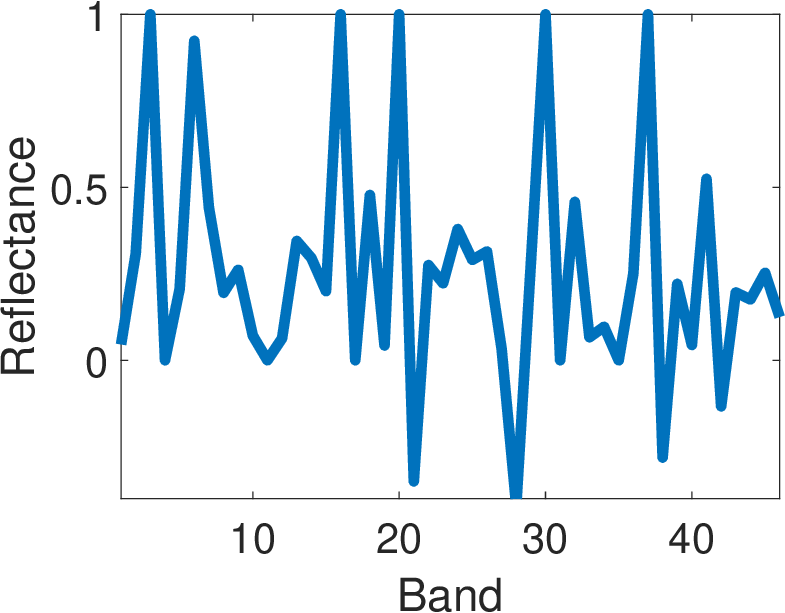}}
   \hspace{-2.1mm}
   \subfigure[BM4D \cite{maggioni2012nonlocal}]{\label{fig:houston_pixel_BM4D}\includegraphics[width=0.124\linewidth]{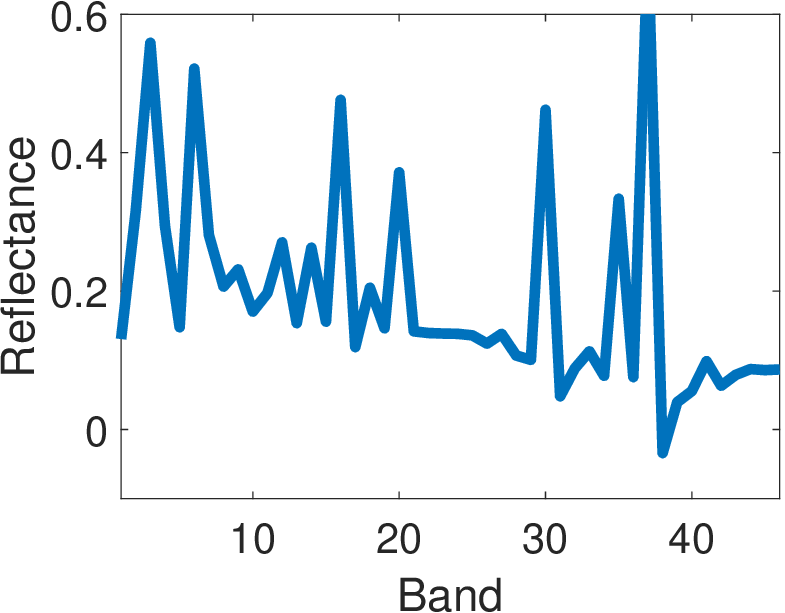}}
   \hspace{-2.1mm}
   \subfigure[MTSNMF \cite{ye2014multitask}]{\label{fig:houston_pixel_MTSNMF}\includegraphics[width=0.124\linewidth]{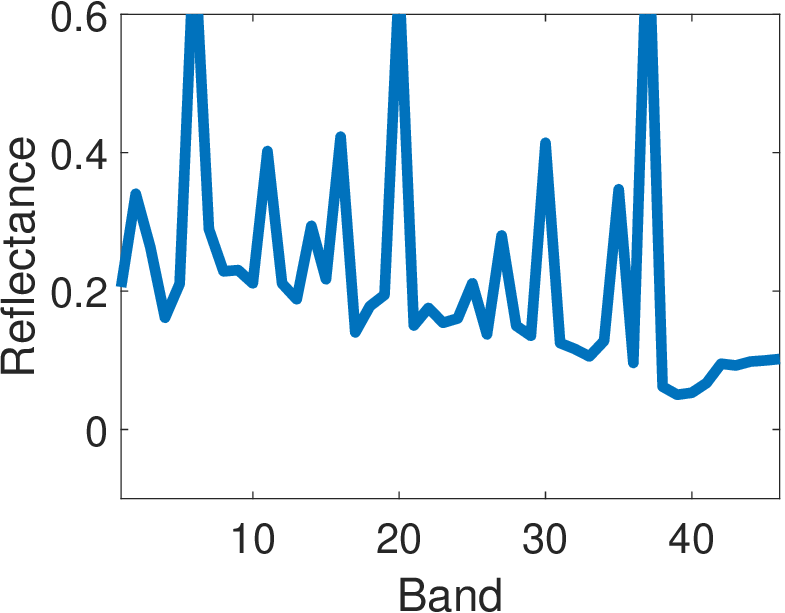}}
   \hspace{-2.1mm}
   \subfigure[LLRT \cite{Chang2017}]{\label{fig:houston_pixel_LLRT}\includegraphics[width=0.124\linewidth]{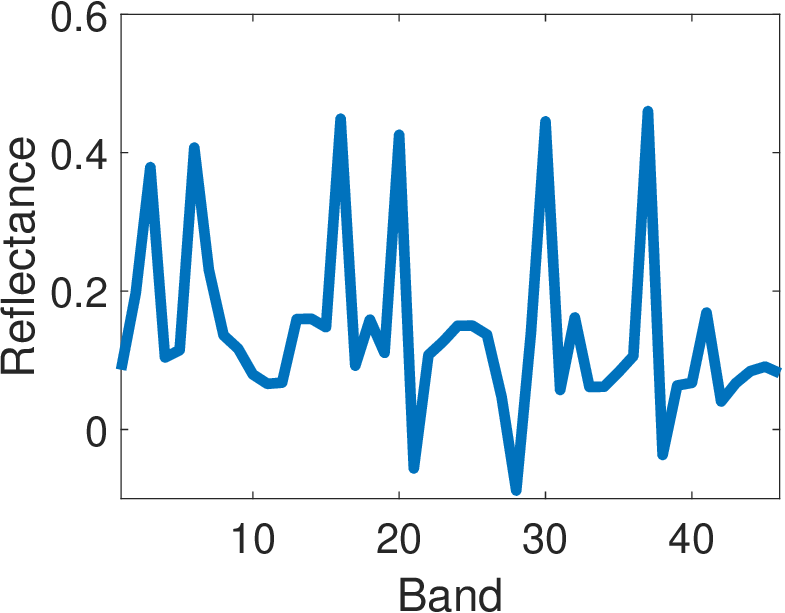}}
   \hspace{-2.1mm}
   \subfigure[NGMeet \cite{He2020}]{\label{fig:houston_pixel_NGMeet}\includegraphics[width=0.124\linewidth]{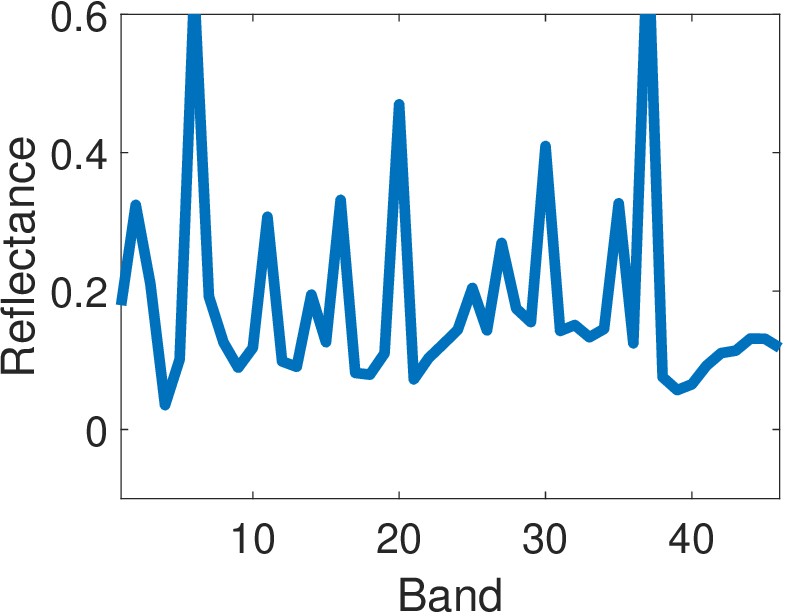}}
   \hspace{-2.1mm}
   \subfigure[LRMR \cite{Zhang2014a}]{\label{fig:houston_pixel_LRMR}\includegraphics[width=0.124\linewidth]{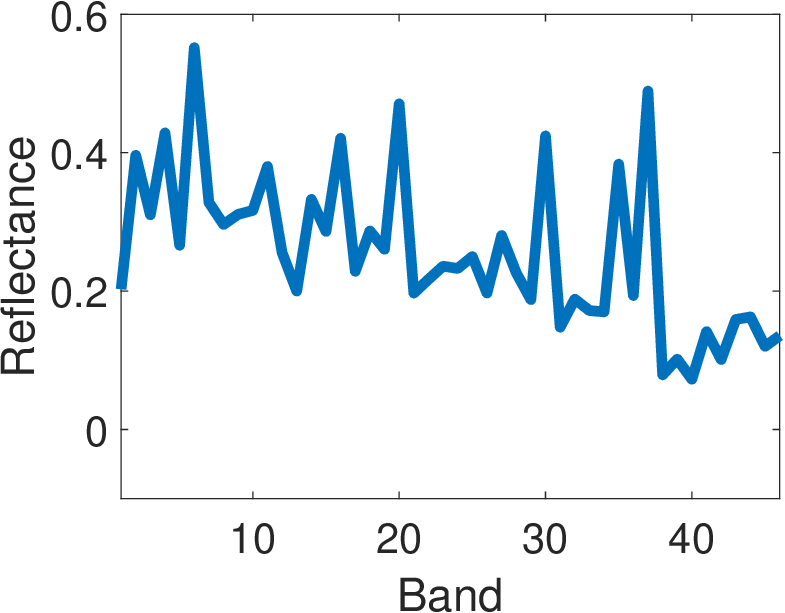}}
   \hspace{-2.1mm}
   \subfigure[FastHyDe \cite{zhuang2018fast}]{\label{fig:houston_pixel_FastHyDe}\includegraphics[width=0.124\linewidth]{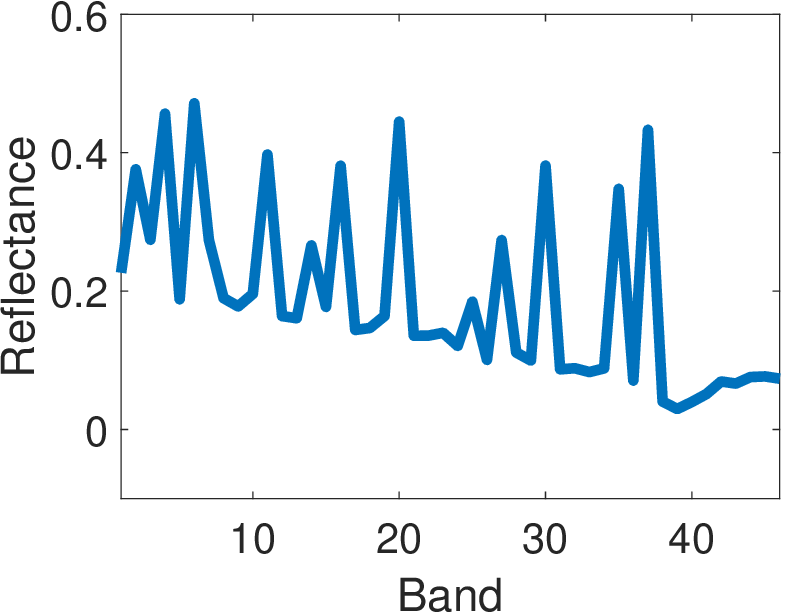}}\\
   \subfigure[LRTF$L_0$ \cite{xiong2019}]{\label{fig:houston_pixel_lrtfl0}\includegraphics[width=0.124\linewidth]{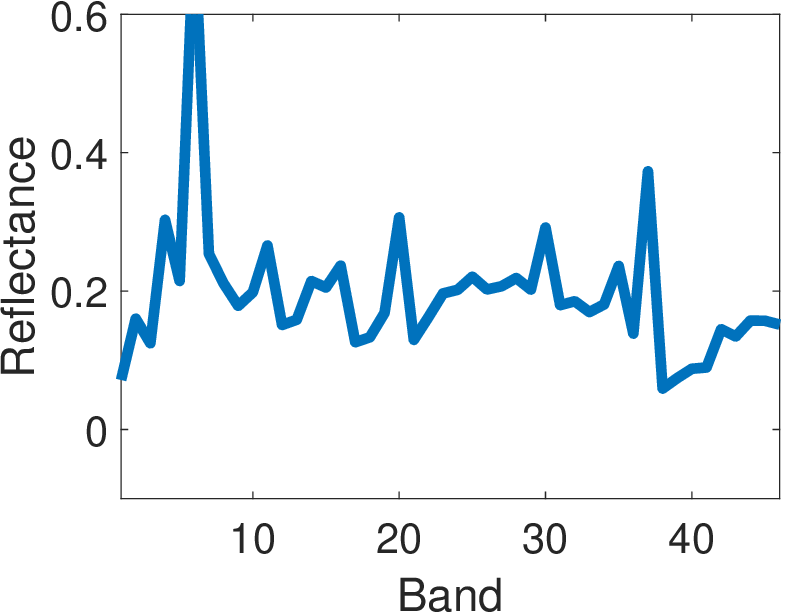}}
   \hspace{-2.1mm}
   \subfigure[E-3DTV \cite{peng2020}]{\label{fig:houston_pixel_e3dtv}\includegraphics[width=0.124\linewidth]{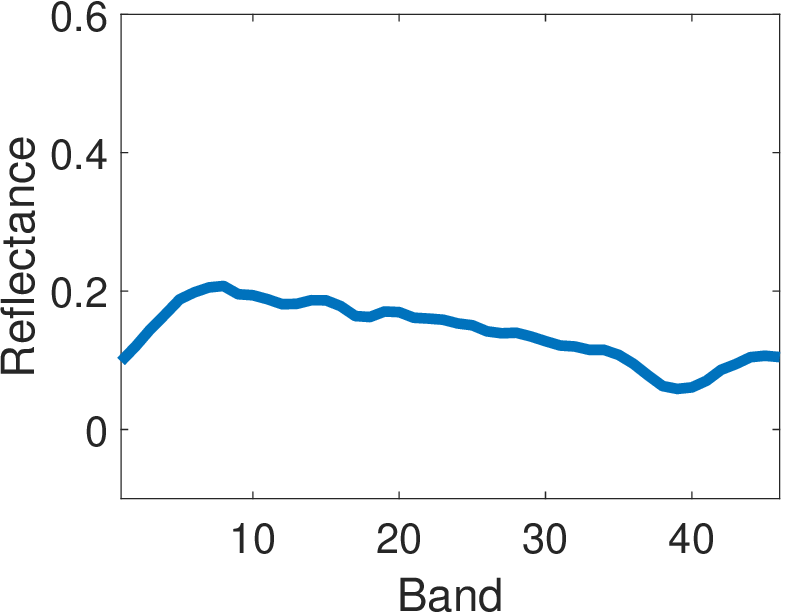}}
   \hspace{-2.1mm}
   \subfigure[T3SC \cite{Bodrito2021}]{\label{fig:houston_pixel_T3SC}\includegraphics[width=0.124\linewidth]{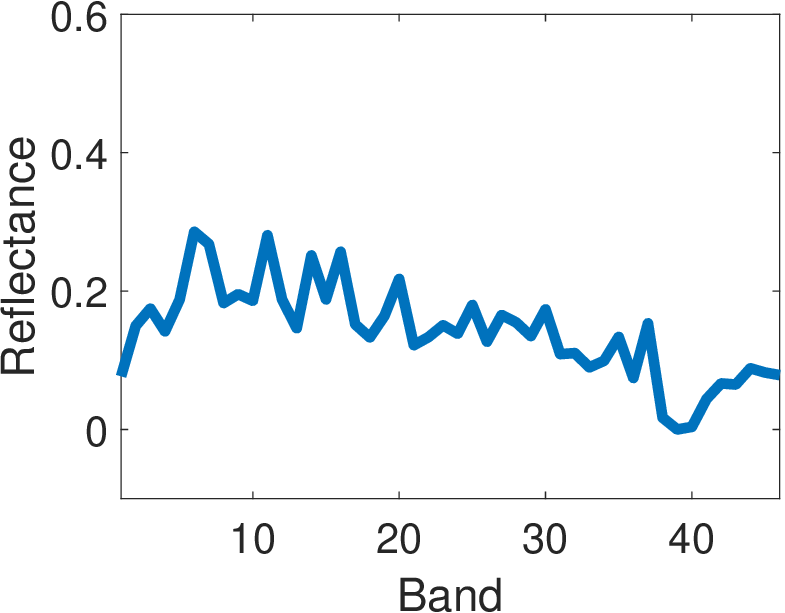}}
   \hspace{-2.1mm}
   \subfigure[MAC-Net \cite{xiong2021mac}]{\label{fig:houston_pixel_MAC-Net}\includegraphics[width=0.124\linewidth]{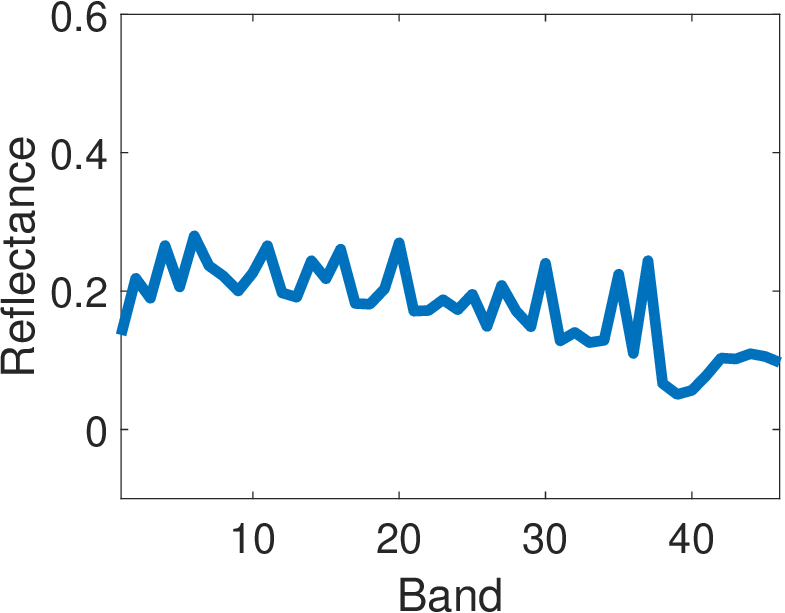}}
   \hspace{-2.1mm}
   \subfigure[NSSNN \cite{guanyiman2022}]{\label{fig:houston_pixel_NSSNN}\includegraphics[width=0.124\linewidth]{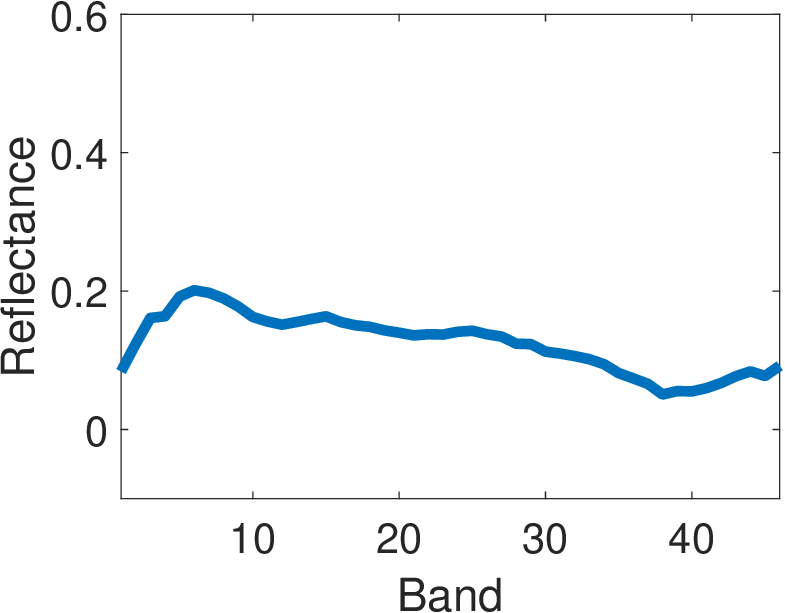}}
   \hspace{-2.1mm}
   \subfigure[TRQ3D \cite{Pang2022}]{\label{fig:houston_pixel_TRQ3D}\includegraphics[width=0.124\linewidth]{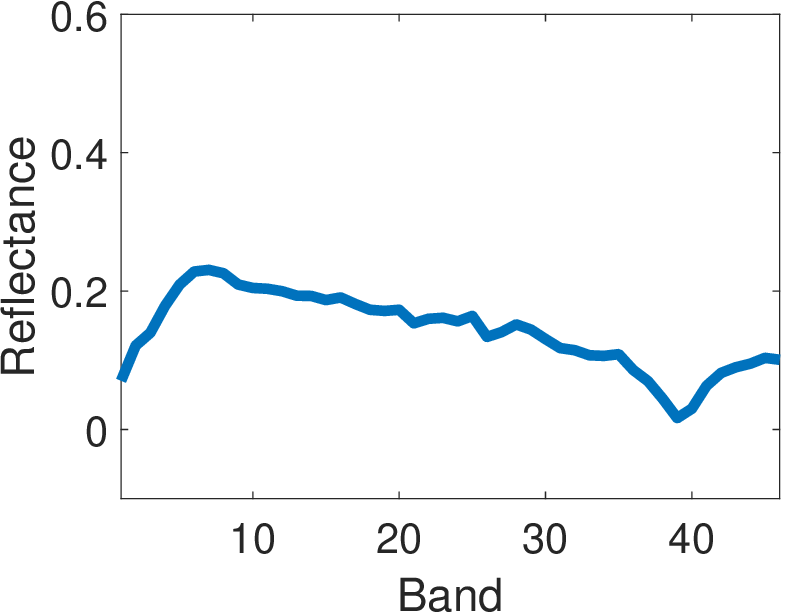}}
   \hspace{-2.1mm}
   \subfigure[SST \cite{li2022spatial}]{\label{fig:houston_pixel_SST}\includegraphics[width=0.124\linewidth]{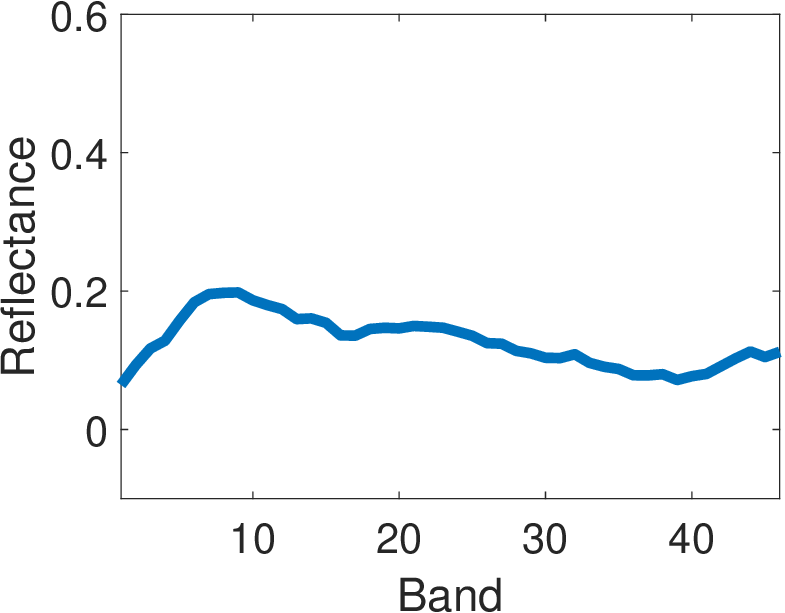}}
   \hspace{-2.1mm}
   \subfigure[\textbf{SSRT-UNet}]{\label{fig:houston_pixel_SSRT}\includegraphics[width=0.124\linewidth]{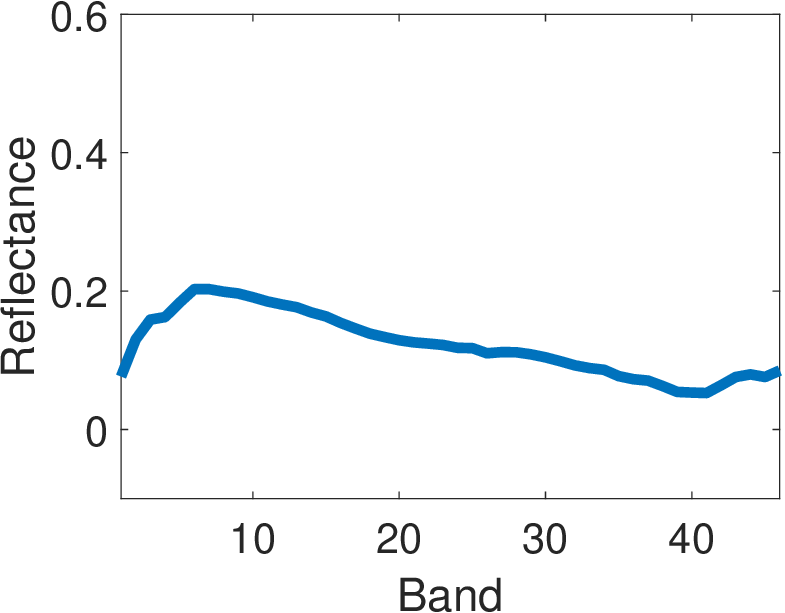}}
      \caption{Reflectance of pixel (417,474) in the Houston 2018 HSI.} \label{fig:houston_pixel}
\end{figure*}

\subsubsection{Houston 2018 HSI} 

The Houston 2018 HSI is a remote sensing HSI acquired with the ITRES CASI hyperspectral imager and contains $1202\times4172$ pixels.  A notable distinction from the HSIs in ICVL lies in the spectral range. While the HSIs in ICVL cover 31 bands with a wavelength range of 400 to 700 nm, the Houston dataset encompasses 48 bands ranging from 380 to 1050 nm. For our experiments, we selected the last 46 bands of the Houston, which are relatively clean, and cropped a $512\times 512$ cube from the center of the HSI for the synthetic experiments. To better simulate the noise in real-world remote sensing images, we introduced mixed noise to the clean HSI, creating the noisy HSI. The deep learning-based models trained on the ICVL dataset were  applied for evaluation. Notably, T3SC, TRQ3D, and SST face limitations in handling HSIs with a different number of bands compared to the training data. To address this, we divided the given HSI into multiple sub-images with 31 bands for testing.

The denoising results are presented in Table~\ref{tab:houston}. It is challenging for deep learning-based methods to handle denoising on an HSI with a different spectral range than the training set. The SST uses a spectral transformer to capture the spectral correlation. However, the spectral transformer lacks the ability to remember previous states, resulting in a significant drop in performance when there are changes in the spectral range.  The recurrent processing in SSRT-UNet enables the propagation of information through recurrent connections, allowing for the full exploitation of global spectral correlation even when the number of bands changes. As a result, both SSRT-UNet and NSSNN outperform other learning-based methods. Additionally, the two interactive branches in our proposed SSRT-UNet enhance knowledge utilization, producing highly competitive performance compared to all other methods.


\begin{table*}[h!]
   \caption{Comparison of Different Methods on Pavia City Center HSI. The Top Three Values Are Marked as \1{Red}, \2{Blue}, and \3{Green}.}\label{tab:paviacity}
   \centering
   \resizebox{\linewidth}{!}{\tablesize{
   \begin{tabular}{c|c|c|c|c|c|c|c|c|c|c|c|c|c|c|c}
      \hline
   &&\multicolumn{8}{c|}{\textbf{Model-based methods}}&\multicolumn{5}{c}{\textbf{Deep learning-based methods}}\\
   \hline
   \multirow{2}*{\makebox[0.02\textwidth][c]{Index}}&\multirow{2}*{\makebox[0.032\textwidth][c]{Noisy}}&\multirow{1}*{\makebox[0.032\textwidth][c]{BM4D}}&\multirow{1}*{\makebox[0.032\textwidth][c]{MTSNMF}}&\multirow{1}*{\makebox[0.032\textwidth][c]{LLRT}}&\multirow{1}*{\makebox[0.032\textwidth][c]{NGMeet}}&\multirow{1}*{\makebox[0.032\textwidth][c]{LRMR}}&\multirow{1}*{\makebox[0.032\textwidth][c]{FastHyDe}}&\multirow{1}*{\makebox[0.032\textwidth][c]{LRTF$L_0$}}&\multirow{1}*{\makebox[0.032\textwidth][c]{E-3DTV}}&\multirow{1}*{\makebox[0.032\textwidth][c]{T3SC}}&\multirow{1}*{\makebox[0.032\textwidth][c]{MAC-Net}}&\multirow{1}*{\makebox[0.032\textwidth][c]{NSSNN}}&\multirow{1}*{\makebox[0.032\textwidth][c]{TRQ3D}}&\multirow{1}*{\makebox[0.032\textwidth][c]{SST}}&\makebox[0.04\textwidth][c]{\textbf{SSRT-UNet}}\\
   &&\cite{maggioni2012nonlocal}&\cite{ye2014multitask}&\cite{Chang2017}&\cite{He2020}&\cite{Zhang2014a}& \cite{zhuang2018fast}&\cite{xiong2019}& \cite{peng2020}& \cite{Bodrito2021}& \cite{xiong2021mac}& \cite{guanyiman2022}& \cite{Pang2022}& \cite{li2022spatial}& (ours)\\
    \hline
  \makebox[0.02\textwidth][c]{PSNR$\uparrow$} & 13.46  & 21.70 &25.15 &15.44 &23.68 &22.72 &26.78 &26.49 &30.44 &28.69 &27.74 &\2{34.85} &\3{33.07 }&31.87 &\1{35.10}   \\
  \makebox[0.02\textwidth][c]{SSIM$\uparrow$} & .2181 &.5128 &.7397 &.2760 &.7176 &.6524 &.8361 &.8147 &.8941 &.8656 &.8724 &\1{.9581} &\3{.9341} &.9234 &\2{.9493}   \\
  \makebox[0.02\textwidth][c]{SAM$\downarrow$} & .8893 &.5297 &.4094 &.7902 &.4760 &.3771 &.4040 &.3703 &\3{.1134} &.2135 &.3222 &\1{.1015} &.1362 &.1676 &\2{.1036}   \\
 \hline
 \end{tabular}}}
 \end{table*}

 \begin{figure*}[htbp]
   \centering
   \subfigure[{Clean}]{\label{fig:pavia_80_clean}\includegraphics[width=0.124\linewidth]{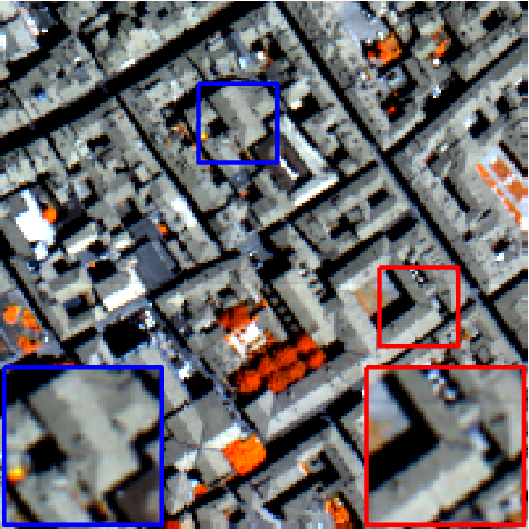}}
   \hspace{-2.1mm}
   \subfigure[{Noisy}]{\label{fig:pavia_80_noisy}\includegraphics[width=0.124\linewidth]{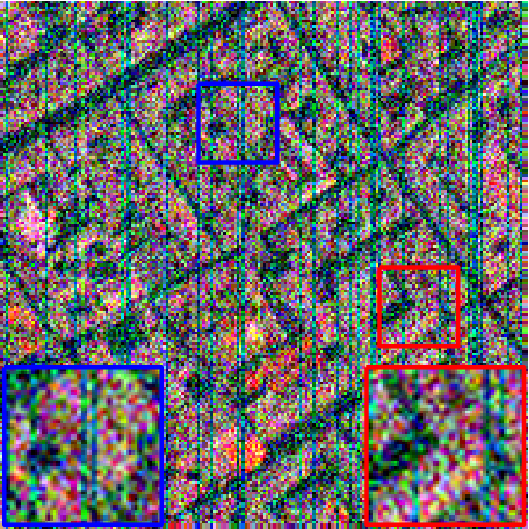}}
   \hspace{-2.1mm}
   \subfigure[{BM4D}~\cite{maggioni2012nonlocal}]{\label{fig:pavia_80_BM4D}\includegraphics[width=0.124\linewidth]{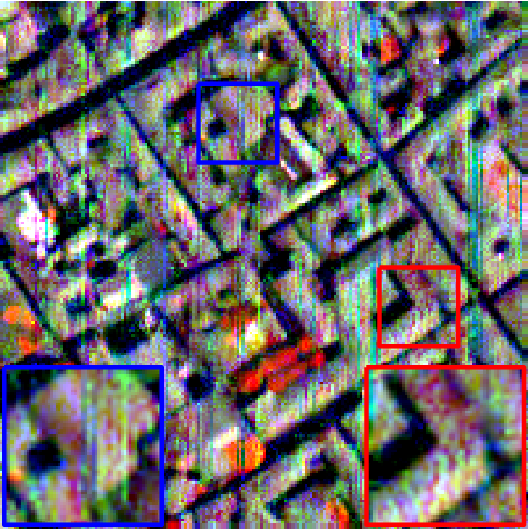}}
   \hspace{-2.1mm}
   \subfigure[{MTSNMF}~\cite{ye2014multitask}]{\label{fig:pavia_80_MTSNMF}\includegraphics[width=0.124\linewidth]{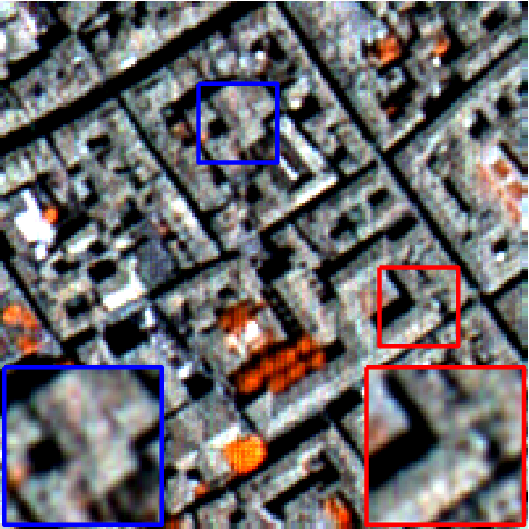}}
   \hspace{-2.1mm}
   \subfigure[{LLRT}~\cite{Chang2017}]{\label{fig:pavia_80_LLRT}\includegraphics[width=0.124\linewidth]{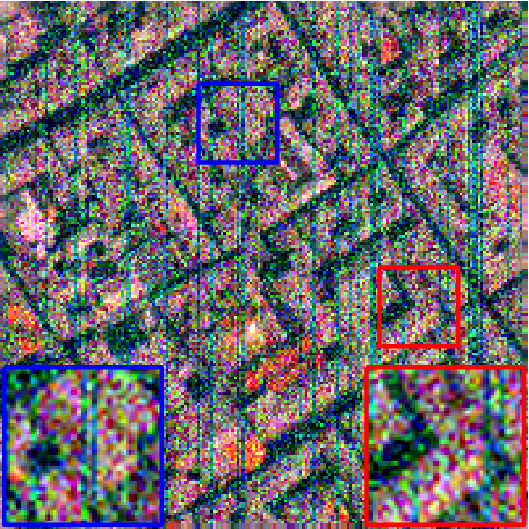}}
   \hspace{-2.1mm}
   \subfigure[{NGMeet}~\cite{He2020}]{\label{fig:pavia_80_NGMeet}\includegraphics[width=0.124\linewidth]{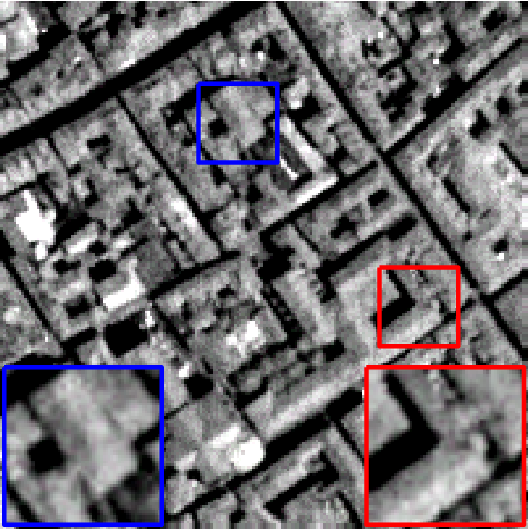}}
   \hspace{-2.1mm}
   \subfigure[{LRMR}~\cite{Zhang2014a}]{\label{fig:pavia_80_LRMR}\includegraphics[width=0.124\linewidth]{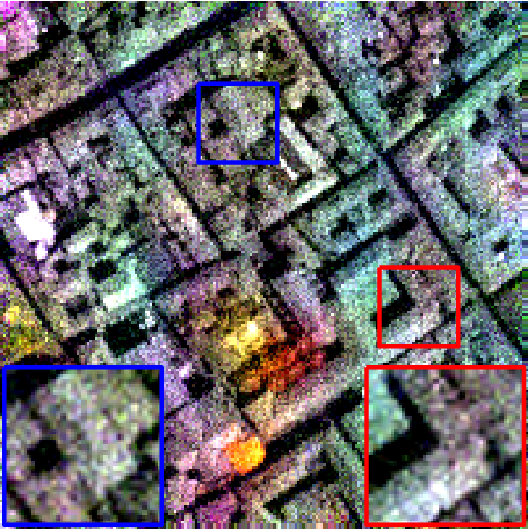}}
   \hspace{-2.1mm}
   \subfigure[{FastHyDe}~\cite{zhuang2018fast}]{\label{fig:pavia_80_FastHyDe}\includegraphics[width=0.124\linewidth]{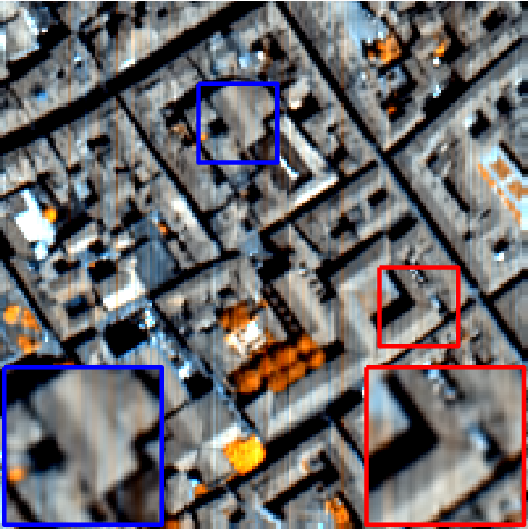}}\\
   \subfigure[{LRTF$L_0$}~\cite{xiong2019}]{\label{fig:pavia_80_lrtfl0}\includegraphics[width=0.124\linewidth]{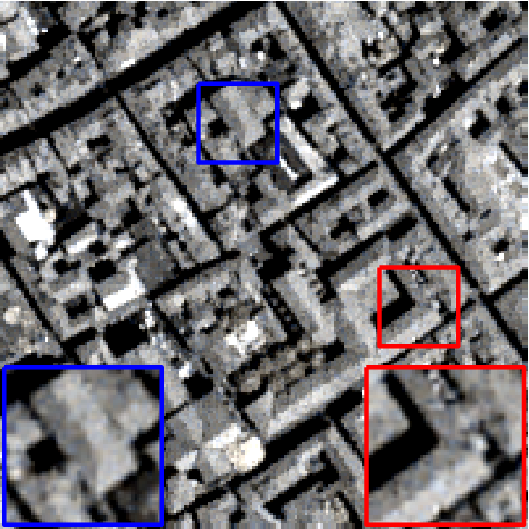}}
   \hspace{-2.1mm}
   \subfigure[{E-3DTV}~\cite{peng2020}]{\label{fig:pavia_80_e3dtv}\includegraphics[width=0.124\linewidth]{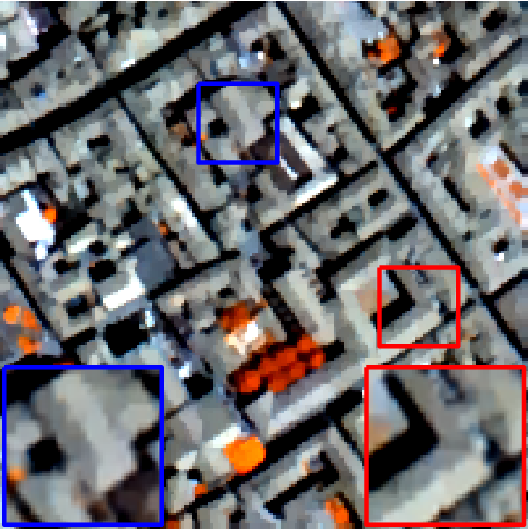}}
   \hspace{-2.1mm}
   \subfigure[{T3SC}~\cite{Bodrito2021}]{\label{fig:pavia_80_T3SC}\includegraphics[width=0.124\linewidth]{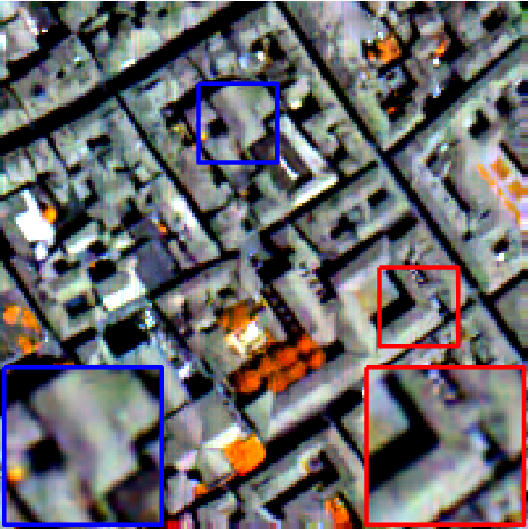}}
   \hspace{-2.1mm}
   \subfigure[{MAC-Net}~\cite{xiong2021mac}]{\label{fig:pavia_80_MAC-Net}\includegraphics[width=0.124\linewidth]{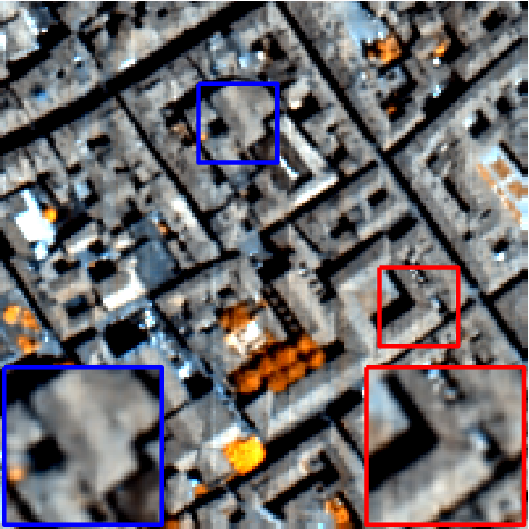}}
   \hspace{-2.1mm}
   \subfigure[{NSSNN}~\cite{guanyiman2022}]{\label{fig:pavia_80_NSSNN}\includegraphics[width=0.124\linewidth]{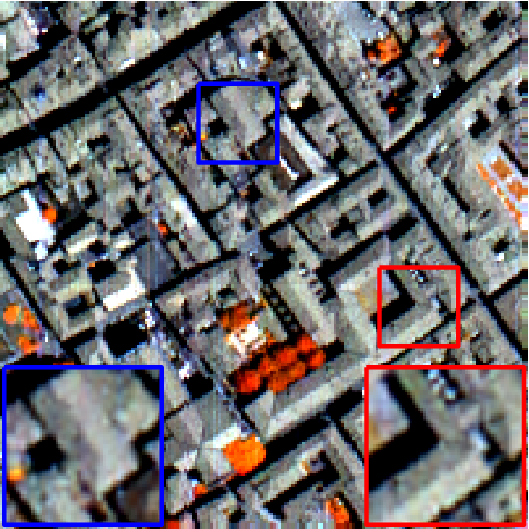}}
   \hspace{-2.1mm}
   \subfigure[{TRQ3D}~\cite{Pang2022}]{\label{fig:pavia_80_TRQ3D}\includegraphics[width=0.124\linewidth]{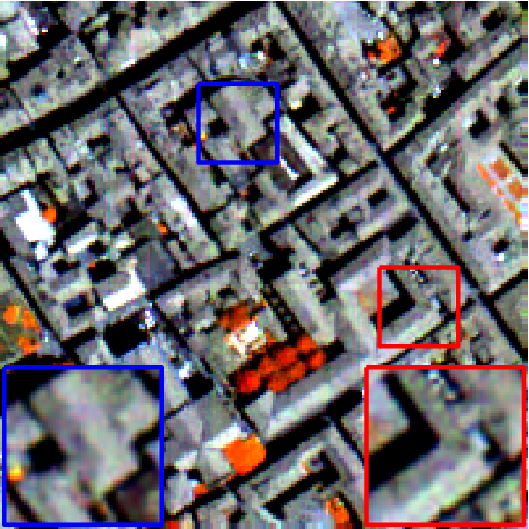}}
   \hspace{-2.1mm}
   \subfigure[{SST}~\cite{li2022spatial}]{\label{fig:pavia_80_SST}\includegraphics[width=0.124\linewidth]{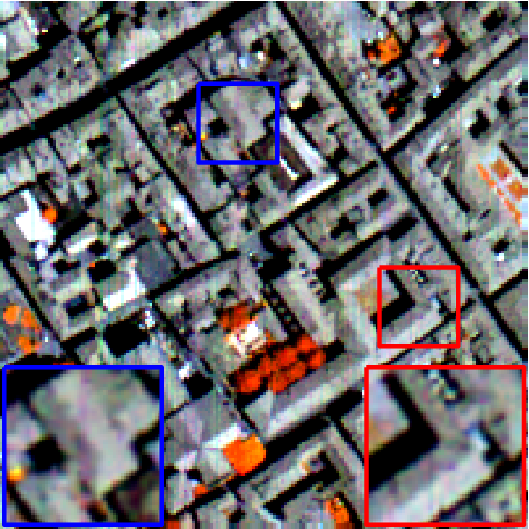}}
   \hspace{-2.1mm}
   \subfigure[{\textbf{SSRT-UNet}}]{\label{fig:pavia_80_SSRT}\includegraphics[width=0.124\linewidth]{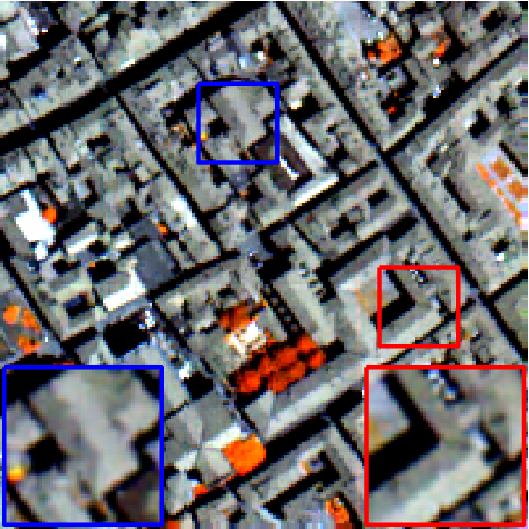}}
      \caption{Denoising results on the Pavia city center HSI with the mixture noise. The false-color images are generated by combining bands 70, 50, and 30.} \label{fig:pavia_80_visual}
\end{figure*}

\begin{figure*}[!ht]
   \centering
   \subfigure[{Clean}]{\label{fig:pavia_80_pixel_clean}\includegraphics[width=0.124\linewidth]{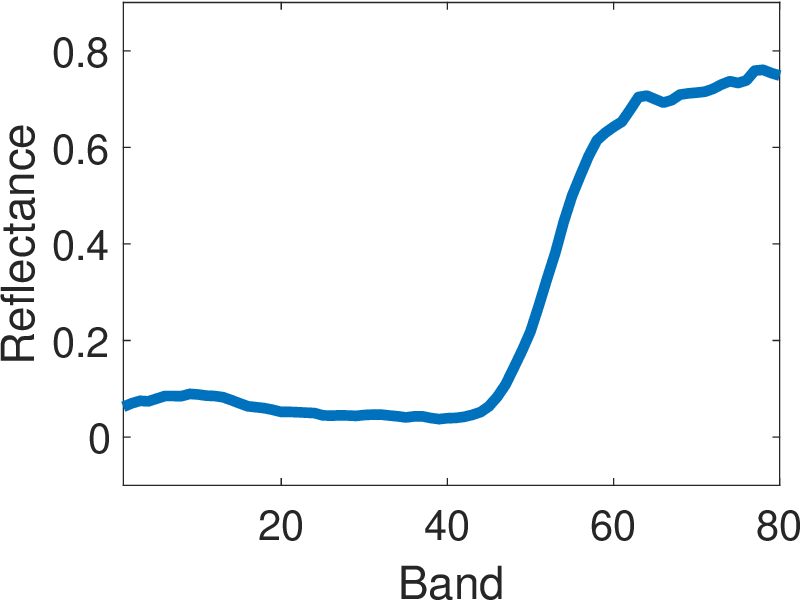}}
   \hspace{-2.1mm}
   \subfigure[{Noisy}]{\label{fig:pavia_80_pixel_noisy}\includegraphics[width=0.124\linewidth]{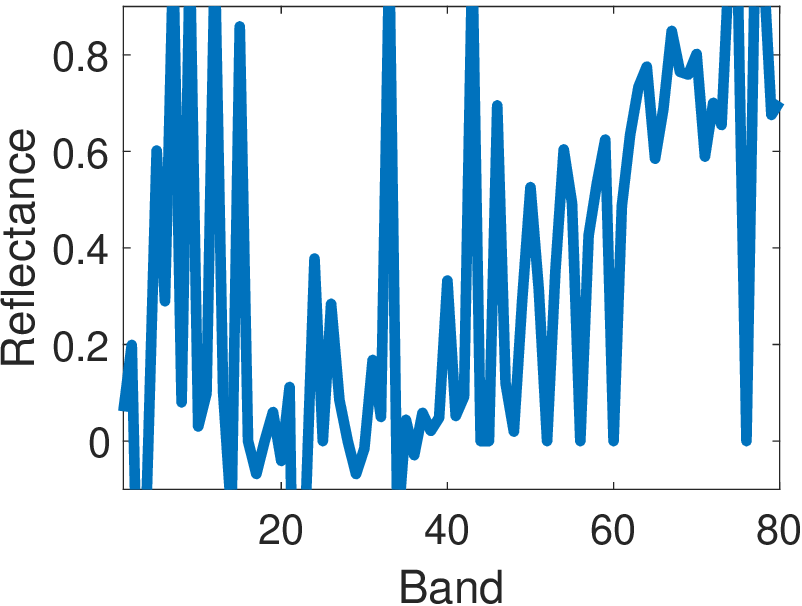}}
   \hspace{-2.1mm}
   \subfigure[{BM4D}~\cite{maggioni2012nonlocal}]{\label{fig:pavia_80_pixel_BM4D}\includegraphics[width=0.124\linewidth]{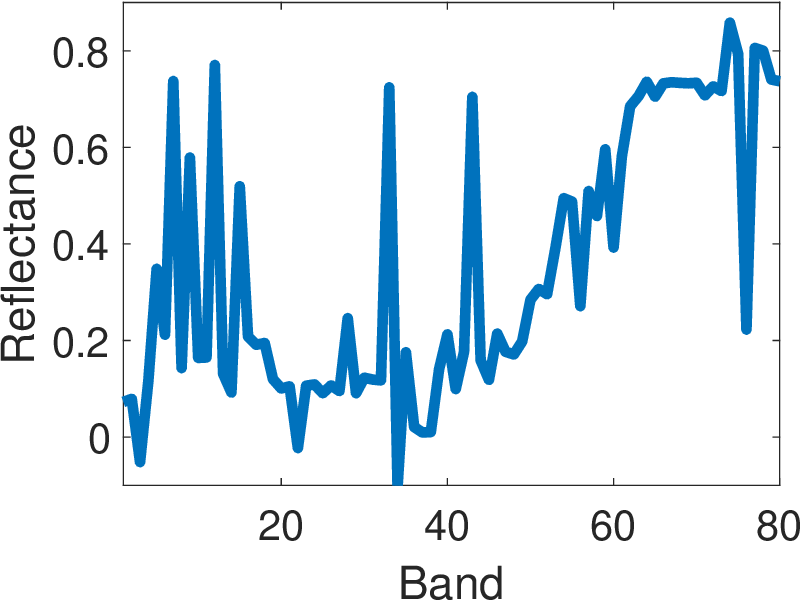}}
   \hspace{-2.1mm}
   \subfigure[{MTSNMF}~\cite{ye2014multitask}]{\label{fig:pavia_80_pixel_MTSNMF}\includegraphics[width=0.124\linewidth]{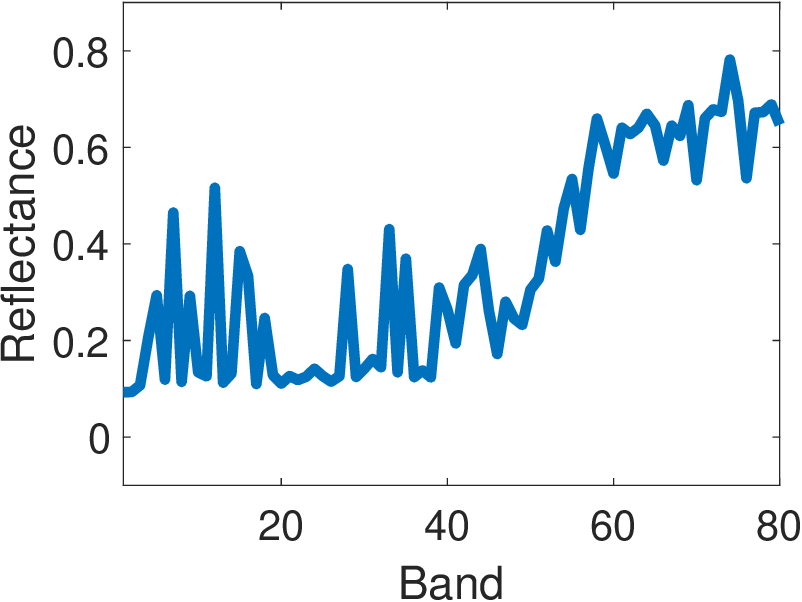}}
   \hspace{-2.1mm}
   \subfigure[{LLRT}~\cite{Chang2017}]{\label{fig:pavia_80_pixel_LLRT}\includegraphics[width=0.124\linewidth]{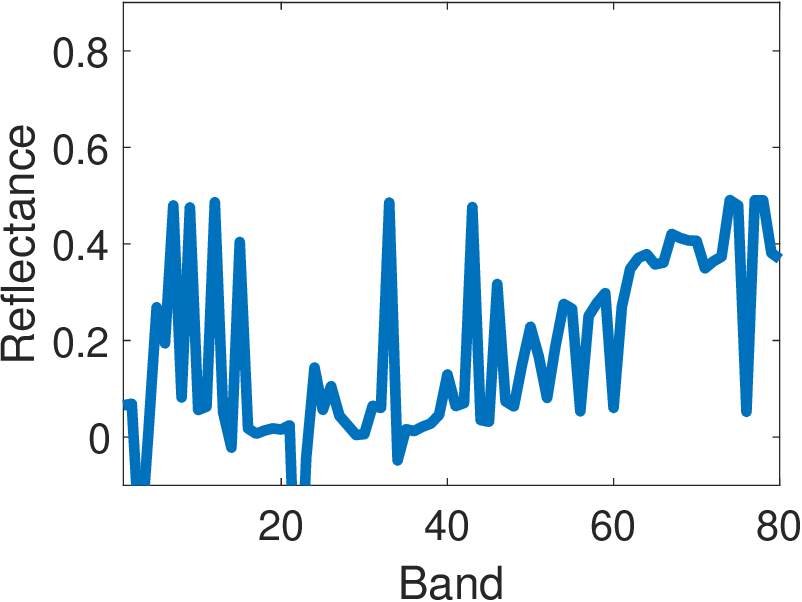}}
   \hspace{-2.1mm}
   \subfigure[{NGMeet}~\cite{He2020}]{\label{fig:pavia_80_pixel_NGMeet}\includegraphics[width=0.124\linewidth]{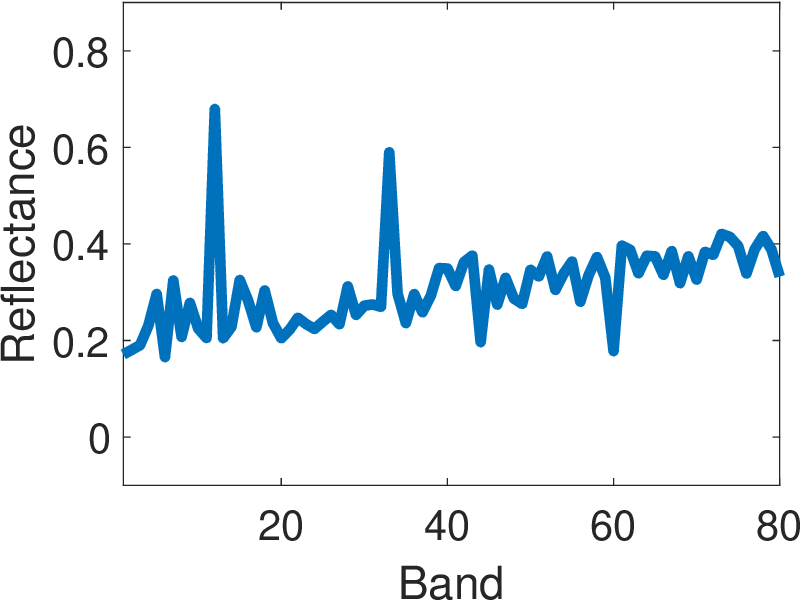}}
   \hspace{-2.1mm}
   \subfigure[{LRMR}~\cite{Zhang2014a}]{\label{fig:pavia_80_pixel_LRMR}\includegraphics[width=0.124\linewidth]{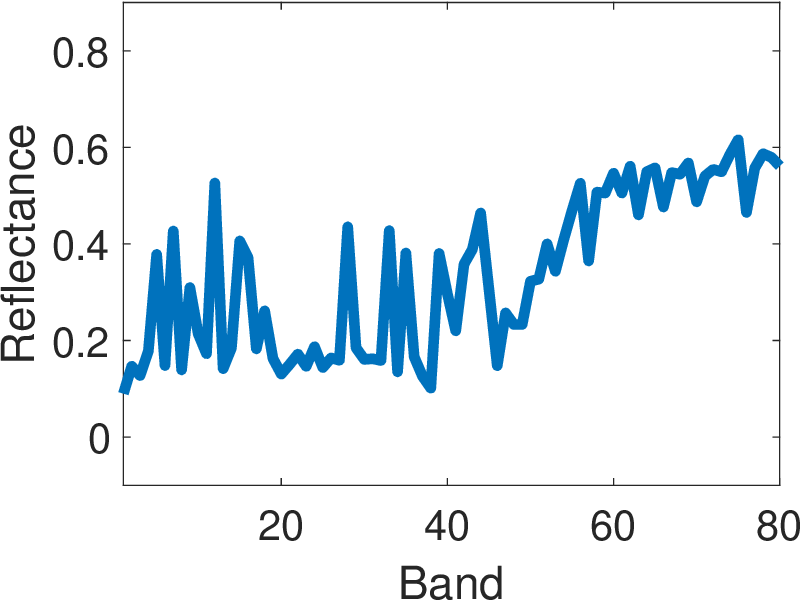}}
   \hspace{-2.1mm}
   \subfigure[{FastHyDe}~\cite{zhuang2018fast}]{\label{fig:pavia_80_pixel_FastHyDe}\includegraphics[width=0.124\linewidth]{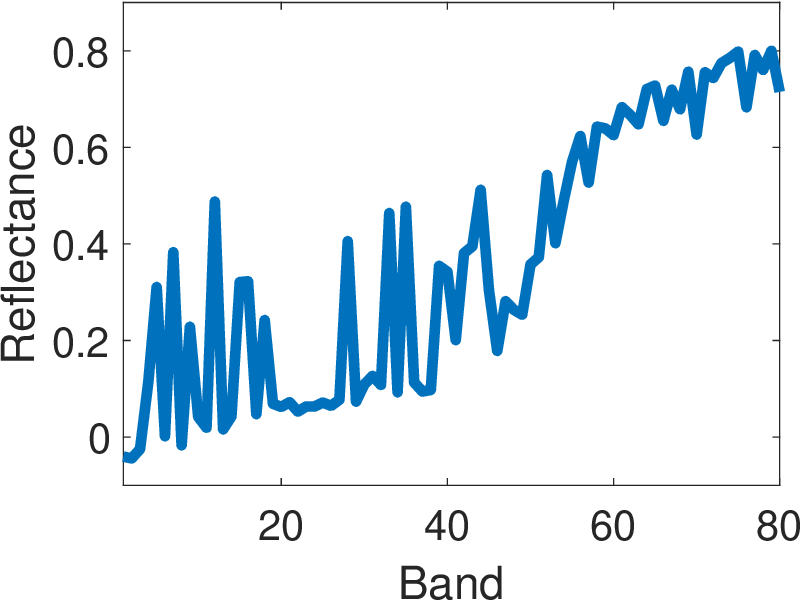}}\\
   \subfigure[{LRTF$L_0$}~\cite{xiong2019}]{\label{fig:pavia_80_pixel_lrtfl0}\includegraphics[width=0.124\linewidth]{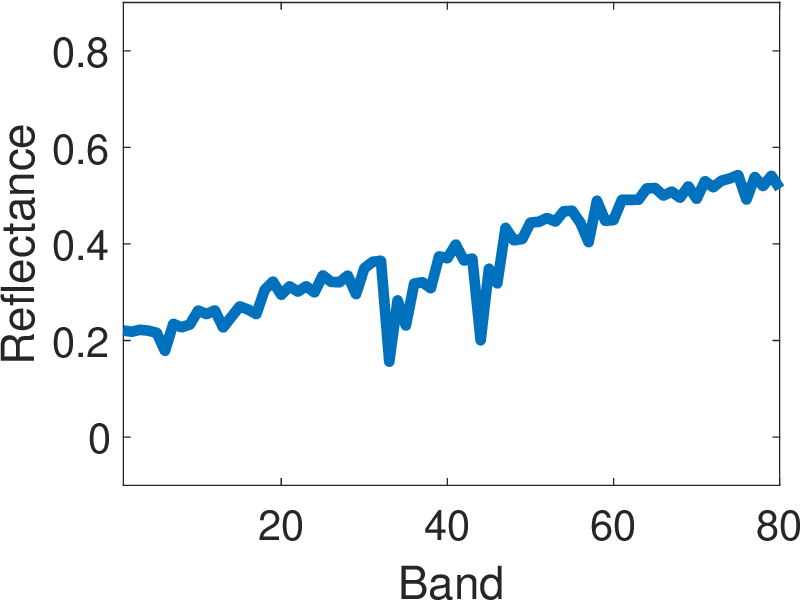}}
   \hspace{-2.1mm}
   \subfigure[{E-3DTV}~\cite{peng2020}]{\label{fig:pavia_80_pixel_e3dtv}\includegraphics[width=0.124\linewidth]{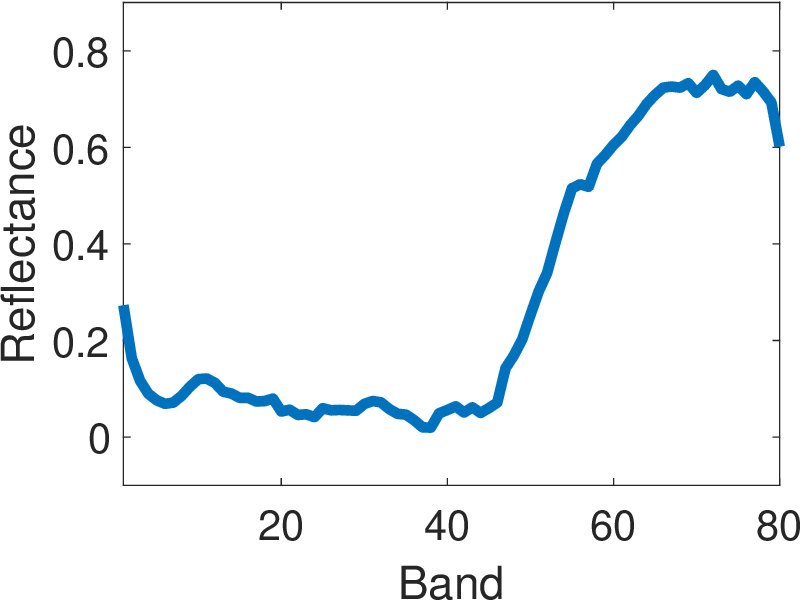}}
   \hspace{-2.1mm}
   \subfigure[{T3SC}~\cite{Bodrito2021}]{\label{fig:pavia_80_pixel_T3SC}\includegraphics[width=0.124\linewidth]{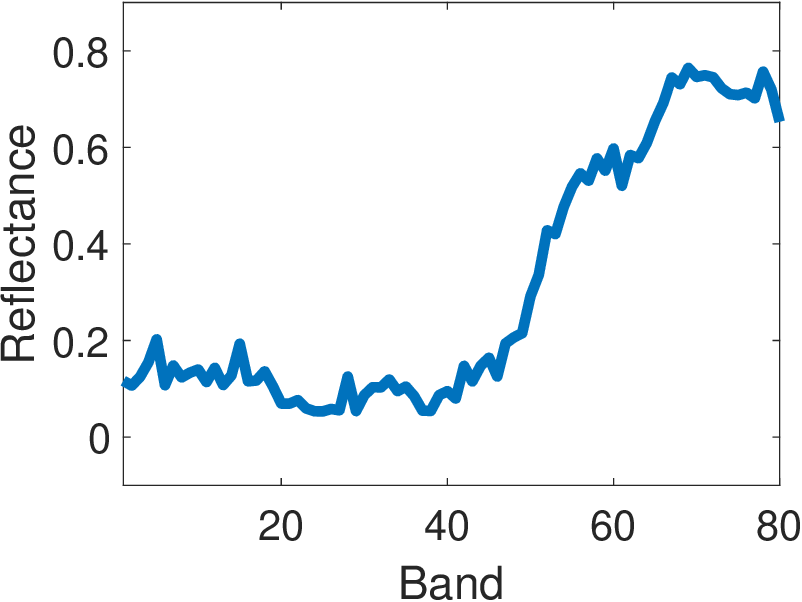}}
   \hspace{-2.1mm}
   \subfigure[{MAC-Net}~\cite{xiong2021mac}]{\label{fig:pavia_80_pixel_MAC-Net}\includegraphics[width=0.124\linewidth]{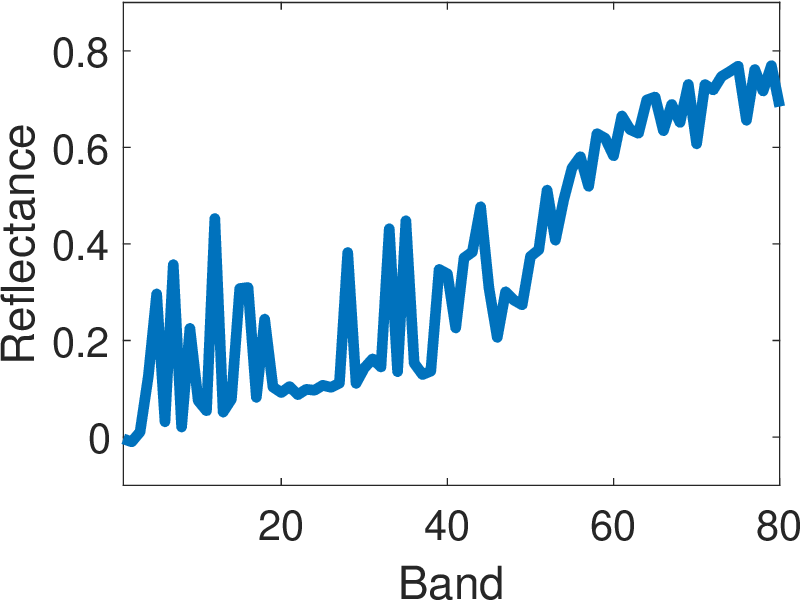}}
   \hspace{-2.1mm}
   \subfigure[{NSSNN}~\cite{guanyiman2022}]{\label{fig:pavia_80_pixel_NSSNN}\includegraphics[width=0.124\linewidth]{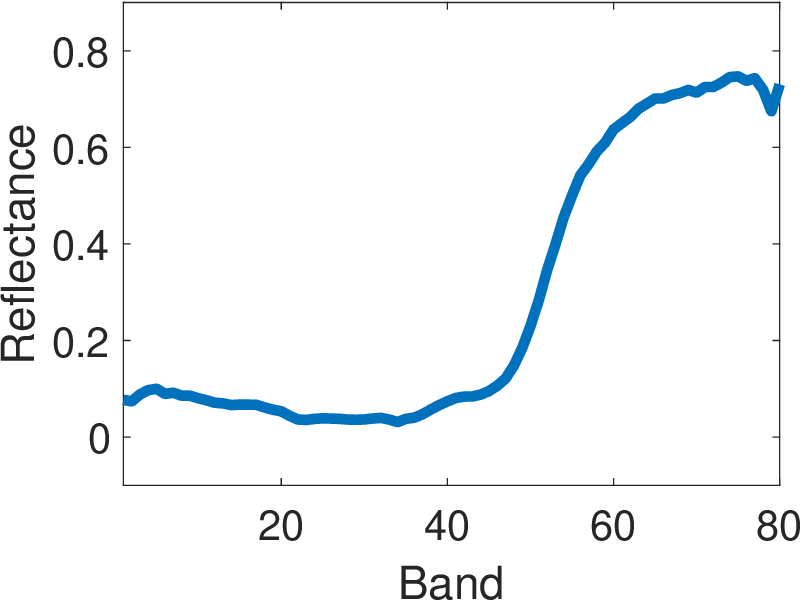}}
   \hspace{-2.1mm}
   \subfigure[{TRQ3D}~\cite{Pang2022}]{\label{fig:pavia_80_pixel_TRQ3D}\includegraphics[width=0.124\linewidth]{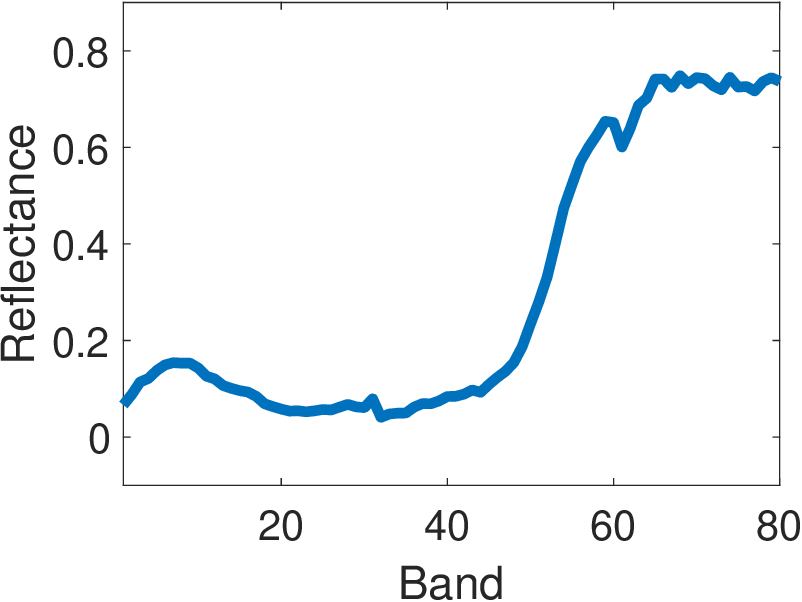}}
   \hspace{-2.1mm}
   \subfigure[{SST}~\cite{li2022spatial}]{\label{fig:pavia_80_pixel_SST}\includegraphics[width=0.124\linewidth]{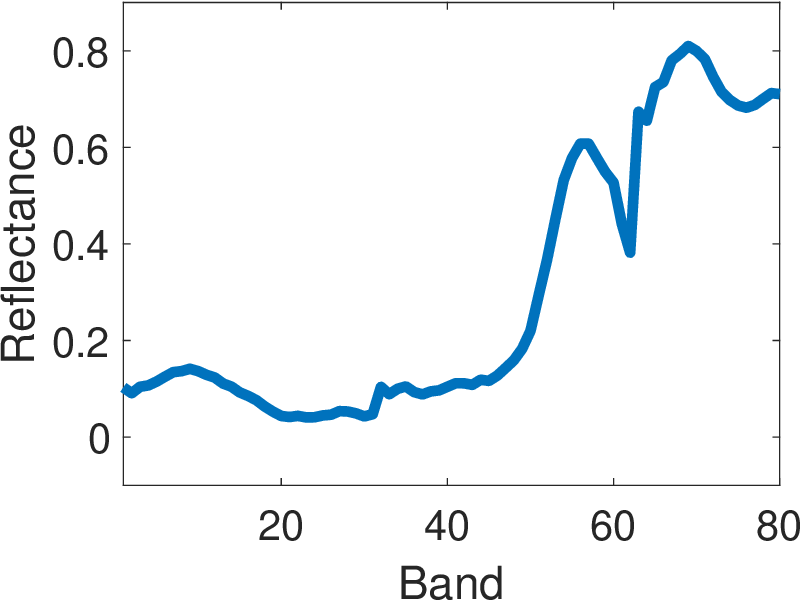}}
   \hspace{-2.1mm}
   \subfigure[{\textbf{SSRT-UNet}}]{\label{fig:pavia_80_pixel_SSRT}\includegraphics[width=0.124\linewidth]{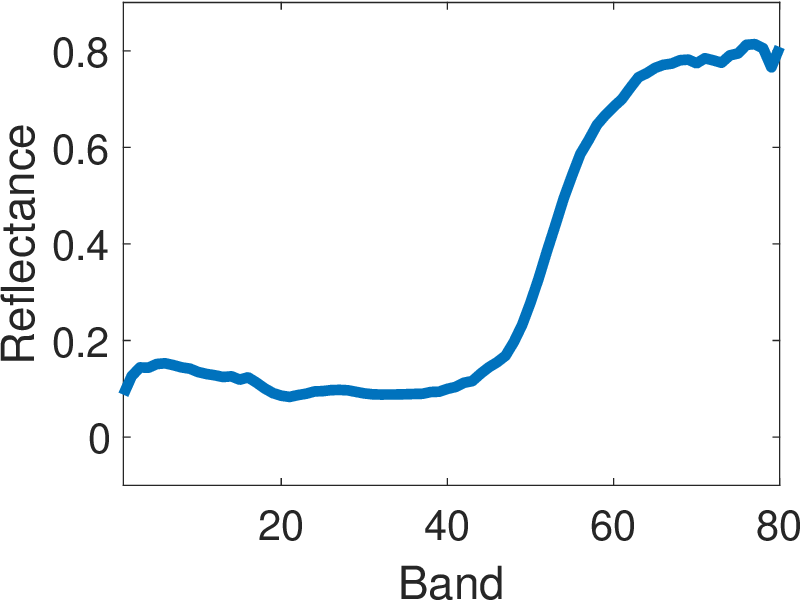}}
   \caption{{Reflectance of pixel (168,97) in the Pavia city center HSI.}} \label{fig:pavia_80_pixel}
\end{figure*}

Fig.~\ref{fig:houston_visual} visualizes  the recovered spatial images of all the bands in the case of mixture noise. Except for E-3DTV, which uses the $l_1$-norm for mixture noise, most model-based methods are designed for Gaussian noise and fail to remove the stripes and deadlines. However, the recovered HSI of E-3DTV has a color distortion, which means inaccuracies in its recovered spectra relative to the original image, indicating an inadequate ability to accurately model the spectral correlation. Due to the variation in the spectral range between Houston and the ICVL training set, the transformers of TRQ3D and SST have a certain degree of failure in the recovery of Houston and retain Gaussian noise and strips. Our SSRT-UNet models the global spectral correlation and non-local spatial self-similarity of HSI by a hybrid RNN and transformer. The more profound and refined domain knowledge it learned allows it to perform better in this challenging case.

\begin{figure*}[ht!]
   \centering
   \subfigure[Original]{\label{fig:shanghai_noisy}\includegraphics[width=0.124\linewidth]{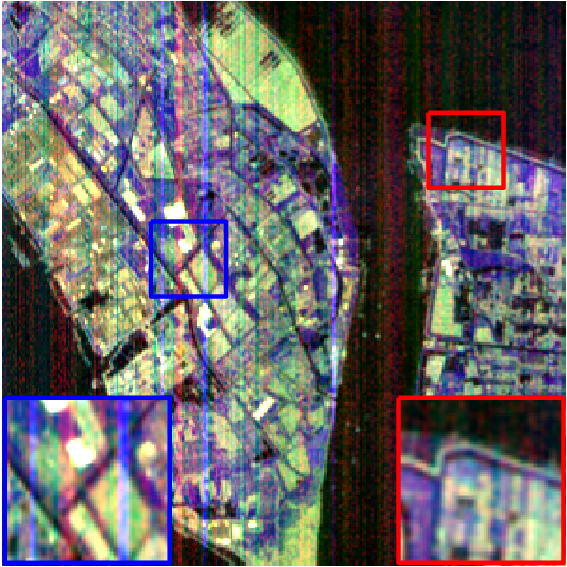}}
   \hspace{-2.1mm}
   \subfigure[BM4D \cite{maggioni2012nonlocal}]{\label{fig:shanghai_BM4D}\includegraphics[width=0.124\linewidth]{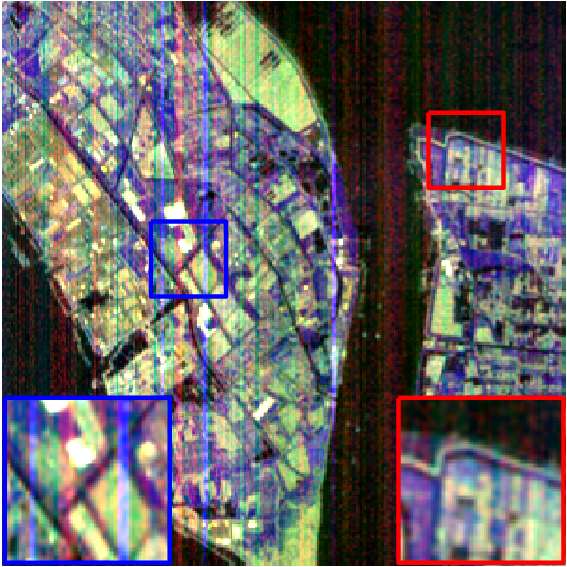}}
   \hspace{-2.1mm}
   \subfigure[MTSNMF \cite{ye2014multitask}]{\label{fig:shanghai_MTSNMF}\includegraphics[width=0.124\linewidth]{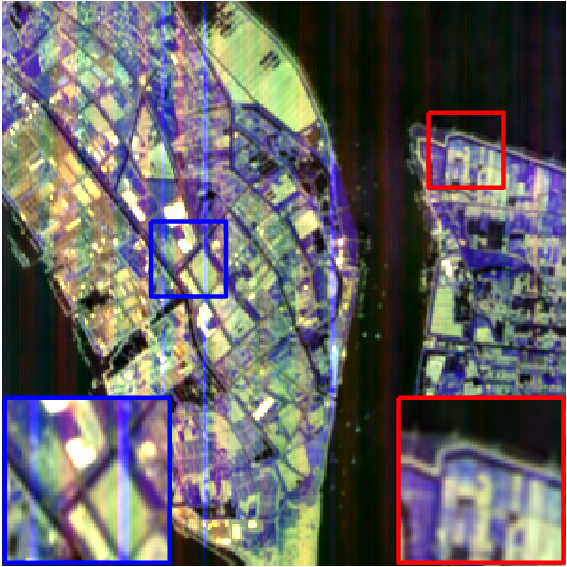}}
   \hspace{-2.1mm}
   \subfigure[LLRT \cite{Chang2017}]{\label{fig:shanghai_LLRT}\includegraphics[width=0.124\linewidth]{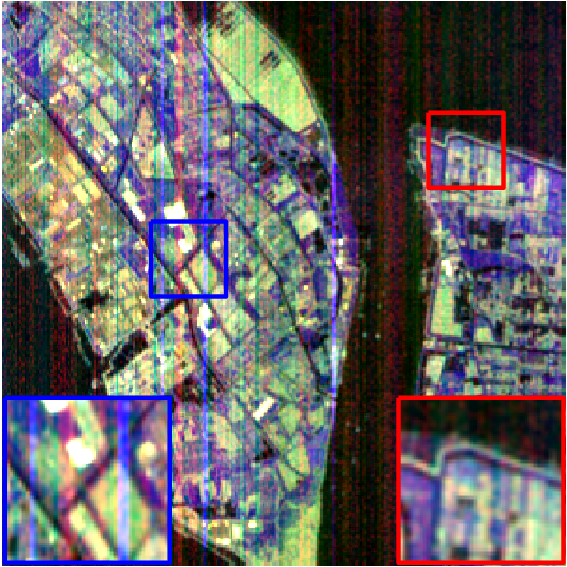}}
   \hspace{-2.1mm}
   \subfigure[NGMeet \cite{He2020}]{\label{fig:shanghai_NGMeet}\includegraphics[width=0.124\linewidth]{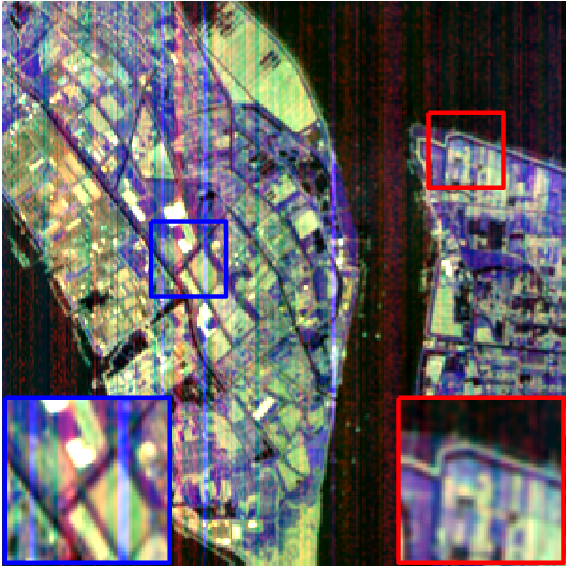}}
   \hspace{-2.1mm}
   \subfigure[LRMR \cite{Zhang2014a}]{\label{fig:shanghai_LRMR}\includegraphics[width=0.124\linewidth]{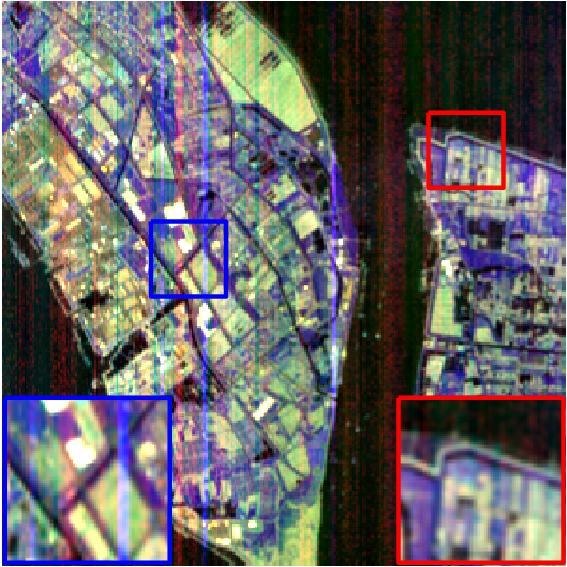}}
   \hspace{-2.1mm}
   \subfigure[FastHyDe \cite{zhuang2018fast}]{\label{fig:shanghai_FastHyDe}\includegraphics[width=0.124\linewidth]{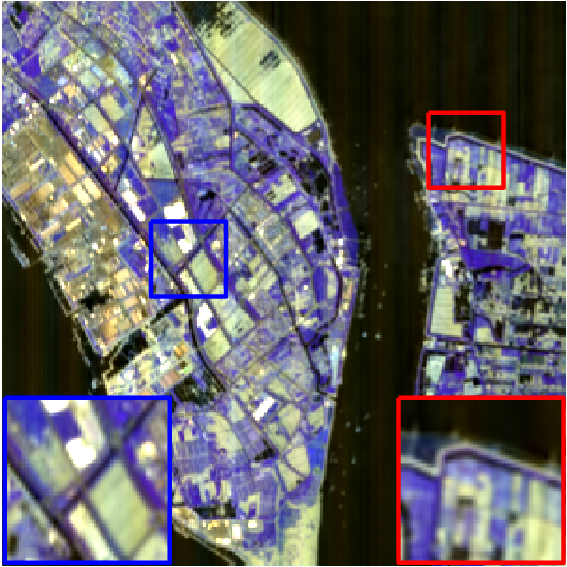}}
   \hspace{-2.1mm}
   \subfigure[LRTF$L_0$ \cite{xiong2019}]{\label{fig:shanghai_lrtfl0}\includegraphics[width=0.124\linewidth]{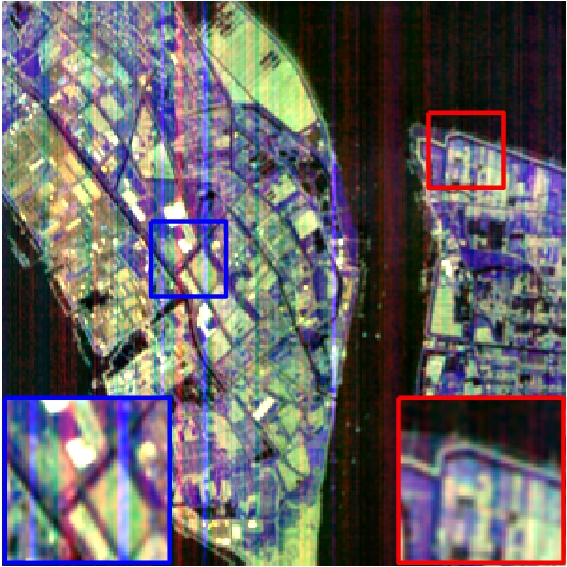}}\\
   \subfigure[E-3DTV \cite{peng2020}]{\label{fig:shanghai_e3dtv}\includegraphics[width=0.142\linewidth]{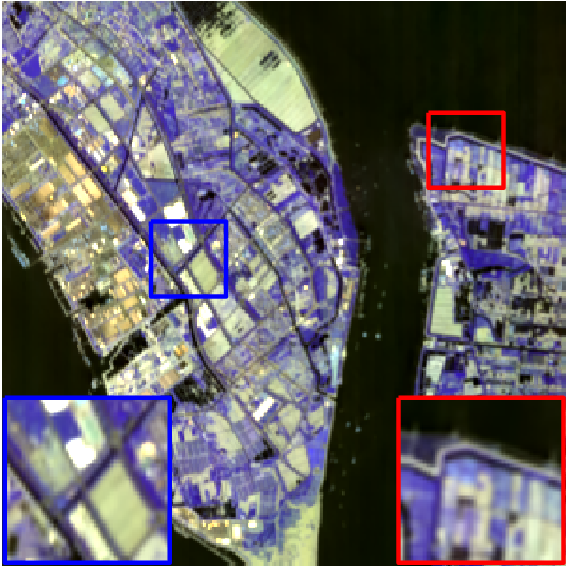}}
   \hspace{-2.1mm}
   \subfigure[T3SC \cite{Bodrito2021}]{\label{fig:shanghai_T3SC}\includegraphics[width=0.142\linewidth]{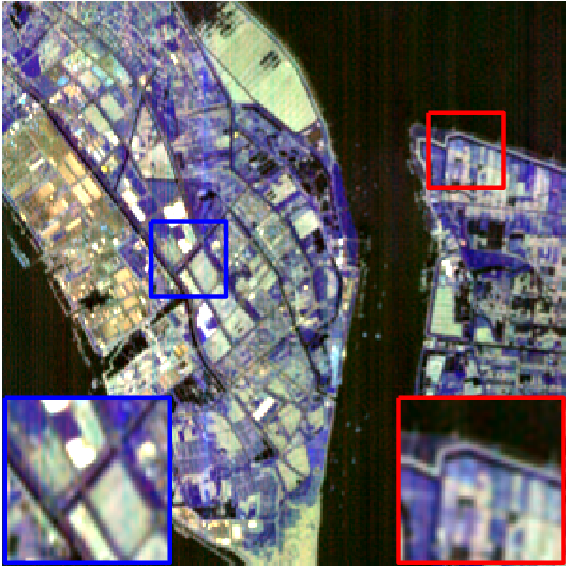}}
   \hspace{-2.1mm}
   \subfigure[MAC-Net \cite{xiong2021mac}]{\label{fig:shanghai_MAC-Net}\includegraphics[width=0.1420\linewidth]{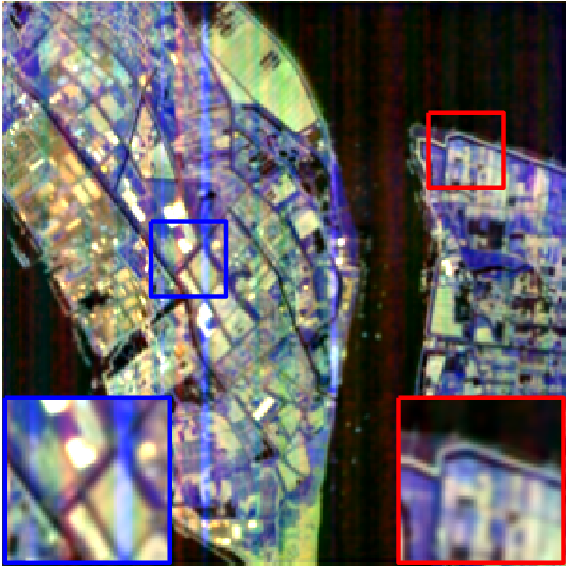}}
   \hspace{-2.1mm}
   \subfigure[NSSNN \cite{guanyiman2022}]{\label{fig:shanghai_NSSNN}\includegraphics[width=0.1420\linewidth]{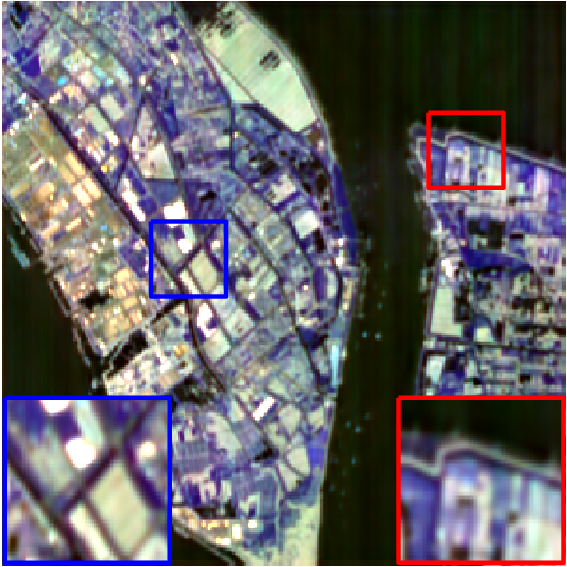}}
   \hspace{-2.1mm}
   \subfigure[TRQ3D \cite{Pang2022}]{\label{fig:shanghai_TRQ3D}\includegraphics[width=0.1420\linewidth]{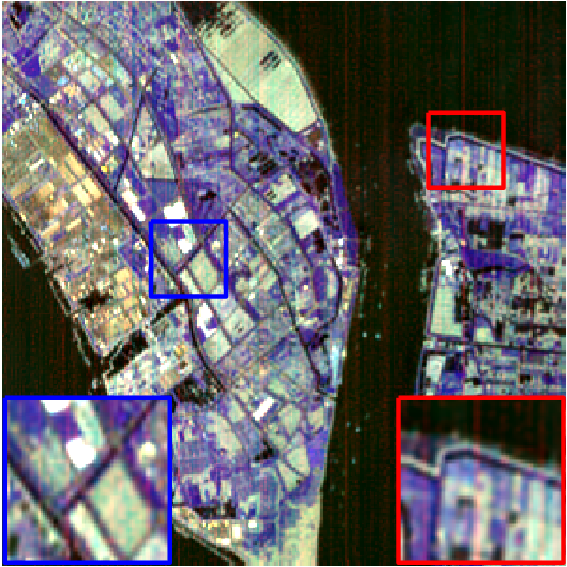}}
   \hspace{-2.1mm}
   \subfigure[SST \cite{li2022spatial}]{\label{fig:shanghai_SST}\includegraphics[width=0.1420\linewidth]{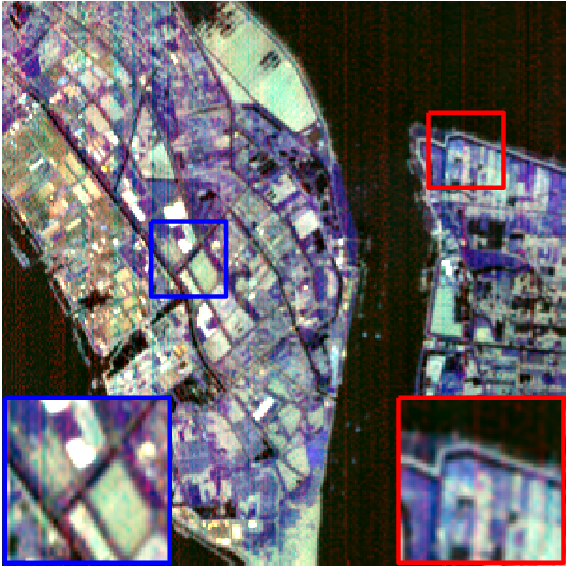}}
   \hspace{-2.1mm}
   \subfigure[\textbf{SSRT-UNet}]{\label{fig:shanghai_SSRT}\includegraphics[width=0.1420\linewidth]{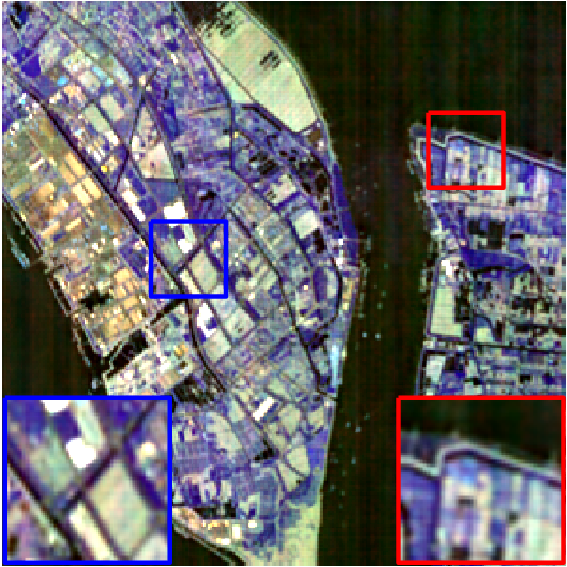}}
 \caption{Denoising results on the GF-5 Shanghai HSI. The false-color images are generated by combining bands 152, 120, and 42.} \label{fig:shanghai_visual}
\end{figure*}

\begin{figure*}[htbp]
   \centering
   \subfigure[Original]{\label{fig:shanghai_pixel_noisy}\includegraphics[width=0.124\linewidth]{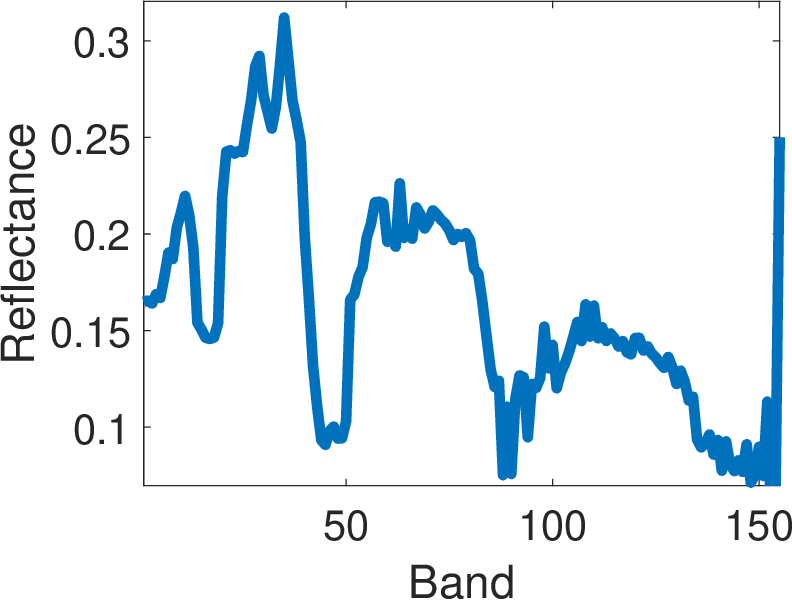}}
   \hspace{-2.1mm}
   \subfigure[BM4D \cite{maggioni2012nonlocal}]{\label{fig:shanghai_pixel_BM4D}\includegraphics[width=0.124\linewidth]{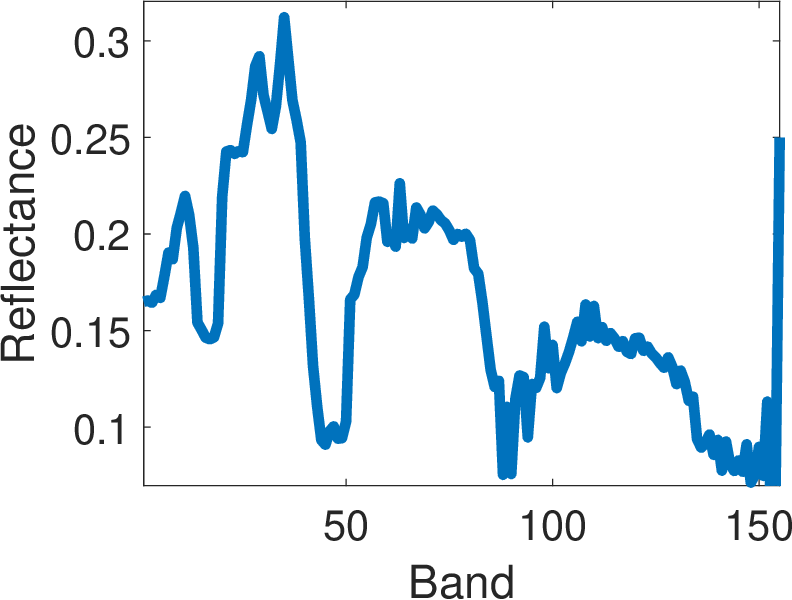}}
   \hspace{-2.1mm}
   \subfigure[MTSNMF \cite{ye2014multitask}]{\label{fig:shanghai_pixel_MTSNMF}\includegraphics[width=0.124\linewidth]{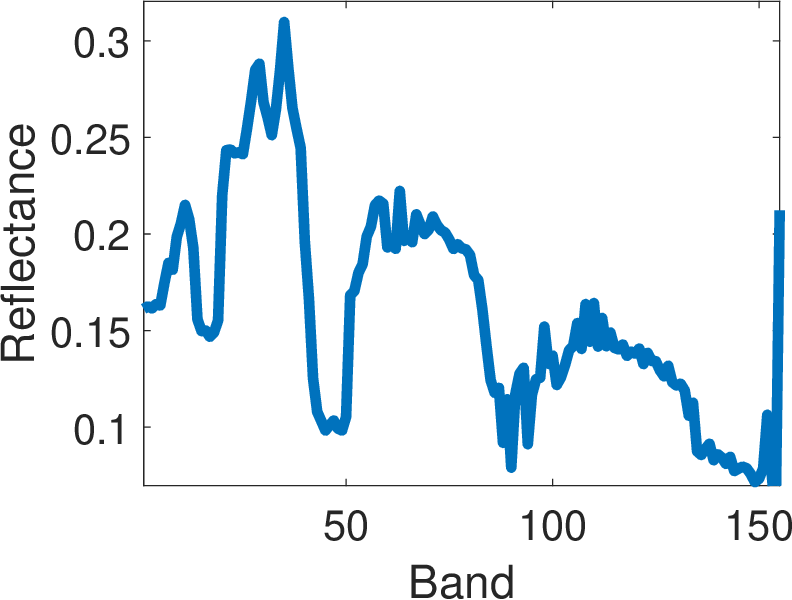}}
   \hspace{-2.1mm}
   \subfigure[LLRT \cite{Chang2017}]{\label{fig:shanghai_pixel_LLRT}\includegraphics[width=0.124\linewidth]{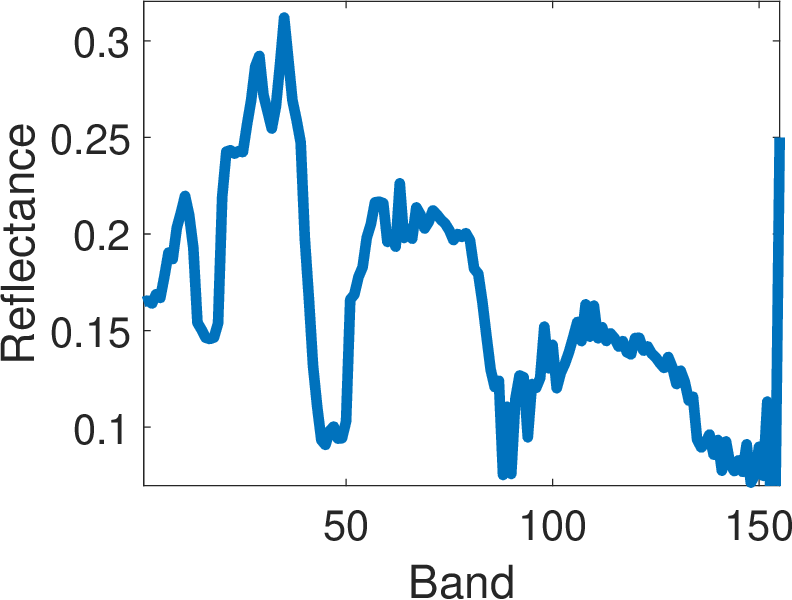}}
   \hspace{-2.1mm}
   \subfigure[NGMeet \cite{He2020}]{\label{fig:shanghai_pixel_NGMeet}\includegraphics[width=0.124\linewidth]{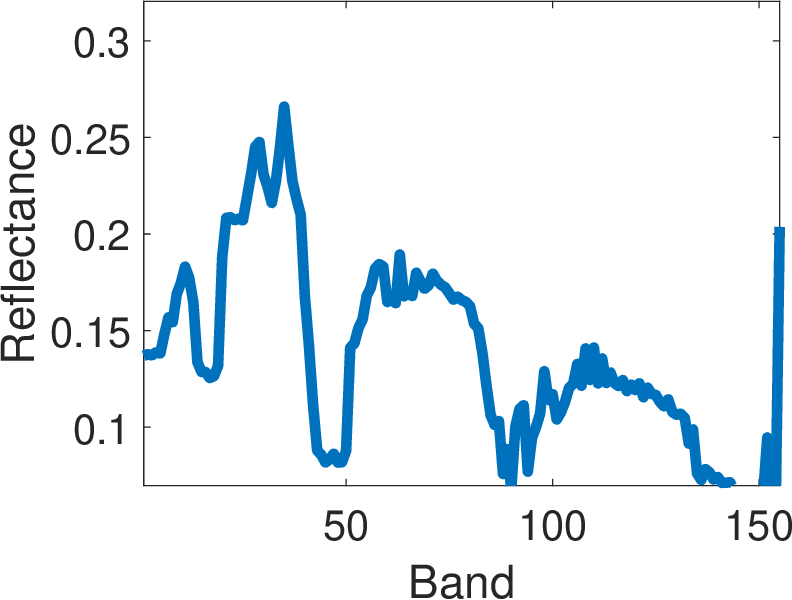}}
   \hspace{-2.1mm}
   \subfigure[LRMR \cite{Zhang2014a}]{\label{fig:shanghai_pixel_LRMR}\includegraphics[width=0.124\linewidth]{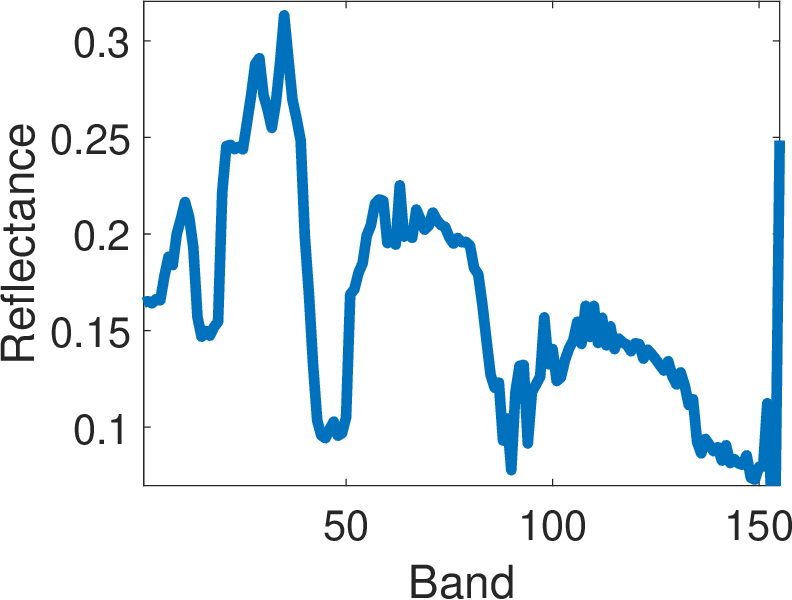}}
   \hspace{-2.1mm}
   \subfigure[FastHyDe \cite{zhuang2018fast}]{\label{fig:shanghai_pixel_FastHyDe}\includegraphics[width=0.124\linewidth]{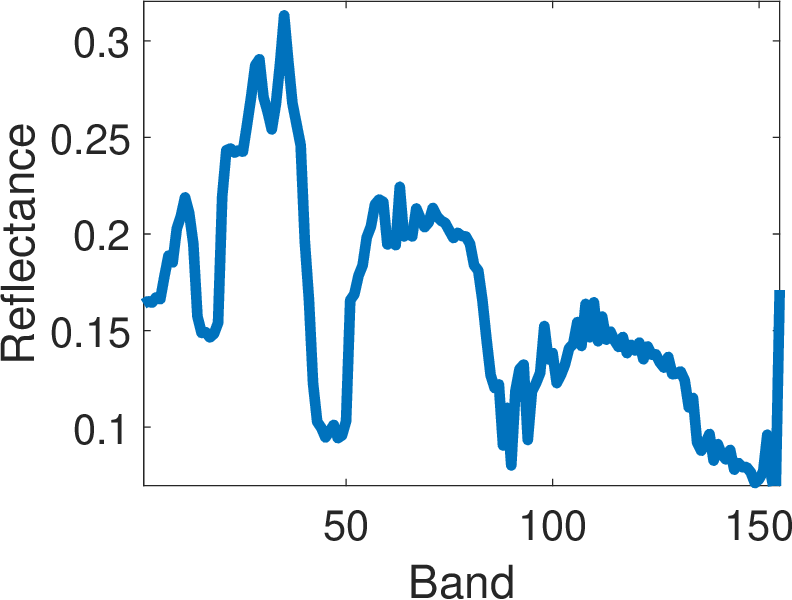}}
   \hspace{-2.1mm} 
   \subfigure[LRTF$L_0$ \cite{xiong2019}]{\label{fig:shanghai_pixel_lrtfl0}\includegraphics[width=0.124\linewidth]{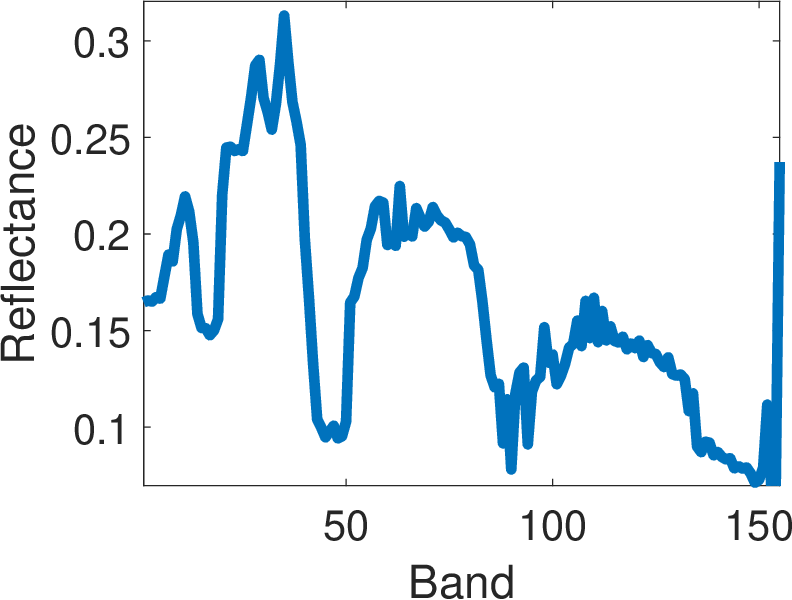}}\\
   \subfigure[E-3DTV \cite{peng2020}]{\label{fig:shanghai_pixel_e3dtv}\includegraphics[width=0.142\linewidth]{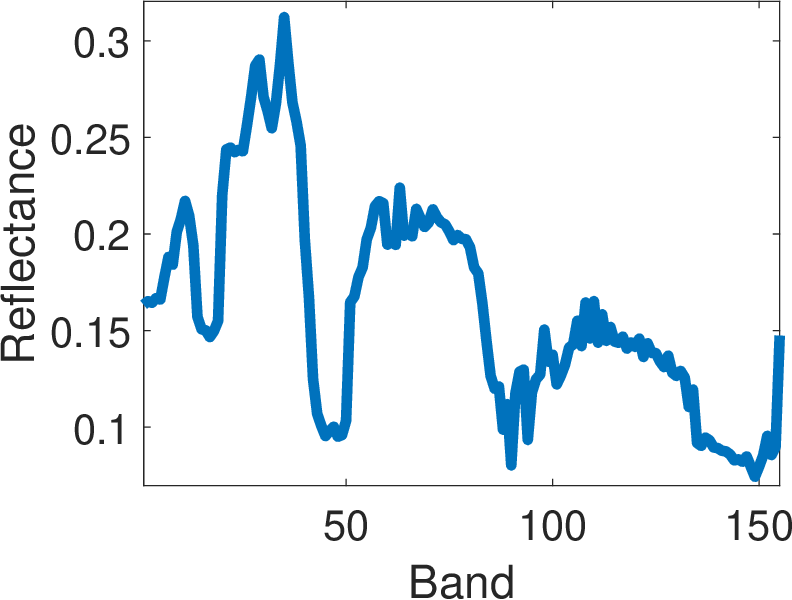}}
   \hspace{-2.1mm}
   \subfigure[T3SC \cite{Bodrito2021}]{\label{fig:shanghai_pixel_T3SC}\includegraphics[width=0.1420\linewidth]{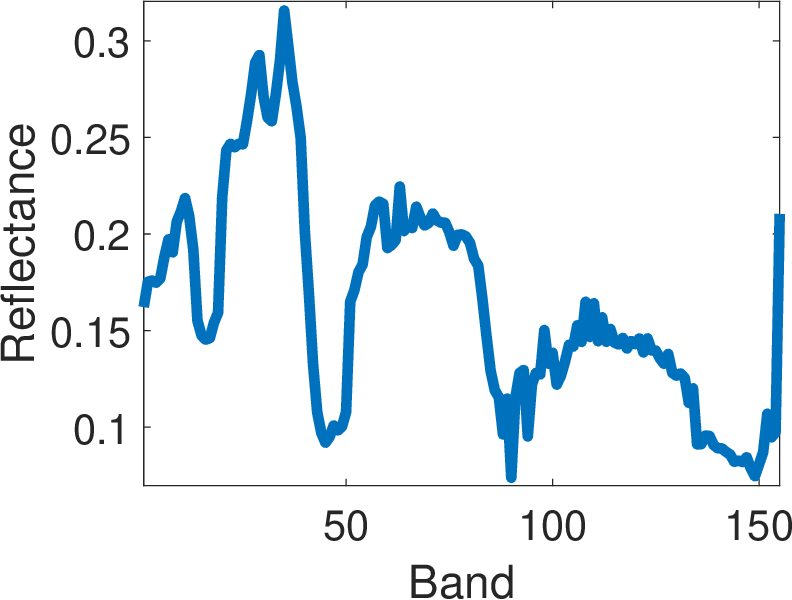}}
   \hspace{-2.1mm}
   \subfigure[MAC-Net \cite{xiong2021mac}]{\label{fig:shanghai_pixel_MAC-Net}\includegraphics[width=0.1420\linewidth]{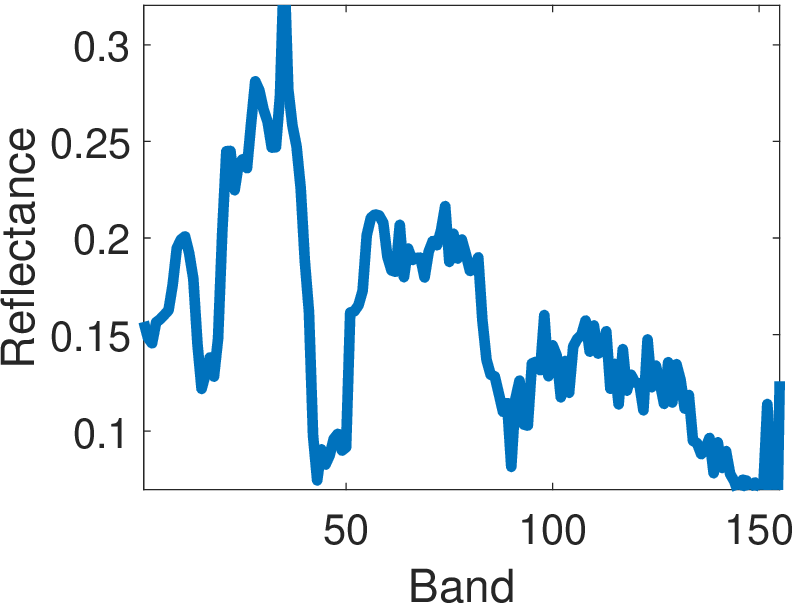}}
   \hspace{-2.1mm}
   \subfigure[NSSNN \cite{guanyiman2022}]{\label{fig:shanghai_pixel_NSSNN}\includegraphics[width=0.1420\linewidth]{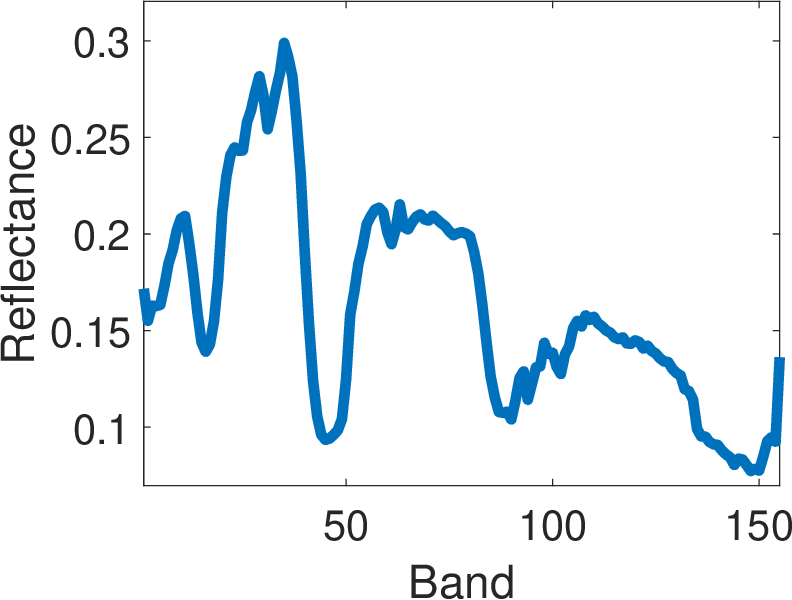}}
   \hspace{-2.1mm}
   \subfigure[TRQ3D \cite{Pang2022}]{\label{fig:shanghai_pixel_TRQ3D}\includegraphics[width=0.1420\linewidth]{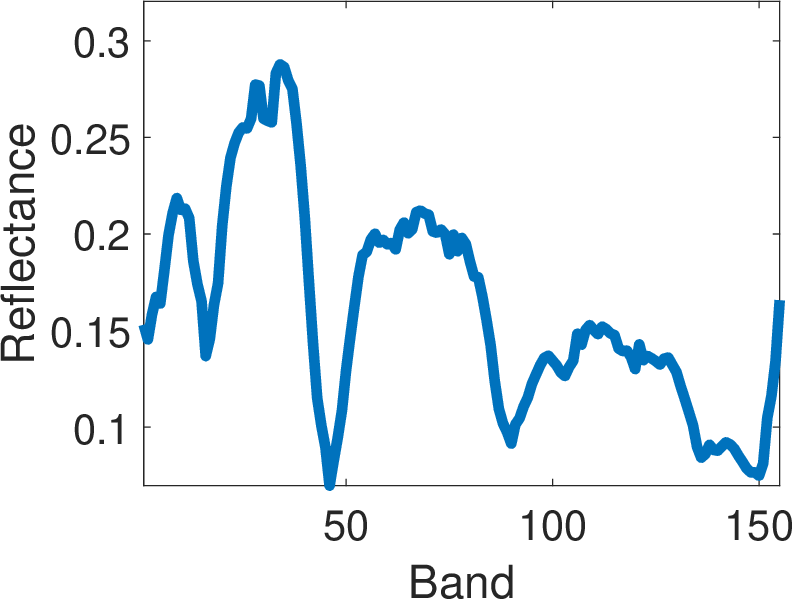}}
   \hspace{-2.1mm}
   \subfigure[SST \cite{li2022spatial}]{\label{fig:shanghai_pixel_SST}\includegraphics[width=0.1420\linewidth]{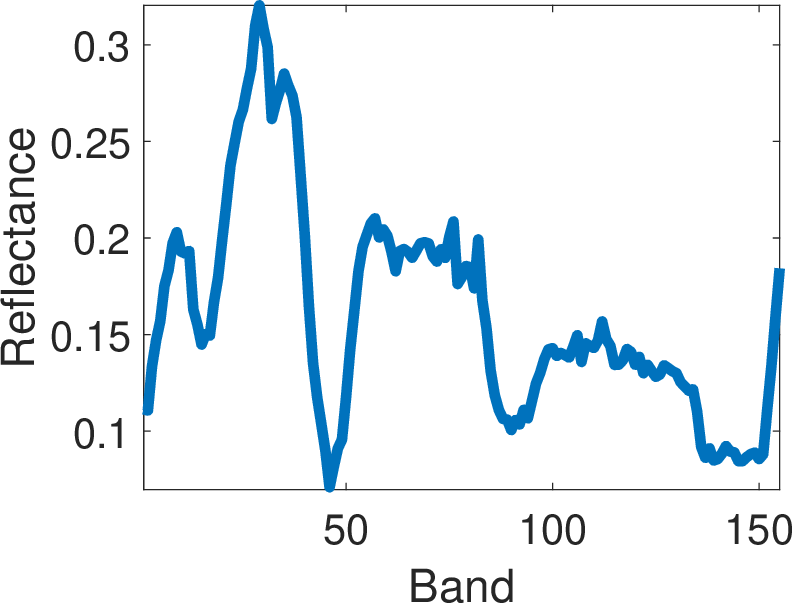}}
   \hspace{-2.1mm}
   \subfigure[\textbf{SSRT-UNet}]{\label{fig:shanghai_pixel_SSRT}\includegraphics[width=0.1420\linewidth]{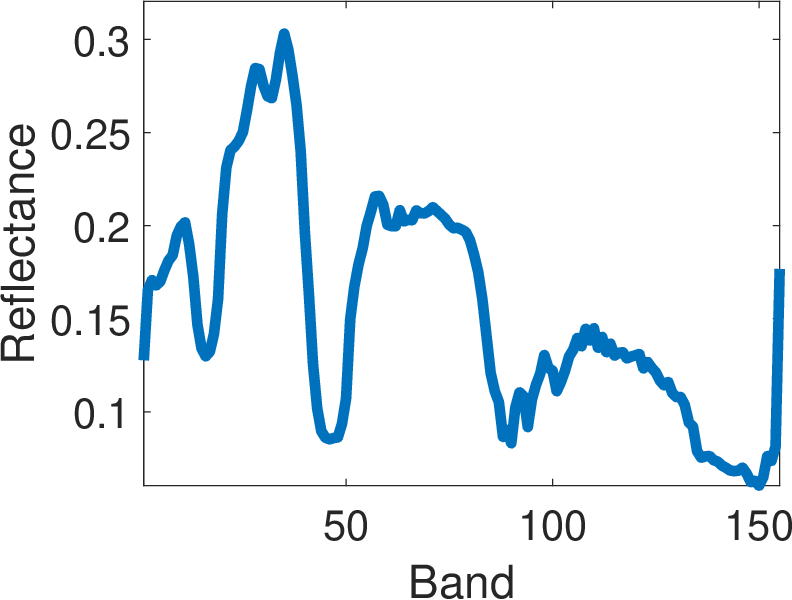}}
      \caption{Reflectance of pixel (257,285) in the GF-5 Shanghai HSI.} \label{fig:shanghai_pixel}
\end{figure*}

The spectral reflectances are shown in Fig.~\ref{fig:houston_pixel}. The addition of mixed noise makes jitters in the recovered spectra reflectances with the model-based methods and the low-rank-based T3SC and MAC-Net. 
From Fig.~\ref{fig:houston_pixel_noisy}, we can see that the  Gaussian noise makes the spectral reflection distortion drastic. Most model-based methods exhibit noticeable spectra distortion in the recovered spectrum. With the consideration of the mixture noise, E-3DTV models the low-rankness of the spatial and spectral gradient map a HSI and yields the best match with the clean HSI among the model-based methods.
TRQ3D and SST cannot be applied directly to images that have a different number of bands with the training set, leading to incomplete GSC and less matched spectral reflectances with clean HSI.  Exploiting the GSC in a refined two-gated recurrent unit, NSSNN can better fit the spectral reflectance than TRQ3D and SST. The transformer-based RNN allows the proposed SSRT-UNet to accurately model the GSC as the transformer does, even when the number of bands changes, and helps the SSRT-UNet achieve the best match with the clean image.

\subsubsection{Pavia City Center HSI}

The remote sensing Pavia city center HSI was obtained using the Reflective Optics System Imaging Spectrometer (ROSIS-03) with a spectral range of 430 to 860 nm and a geometric resolution of 1.3 meters. Following~\cite{He2015}, the first 22 noisy bands were removed, and a $200\times200$ clean sub-image with 80 bands was selected as the ground truth HSI. Mixture noise was subsequently added to the clean HSI to generate the noisy counterpart. Other experimental settings were maintained in line with those used for the Houston 2018 HSI.

Table~\ref{tab:paviacity} presents the denoising performance. Thanks to the low-rank representation in the gradient domain, which facilitates global spectral correlation modelling, E-3DTV outperforms all the model-based methods.  Leveraging the recurrent processing scheme for comprehensive utilization of global spectral correlation, SSRT-UNet and NSSNN outperform other learning-based methods, even with an increased number of bands.  However, it is noteworthy that SSRT-UNet falls behind in the SAM index. The possible reason is that the Pavia Centre HSI contains 80 bands, which is significantly more than the training set. Using recurrent computation alone may not fully capture the global spectral correlation with dramatically more bands than the training set.

The visualization results are presented in Fig.~\ref{fig:pavia_80_visual} and Fig.~\ref{fig:pavia_80_pixel}. Among the model-based methods, NGMeet and LRTF$L_0$ can remove most of the noise. However, they failed to recover HSIs with the correct color and matched reflectance with the clean HSI. By combining the TV regularization term with low rank prior, E-3DTV provides superior HSI recovery in both spatial and spectral domains. Regarding to deep learning-based methods, TRQ3D and SST, employing transformer-based modules to model non-local spatial self-similarity, produce sharper edges compared to MAC-Net and T3SC. However, as the number of bands increases, their inability to model global spectral correlation hinders them from completely eliminating stripe noise and accurately recovering spectral reflectances matching the clean HSI. Thanks to the exploit of global spectral correlation with recurrent computation along bands, NSSNN can recover the spectra reflectance more precisely. The proposed SSRT-UNet utilizes a two-branch transformer to jointly exploit global spectral correlation and non-local spatial self-similarity. This enables the model to learn robust domain knowledge and achieve superior spatial and spectral recovery.

\begin{figure*}[htbp]
  \centering
  \subfigure[{Original}]{\label{fig:eo1_noisy}\includegraphics[width=0.124\linewidth]{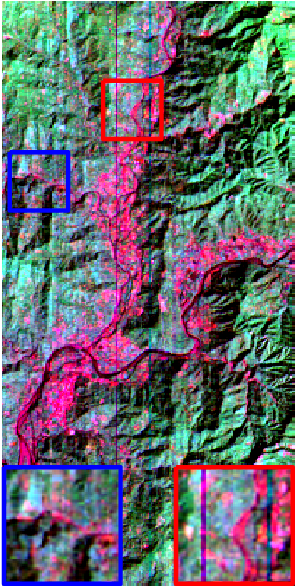}}
  \hspace{-2.1mm}
  \subfigure[{BM4D} \cite{maggioni2012nonlocal}]{\label{fig:eo1_BM4D}\includegraphics[width=0.124\linewidth]{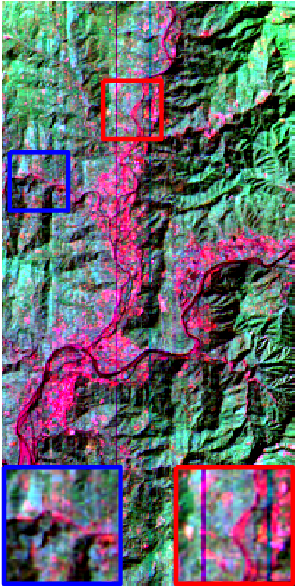}}
  \hspace{-2.1mm}
  \subfigure[{MTSNMF} \cite{ye2014multitask}]{\label{fig:eo1_MTSNMF}\includegraphics[width=0.124\linewidth]{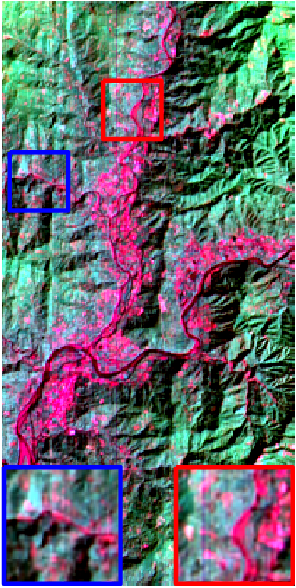}}
  \hspace{-2.1mm}
  \subfigure[{LLRT} \cite{Chang2017}]{\label{fig:eo1_LLRT}\includegraphics[width=0.124\linewidth]{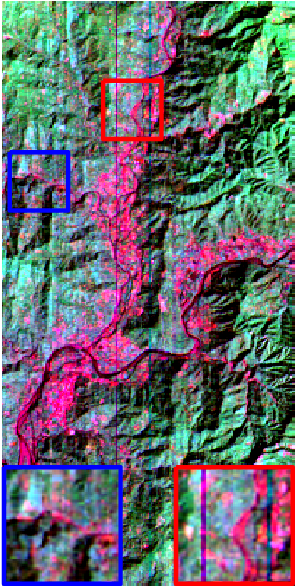}}
  \hspace{-2.1mm}
  \subfigure[{NGMeet} \cite{He2020}]{\label{fig:eo1_NGMeet}\includegraphics[width=0.124\linewidth]{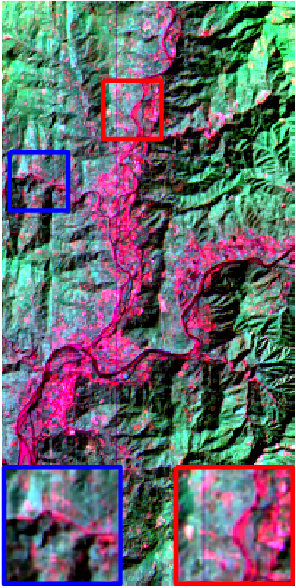}}
  \hspace{-2.1mm}
  \subfigure[{LRMR} \cite{Zhang2014a}]{\label{fig:eo1_LRMR}\includegraphics[width=0.124\linewidth]{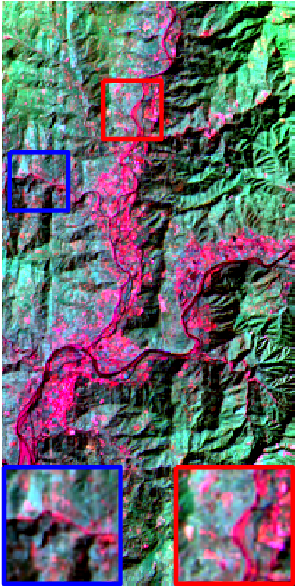}}
  \hspace{-2.1mm}
  \subfigure[{FastHyDe} \cite{zhuang2018fast}]{\label{fig:eo1_FastHyDe}\includegraphics[width=0.124\linewidth]{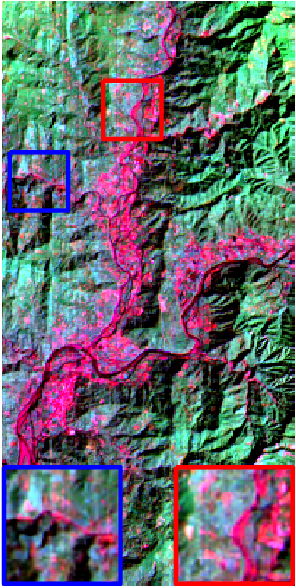}}
  \hspace{-2.1mm}
  \subfigure[{LRTF$L_0$} \cite{xiong2019}]{\label{fig:eo1_lrtfl0}\includegraphics[width=0.124\linewidth]{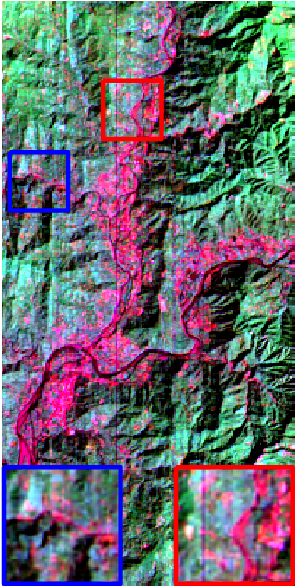}}\\
  \subfigure[{E-3DTV} \cite{peng2020}]{\label{fig:eo1_e3dtv}\includegraphics[width=0.142\linewidth]{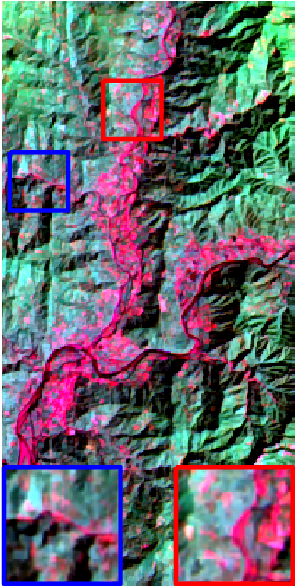}}
  \hspace{-2.1mm}
  \subfigure[{T3SC} \cite{Bodrito2021}]{\label{fig:eo1_T3SC}\includegraphics[width=0.1420\linewidth]{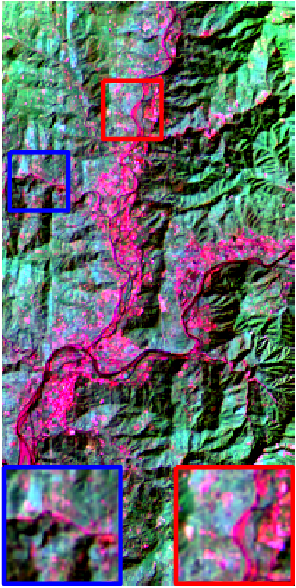}}
  \hspace{-2.1mm}
  \subfigure[{MAC-Net} \cite{xiong2021mac}]{\label{fig:eo1_MAC-Net}\includegraphics[width=0.1421\linewidth]{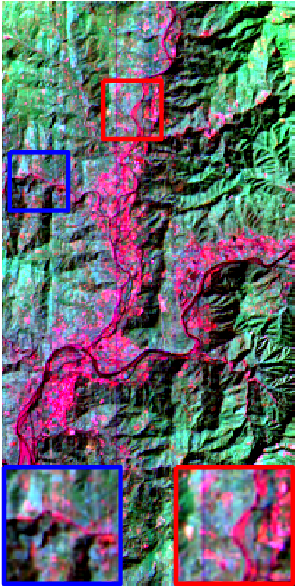}}
  \hspace{-2.1mm}
  \subfigure[{NSSNN} \cite{guanyiman2022}]{\label{fig:eo1_NSSNN}\includegraphics[width=0.1420\linewidth]{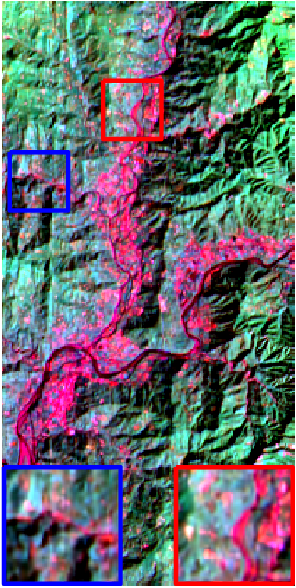}}
  \hspace{-2.1mm}
  \subfigure[{TRQ3D} \cite{Pang2022}]{\label{fig:eo1_TRQ3D}\includegraphics[width=0.1420\linewidth]{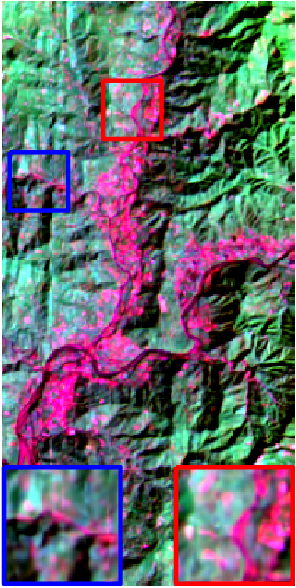}}
  \hspace{-2.1mm}
  \subfigure[{SST} \cite{li2022spatial}]{\label{fig:eo1_SST}\includegraphics[width=0.1420\linewidth]{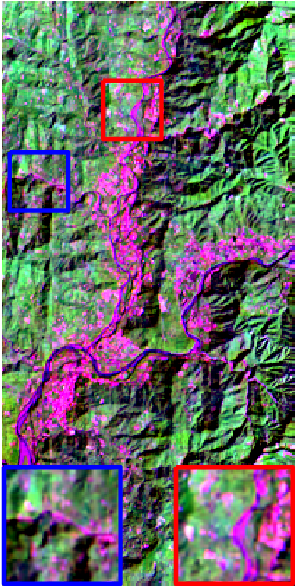}}
  \hspace{-2.1mm}
  \subfigure[{\textbf{SSRT-UNet}}]{\label{fig:eo1_SSRT}\includegraphics[width=0.1420\linewidth]{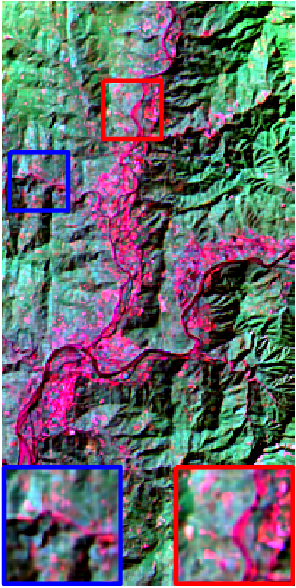}}
\caption{{Denoising results on the EO-1 HSI. The false-color images are generated by combining bands 148, 73, and 30.}} \label{fig:eo1_visual}
\end{figure*}

\begin{figure*}[htbp]
  \centering
  \subfigure[{Original}]{\label{fig:eo1_pixel_noisy}\includegraphics[width=0.124\linewidth]{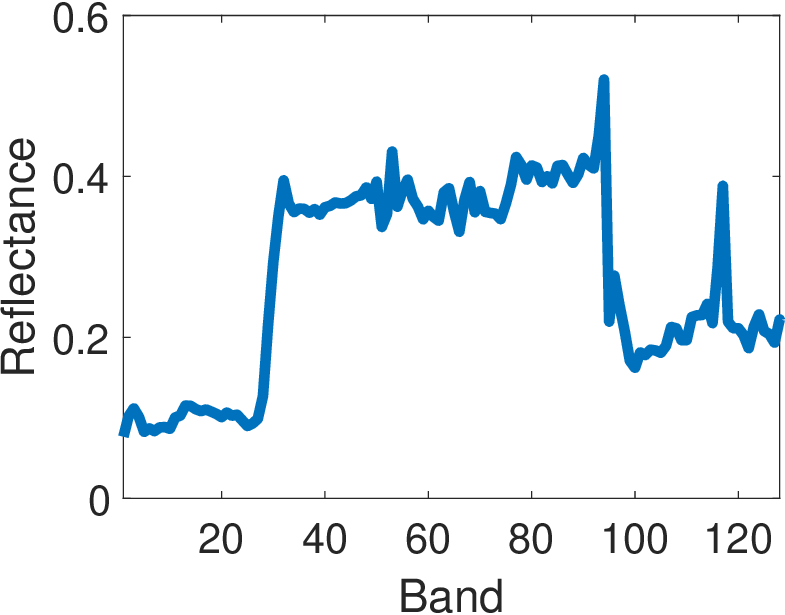}}
  \hspace{-2.1mm}
  \subfigure[{BM4D \cite{maggioni2012nonlocal}}]{\label{fig:eo1_pixel_BM4D}\includegraphics[width=0.124\linewidth]{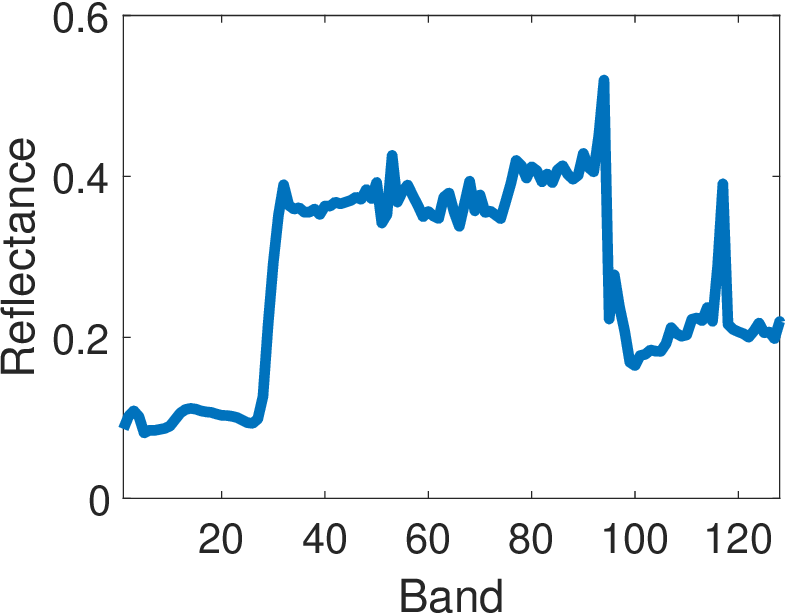}}
  \hspace{-2.1mm}
  \subfigure[{MTSNMF \cite{ye2014multitask}}]{\label{fig:eo1_pixel_MTSNMF}\includegraphics[width=0.124\linewidth]{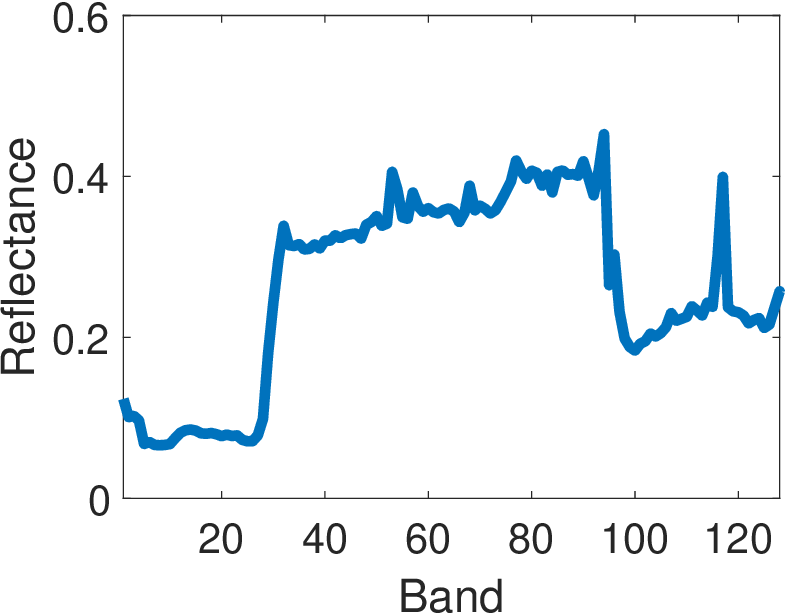}}
  \hspace{-2.1mm}
  \subfigure[{LLRT \cite{Chang2017}}]{\label{fig:eo1_pixel_LLRT}\includegraphics[width=0.124\linewidth]{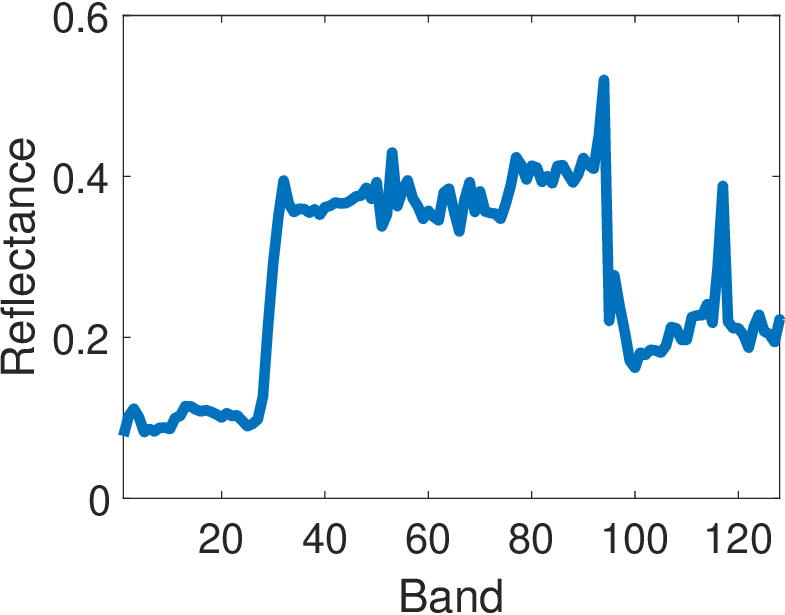}}
  \hspace{-2.1mm}
  \subfigure[{NGMeet \cite{He2020}}]{\label{fig:eo1_pixel_NGMeet}\includegraphics[width=0.124\linewidth]{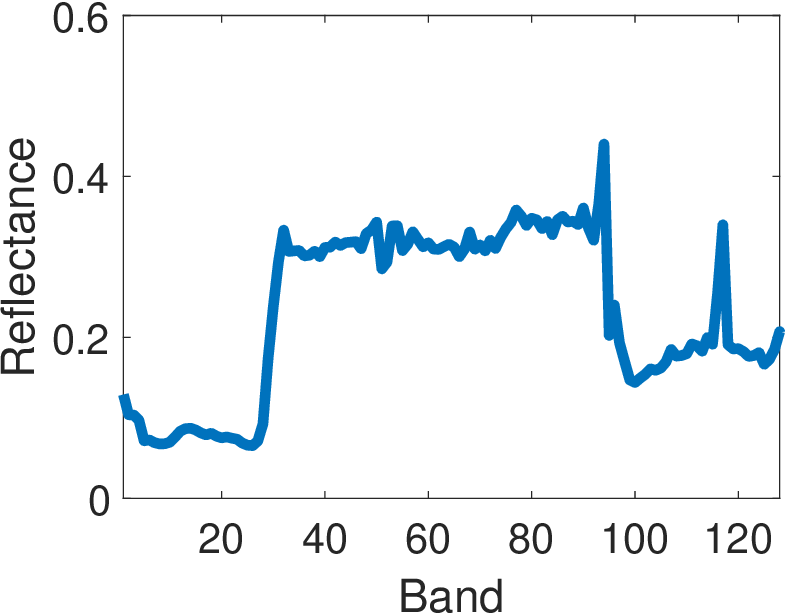}}
  \hspace{-2.1mm}
  \subfigure[{LRMR \cite{Zhang2014a}}]{\label{fig:eo1_pixel_LRMR}\includegraphics[width=0.124\linewidth]{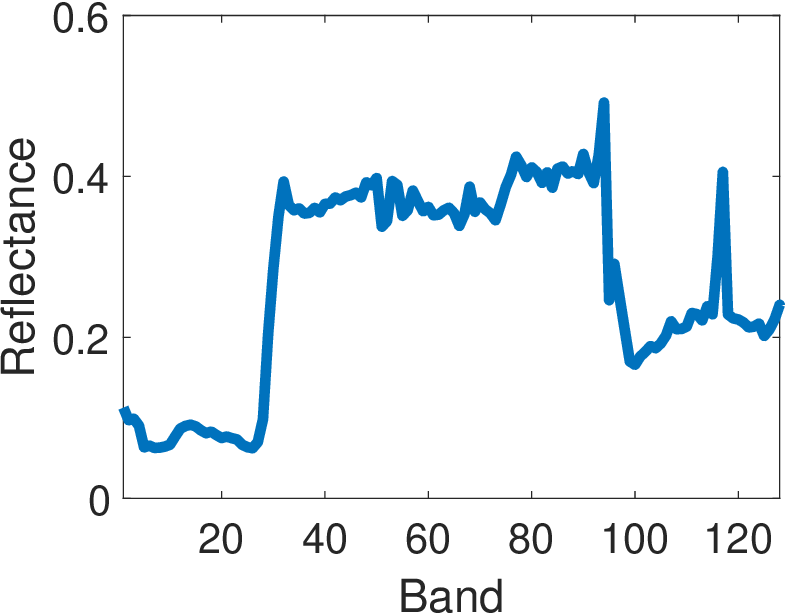}}\hspace{-2.1mm}
  \subfigure[{FastHyDe \cite{zhuang2018fast}}]{\label{fig:eo1_pixel_FastHyDe}\includegraphics[width=0.124\linewidth]{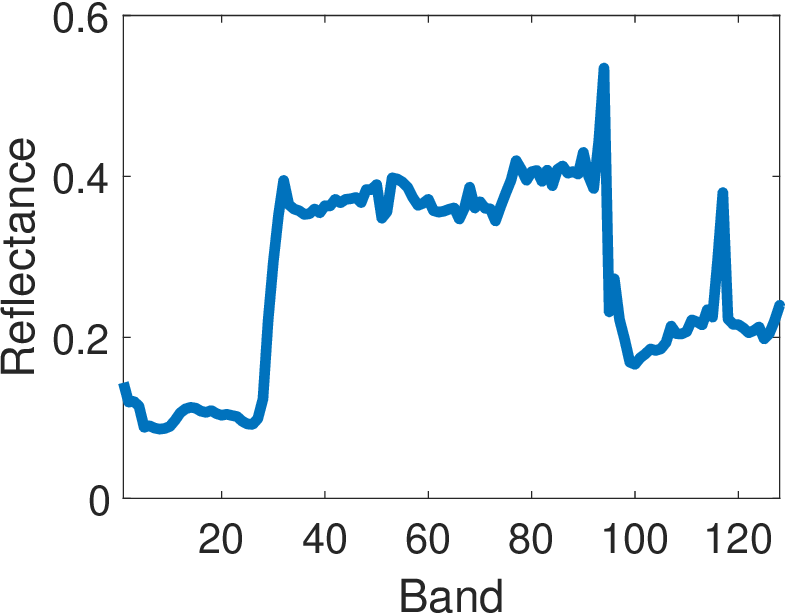}}
  \hspace{-2.1mm} 
  \subfigure[{LRTF$L_0$ \cite{xiong2019}}]{\label{fig:eo1_pixel_lrtfl0}\includegraphics[width=0.124\linewidth]{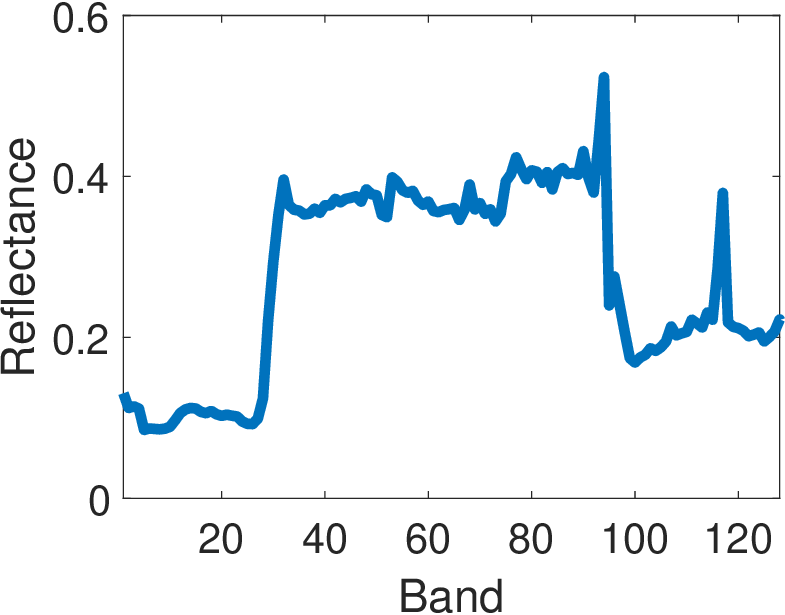}}\\
  \subfigure[{E-3DTV \cite{peng2020}}]{\label{fig:eo1_pixel_e3dtv}\includegraphics[width=0.142\linewidth]{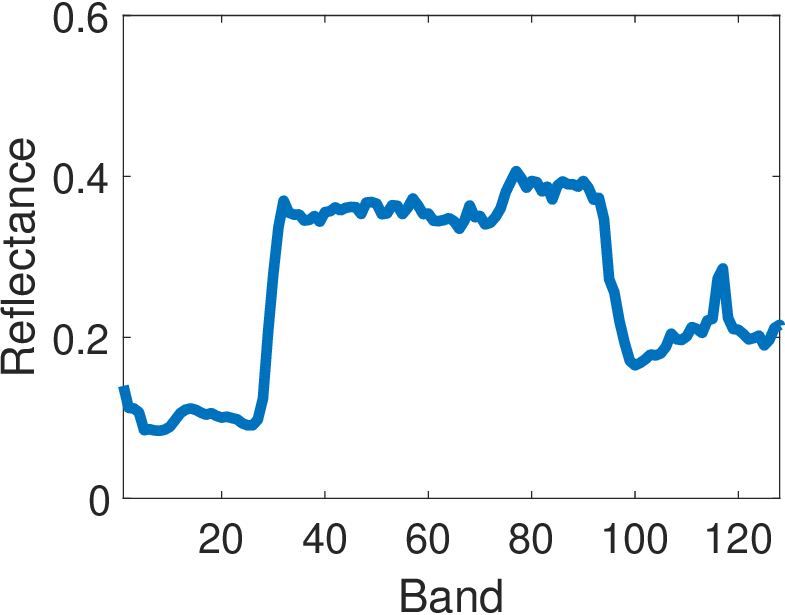}}
  \hspace{-2.1mm}
  \subfigure[{T3SC \cite{Bodrito2021}}]{\label{fig:eo1_pixel_T3SC}\includegraphics[width=0.1420\linewidth]{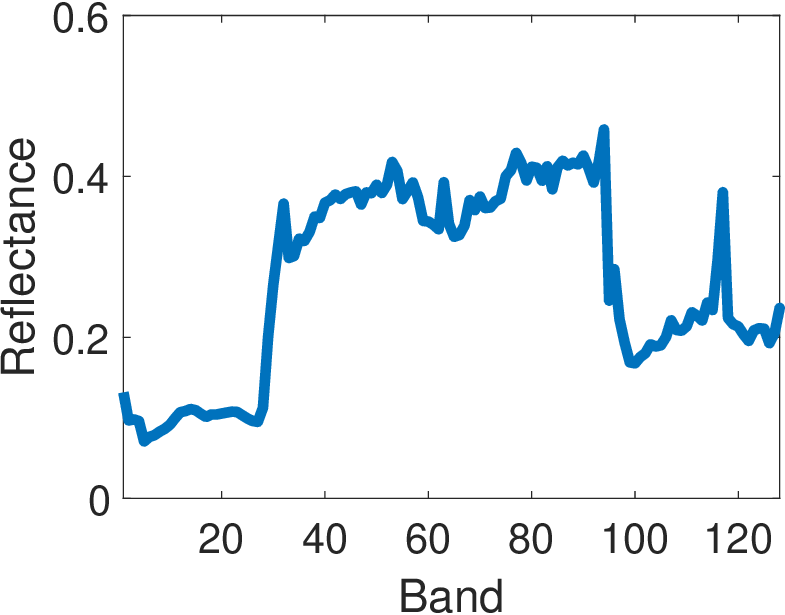}}
  \hspace{-2.1mm}
  \subfigure[{MAC-Net \cite{xiong2021mac}}]{\label{fig:eo1_pixel_MAC-Net}\includegraphics[width=0.1420\linewidth]{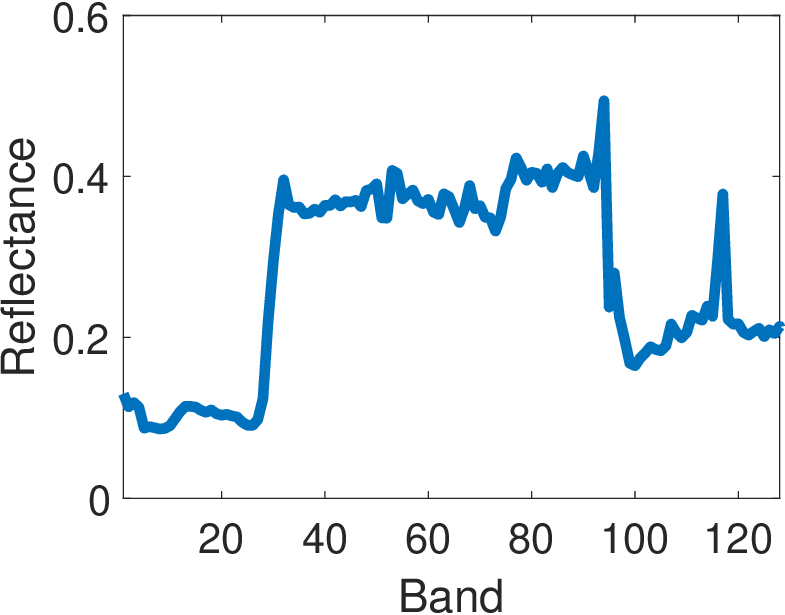}}
  \hspace{-2.1mm}
  \subfigure[{NSSNN \cite{guanyiman2022}}]{\label{fig:eo1_pixel_NSSNN}\includegraphics[width=0.1420\linewidth]{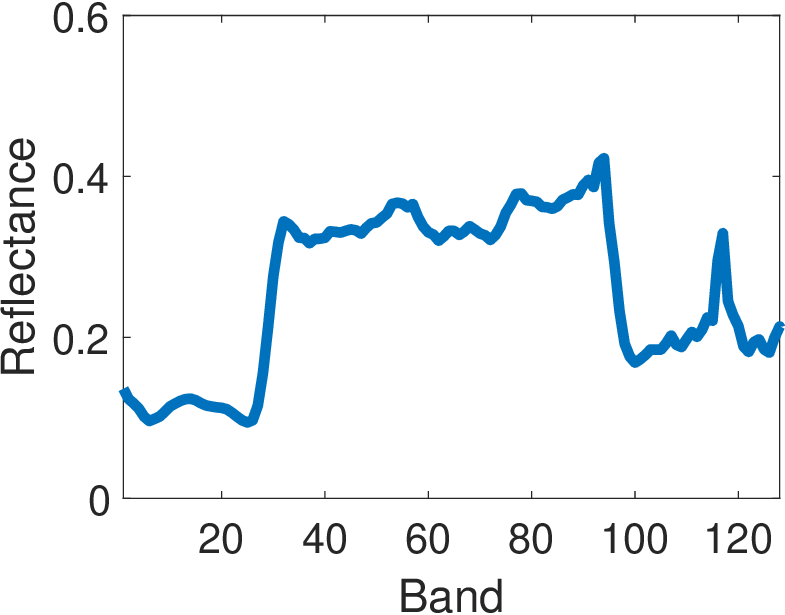}}
  \hspace{-2.1mm}
  \subfigure[{TRQ3D \cite{Pang2022}}]{\label{fig:eo1_pixel_TRQ3D}\includegraphics[width=0.1420\linewidth]{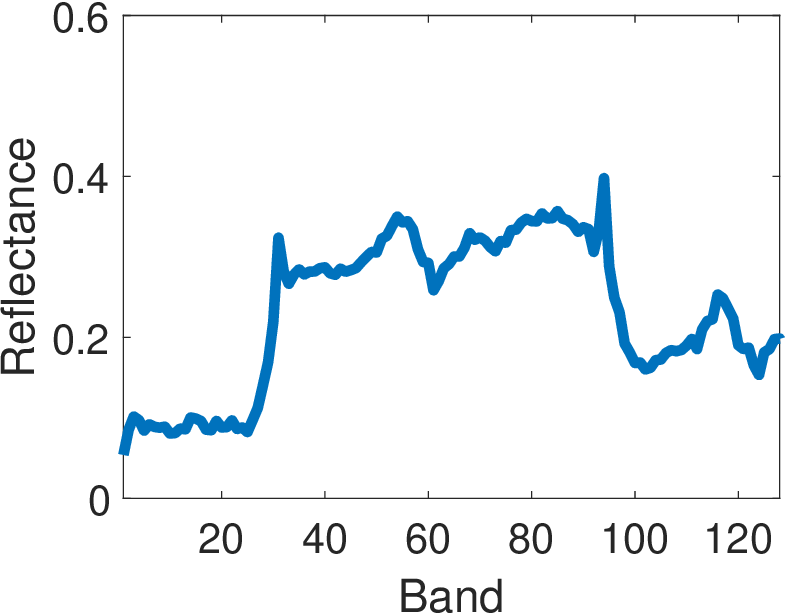}}
  \hspace{-2.1mm}
  \subfigure[{SST \cite{li2022spatial}}]{\label{fig:eo1_pixel_SST}\includegraphics[width=0.1420\linewidth]{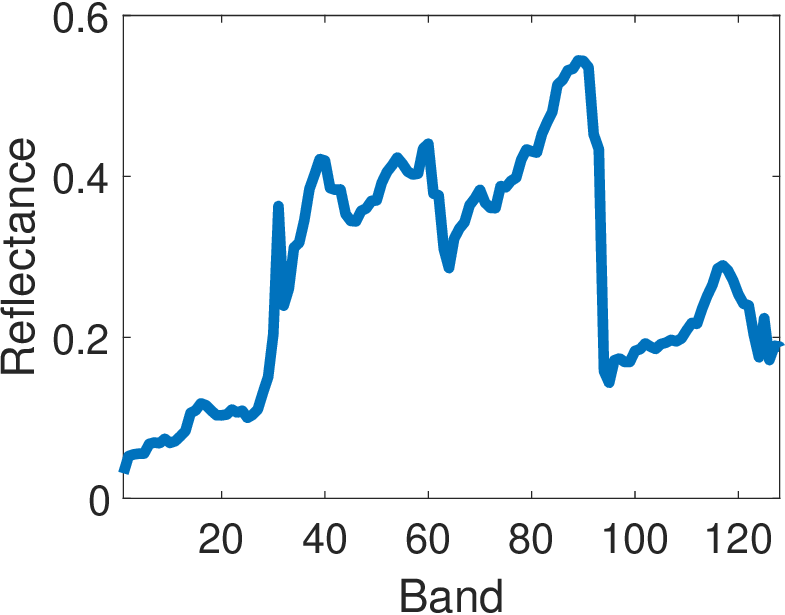}}
  \hspace{-2.1mm}
  \subfigure[{\textbf{SSRT-UNet}}]{\label{fig:eo1_pixel_SSRT}\includegraphics[width=0.1420\linewidth]{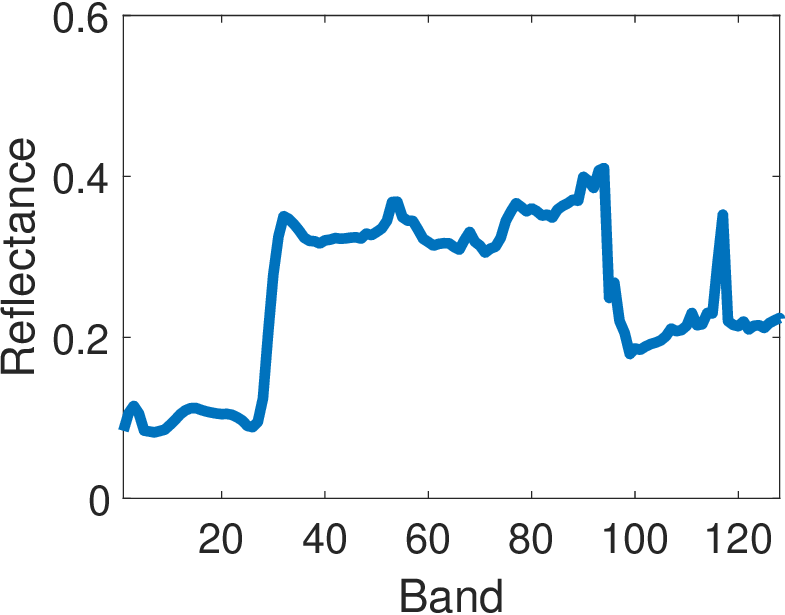}}
     \caption{{Reflectance of pixel (112,156) in the EO-1 HSI.}} \label{fig:eo1_pixel}
\end{figure*}

\subsection{Comparison on Real-world HSIs}

We conducted additional experiments on real-world remote sensing  HSIs to comprehensively assess denoising efficacy across all the competing methods. A qualitative comparative analysis is provided to evaluate the denoising outcomes. 

\subsubsection{Gaofen-5 Shanghai HSI}

Gaofen-5 (GF-5) Shanghai HSI was captured by the Advanced Hyperspectral Imager (AHSI) and cropped into $300\times 300$ with 155 bands. The visual comparison is shown in Fig.~\ref{fig:shanghai_visual}. As shown in Fig.~\ref{fig:shanghai_noisy}, the original HSI includes heavy impulse noise and stripes. FastHyDe and E-3DTV achieve more promising denoising  compare to other model-based methods. Among the deep learning-based methods,  T3SC and MAC-Net retain stripes because the physical models are designed for Gaussian noise. Similar to the experiments with Houston 2018 HSI, the inability of TRQ3D and SST to model the complete GSC in its entirety resulted in a weaker denoising ability in 155 bands of the Shanghai HSI for them trained in a 31-band training set. Our proposed SSRT-UNet and NSSNN can recover the HSI by exploiting all the bands without splitting them and removing most of the stripes by using recurrent schemes to obtain GSC. Additionally, the proposed SSRT restores the HSI with sharper edges by exploiting NSS with the spatial branch. Fig.~\ref{fig:shanghai_pixel} shows the spectral reflectances of all the compared methods, where the proposed SSRT-UNet produces smoother curves that are in accordance with the original shape.

\subsubsection{Earth Observing-1 HSI}
Earth Observing-1 (EO-1) HSI was captured by the Hyperion with a size of $400 \times 1000 \times 242$ with a spectral range of  400 to 2500 nm. Following~\cite{Zhang2014a},  a sub-image with the size of $200\times400\times166$ is used for experiment. Fig.~\ref{fig:eo1_visual} provides a visual comparison of all the methods considered. Notably, the original HSI exhibits heavy deadlines and stripes.  E-3DTV removes most of the stripes, but the denoised HSI is blurred. T3SC and MAC-Net struggle to remove some stripes. Transformer-based methods TRQ3D and SST succeed in eliminating most stripes and deadlines but fall short in capturing complete GSC, resulting in blurred and color distortion. In contrast, SSRT-UNet and NSSNN offer superior recoveries, benefiting from their ability to fully exploit GSC. The spectral reflectances presented in Fig.~\ref{fig:eo1_pixel} show that our SSRT-UNet achieves a more continuous spectrum.

\subsection{Ablation Study}

In this section, we conduct experiments on the impact of some components and network width in our SSRT-UNet.  All the models are evaluated on the ICVL testing set  with noise variances $\sigma \in  [0,95]$.

\begin{table}[htbp]
   \centering
   \caption{Ablation Study on The Effect of Introduced Components.}
   \label{tab:ab_module}
   \begin{tabular}{ c  |c |c |c |c |c}
      \hline
   \makebox[0.05\textwidth][c]{MLP}&\makebox[0.05\textwidth][c]{Gates}&\makebox[0.05\textwidth][c]{Self-attns}&\makebox[0.03\textwidth][c]{PSNR$\uparrow$}&\makebox[0.03\textwidth][c]{SSIM$\uparrow$}&\makebox[0.03\textwidth][c]{SAM$\downarrow$}\\
   \hline
   \Checkmark&\Checkmark&\Checkmark&\textbf{46.20}&.9862&\textbf{.0375}\\
   \XSolidBrush&\Checkmark&\Checkmark&46.11&\textbf{.9865}&.0381\\
   \Checkmark&\XSolidBrush&\Checkmark&46.14&.9863&.0378\\
   \Checkmark&\XSolidBrush&\XSolidBrush&45.42&.9853&.0400\\
   \hline
\end{tabular}
\end{table}

\subsubsection{Effectiveness of Different Modules} Here, we perform an ablation study on three key components in our SSRT, namely, residual MLP (MLP), SRU gates (Gates), and self-attentions (Self-attns), to demonstrate their contributions to denoising. The results are presented in Table~\ref{tab:ab_module}. The residual MLP, being a crucial component in the transformer, is utilized for additional feature transformation. Consequently, the removal of MLP results in performance degradation. The inclusion of SRU gates in the SSRT enables the model to leverage long-range spectral correlation, leading to enhanced  recovery. Additionally, the self-attention modules in the SSRT-UNet facilitate the effective capture of non-local spatial correlations between pixels, further contributing to the improved performance.

    \begin{table}[htbp]
      \centering
      \caption{Study on The Impact of Network Width (Channels).}
      \label{tab:ab_width}
      \begin{tabular}{ c | c| c| c |c| c| c}
      \hline
      Channels&\#Param.&FLOPs&Time&PSNR$\uparrow$&SSIM$\uparrow$&SAM$\downarrow$\\
      \hline
      48 &10.2M&11.39T&32.80s&\textbf{46.20}&\textbf{.9862}&\textbf{.0375}\\
      36&5.8M&6.46T&27.89s&45.86&.9855&.0386\\
      24&2.6M&2.93T&25.37s&46.04&.9859&.0392\\
      12&0.6M&0.80T&22.14s&45.34&.9838&.0415\\
      \hline
      \end{tabular}
   \end{table}



\subsubsection{Impact of Network Width} The width of the network determines the number of parameters and has a significant impact on the network's representation ability. In our experiment, we tested our SSRT-UNet with different channel numbers: 12, 24, 36, and 48. The results in Table~\ref{tab:ab_width} show that increasing the width of the network generally improves denoising performance. However, this comes at the cost of increased computational resources, as indicated by the floating point operations (FLOPs) and running time.

  \begin{table}[htbp]
   \centering
   \caption{Comparison of Parameters, FLOPs and Running Time.}
   \label{tab:ab_width2}
   \begin{tabular}{ c | c| c| c |c| c| c}
   \hline
   Methods&\#Param.&FLOPs&Time&PSNR$\uparrow$&SSIM$\uparrow$&SAM$\downarrow$\\
   \hline
   T3SC     &0.83M  &-     &1.07s&43.10&.9734&.0747\\
   MAC-Net  &0.43M  &0.06T &3.44s&41.24&.9577&.0841\\
   NSSNN    &1.78M  &2.64T &2.06s&44.42&.9809&.0524\\
   TRQ3D    &0.68M  &1.07T &1.03s&43.54&.9806&.0523\\
   SST      &4.14M  &3.24T &4.58s&44.83&.9838&.0513\\
   \hline
   \end{tabular}
\end{table}

We additionally provide information on the FLOPs, running time, and number of parameters for alternative deep methods in Table~\ref{tab:ab_width2}. As evident from both Table~\ref{tab:ab_width} and Table~\ref{tab:ab_width2}, even with only 12 channels and 0.6M parameters, our method consistently outperforms alternative deep learning-based methods that demand significantly more parameters. It is noteworthy that our method exhibits higher complexity, as indicated by longer running time. It is important to highlight that, given the typical offline nature of denoising processes, the additional time invested in our method's higher complexity is generally justified by the resulting improved performance. 

\section{Conclusion}\label{sec:conclusion}
This paper introduces a spatial-spectral recurrent transformer  block-based  U-Net for HSI denoising. Associating the spatial branch and spectral branch with shared \emph{keys} and \emph{values}, the non-local spatial self-similarity (NSS) and global spectral correlation (GSC) can be simultaneously captured with a single spatial-spectral recurrent transformer  block.  Benefiting  from the transformer and RNN, the GSC can be more effectively modeled with the spectral branch of the SSRT beyond a fixed number of bands. Under the guidance of the GSC from the spectral branch, the spatial branch is capable of exploiting the NSS in each band to recover the HSI. Experimental results demonstrate that our proposed method achieves state-of-the-art performance, and the ablation study confirms the effectiveness of each component. In our future work, we plan to explore advanced methods to better capture and depict spectral correlations within a significantly larger number of bands.

		\appendices
		\bibliographystyle{IEEEtran}
		\bibliography{IEEEabrv,ssrt}
	\end{document}